\documentclass[superscriptaddress,twocolumn,showpacs,preprintnumbers,amsmath,amssymb,aps]{revtex4}
\usepackage{epsfig}
\pdfoutput=1

\usepackage{amsmath,amssymb,amsthm,amsfonts}
\usepackage{MnSymbol}
\usepackage{ mathrsfs }
\usepackage{slashed}
\usepackage{graphicx}
\usepackage{subfigure}
\usepackage{color,rotating}
\usepackage{dsfont}
\usepackage{setspace}
\usepackage{verbatim}
\usepackage{fancyhdr}
\usepackage{hyperref}
\newcommand{\beq}{\begin{equation}}
\newcommand{\eeq}{\end{equation}}
\newcommand{\beqa}{\begin{eqnarray}}
\newcommand{\eeqa}{\end{eqnarray}}
\newcommand{\bfc}{\begin{figure}[h!]\begin{center}}
\newcommand{\efc}{\end{center}\end{figure}}
\newcommand{\nn}{\nonumber}

\newcommand{\wrt}{with respect to }
\def\Fig#1{Fig.~\ref{#1}}

\graphicspath{{./figures/}}

\newcommand{\bea}{\begin{eqnarray}}
\newcommand{\eea}{\end{eqnarray}}
\newcommand{\Eq}[1]{Eq.~(\ref{#1})}

\newcommand*{\mn}{{\mu\nu}}

\def\eq#1{(\ref{#1})}
\def\Eq#1{Eq.~(\ref{#1})}
\newcommand {\apgt} {\ {\raise-.5ex\hbox{$\buildrel>\over\sim$}}\ }
\newcommand {\aplt} {\ {\raise-.5ex\hbox{$\buildrel<\over\sim$}}\ } 

\def\id{1\!\mbox{l}}
\def\s0#1#2{\mbox{\small{$ \frac{#1}{#2} $}}}
\def\0#1#2{\frac{#1}{#2}}

\def\dr{{D\!\llap{/}}\,}

\def\Cz{{\mathcal Z}}

\def\CP{{\mathcal P}}

\def\CV{{\mathcal V}}

\newcommand{\Tr}{\mathrm{Tr}}

\newcommand{\tr}{\mathrm{tr}}
\newcommand{\I}{\mathrm{i}}
\newcommand{\be}{\begin{eqnarray}}
\newcommand{\ee}{\end{eqnarray}}

\newcommand{\Nc}{N_{\rm{c}}}

\begin{document}
\title{Confinement from Correlation Functions} \vspace{1.5 true cm}
\author{Leonard Fister} 

\affiliation{Department for Mathematical
  Physics, National University of Ireland Maynooth, Maynooth, Ireland}
\affiliation{Institut f\"ur Theoretische Physik, Universit\"at
  Heidelberg, Philosophenweg 16, 69120 Heidelberg, Germany}
\author{Jan M. Pawlowski}
\affiliation{Institut f\"ur Theoretische
  Physik, Universit\"at Heidelberg, Philosophenweg 16, 69120
  Heidelberg, Germany} \affiliation{ExtreMe Matter Institute EMMI, GSI
  Helmholtzzentrum f\"ur Schwerionenforschung mbH, Planckstr. 1,
  D-64291 Darmstadt, Germany}

\begin{abstract}
  We compute the Polyakov loop potential in Yang--Mills theory from
  the fully dressed primitively divergent correlation functions
  only. This is done in a variety of functional approaches ranging
  from functional renormalisation group equations over
  Dyson--Schwinger equations to two-particle irreducible functionals.
  We present a confinement criterion that links the infrared behaviour
  of propagators and vertices to the Polyakov loop expectation
  value. The present work extends the works of
  \cite{Braun:2007bx,Marhauser:2008fz, Braun:2010cy} to general
  functional methods and sharpens the confinement criterion presented
  there. The computations are based on the thermal correlation
  functions in the Landau gauge calculated in
  \cite{Fister:2011uw,Fister:2011um, Fister:Diss}. 
\end{abstract}


\maketitle

\section{Introduction}
In recent years much progress has been made in our understanding of
the strongly-correlated low energy regime of QCD in terms of
gauge-fixed correlation functions, for reviews see e.g.
\cite{Alkofer:2000wg,Roberts:2000aa,Fischer:2006ub,Fischer:2008uz,%
  Binosi:2009qm,Pawlowski:2010ht,Maas:2011se,vonSmekal:2012vx}. This
progress is tightly linked to the advancement in our understanding of
the basic phenomena of low energy QCD, strong chiral symmetry breaking
and confinement. While chiral symmetry breaking allows for a simple
description in terms of the related order parameter, the chiral
condensate, and its effective potential, our understanding of the
confinement-deconfinement phase transition and the mechanism behind is
still less developed.

For static quarks with infinite masses, confinement can be thought of
in terms of the free energy of a single quark $F_{q}$. Removing the
antiquark to infinity in a colourless system with a quark-antiquark
pair requires an infinite amount of energy in a confined system. The 
corresponding free energy difference can be related to the free energy
of a single quark. Indeed, the gauge field part of such an operator is the
Polyakov loop $L$, 
\beq L=\frac{1}{\Nc} \tr_{\rm f}\, P(\vec x)\,,\quad {\rm with}\quad 
P(\vec x)= \CP\, e^{\I g \int_0^\beta dx_0\,
    {A}_0(x_0,x)}\,,
\label{eq:Polloop}
\eeq 
the trace in the fundamental representation of $SU(N_c)$ of the closed
Wilson line $P(\vec x)$ in time direction. Here, $\CP$ denotes path
ordering, $A$ is the gauge field and the inverse temperature
$\beta=1/T$. The related quark current comprises the worldline of a
static quark. The free energy of such a state is proportional to the
expectation value of $L$,
\begin{equation}\label{eq:expL}
\langle L \rangle \sim {\rm exp}^{-F_{q}/T}\,. 
\end{equation}
Hence, $\langle L\rangle$ is an order parameter for confinement: it is
strictly zero in the confined phase but non-zero in the deconfined
phase. This links the confinement-deconfinement phase transition in
the Yang--Mills system to the order-disorder phase transition of center
symmetry $Z_N$ in $SU(N_c)$: Under center transformations $z\in Z_N$ the
Polyakov loop transforms with $L\to z\, L$ in the fundamental
representation. We conclude that in the center-symmetric, confining
phase we have $\langle L\rangle=0$, while in the center-broken,
deconfined phase we have $\langle L\rangle\neq 0$.

In \cite{Braun:2007bx,Marhauser:2008fz} it has been shown that also
$L[\langle A_0\rangle] \geq \langle L[A_0] \rangle$ for constant
fields is an order parameter for static quark confinement. Note that
the expectation value $\langle g\beta\, A_0\rangle$ relates to the
eigenvalues of $\phi$ with $P(\vec x)=\exp i\phi$ and is
gauge-invariant and gauge-independent. 

The order parameter $L[\langle A_0\rangle]$ has the advantageous
property that its full effective potential $V[A_0]$ can be computed
straightforwardly with functional continuum methods. Within the
functional renormalisation group (FRG) approach it has been shown
\cite{Braun:2007bx, Marhauser:2008fz,Braun:2009gm}, that the
computation of $V[A_0]$ has a closed representation in terms of the full
propagators of gluon, ghosts (and quarks) in constant
$A_0$-backgrounds.  This link of the propagators to the Polyakov loop
potential also allowed to put forward a confinement criterion for the
infrared behaviour of the ghost and gluon propagators
\cite{Braun:2007bx}. 

As similar link of confinement to the infrared behaviour of
(gauge-fixed) correlation functions has been put forward more recently
with dual order parameters, \cite{Gattringer:2006ci,%
  Synatschke:2007bz, Bilgici:2008qy,Fischer:2009wc,Fischer:2009gk,%
  Braun:2009gm,Fischer:2011mz,Fischer:2012vc,Hopfer:2012qr}. The
latter class of order parameters is directly sensitive to spectral
properties of the Dirac operator and hence is tightly linked to quark
correlation functions. The former one, $L[\langle A_0\rangle]$, is
directly sensitive to the gluon and ghost correlation functions as
well as to the quarks. Nonetheless, there is a close relation between
the two classes of order parameters, which has been discussed in
\cite{Braun:2009gm} for fully dynamical two-flavour QCD at finite
temperature. There it has been shown that the two order parameters
concur for the case of the dual density or dual pressure.

By now $\langle L[A_0] \rangle$ and the effective Polyakov loop
potential $V[A_0]$ has been computed in Yang--Mills theory in the
Landau gauge \cite{Braun:2007bx,Braun:2010cy,Eichhorn:2011gc}, the
Polyakov gauge \cite{Marhauser:2008fz} and in the Coulomb gauge
\cite{Reinhardt:2012qe}. A comparison of $\langle g\beta\, A_0\rangle$
in different gauges has been made in \cite{Marhauser:2008fz} which
confirms the formal results of gauge independence. More recently,
there have been also first computations of the Polyakov loop potential
in Yang-Mills theory on the lattice, see
\cite{Diakonov:2012dx,Greensite:2012dy}. In \cite{Braun:2009gm} the
Yang--Mills studies have been extended to fully dynamical two-flavour
QCD. In \cite{Fukushima:2012qa,Kashiwa:2012td} the Polyakov loop
potential $V[A_0]$ is used in an effective model approach to QCD with
interesting applications to thermodynamic observables.

For a quantitative computation of the Polyakov loop potential, the
temperature dependence of the order parameter and in particular the
critical temperature a good grip on the thermal ghost and gluon
propagtors is required. In Landau gauge they have been computed on the
lattice, \cite{Cucchieri:2007ta, Fischer:2010fx, Maas:2011ez,
  Bornyakov:2011jm,Cucchieri:2011di,%
  Maas:2011se, Aouane:2011fv,Aouane:2012bk}, and in the continuum,
\cite{Fister:2011uw,Fister:2011um, Fister:Diss} with FRG-methods,
extending previous studies in extreme temperature limits in the Dyson--Schwinger
framework \cite{Maas:2004se, Gruter:2004bb, Maas:2005hs, Cucchieri:2007ta}.

In the present work we compute the Polyakov loop effective potential
$V[A_0]$ in the background field formalism \cite{Abbott:1980hw} in
Landau--deWitt gauge within different functional approaches. In
Section~\ref{sec:funPolPot} we derive representations for the Polyakov
loop potential $V[A_0]$ within the FRG-approach, Dyson-Schwinger
equations (DSEs) and the two-particle irreducible (2PI-) effective action. In
Section~\ref{sec:confcrit} we extend and sharpen the confinement
criterion of \cite{Braun:2007bx} in terms of the propagators: Infrared
suppression of gluons but non-suppression of ghosts suffices to
confine static quarks. In Section~\ref{sec:matter} the criterion is
applied to general Yang--Mills-matter systems. In
Section~\ref{sec:resultsPolPot} the Polyakov loop potential is computed
in Yang--Mills theory on the basis of the finite temperature
propagators computed within the FRG in
\cite{Fister:2011uw,Fister:2011um, Fister:Diss}. The results for the
different functional methods are in quantitative agreement in the
confinement-deconfinement regime.

\section{Polyakov Loop Potential from Functional
  Methods}\label{sec:funPolPot}
In this section we discuss different representations of the Polyakov
loop potential derived from the FRG, DSEs and
2PI-functionals. The Polyakov loop potential $V$ is simply the free
energy, or one-particle irreducible (1PI) effective action $\Gamma$,
evaluated on constant backgrounds $A_0$.

\subsection{Polyakov Loop Potential}
Most functional approaches are based on closed expressions for the
effective action or derivatives thereof in terms of full correlation
functions. Hence, the knowledge of the latter in constant
$A_0$-backgrounds allows us to compute the Polyakov loop potential
$V[A_0]$. In turn, confinement requires the Polyakov loop potential to
have minima at the confining values for $A_0$. In $SU(\Nc)$ these are
the center-symmetric points. This restricts the infrared behaviour of
the correlation functions computed in the constant $A_{0}$-background.  

Gauge covariance of the correlation functions and gauge invariance of
the effective action and, hence, the effective potential is achieved
within the background field approach \cite{Abbott:1980hw}. We split
the gauge field $A$ in a background $\bar A$ and fluctuations $a$
about the background, $A=\bar A+a\,$.  This split allows us to define
background field-dependent gauges that transform covariantly under
gauge transformations of both the background and the full gauge
field,
\begin{equation}\label{eq:backgauge}
  \bar D_\mu a_\mu=0\,,\qquad {\rm with}\qquad 
D_\mu(A)=\partial_\mu -i\,g\,A_\mu\,,
\end{equation}
and $\bar D=D(\bar A)$. As a consequence all correlation functions
transform covariantly under combined gauge transformations of $A$ and
$\bar A$. Hence, the effective action $\Gamma$ is invariant under
combined gauge transformations.  However, due to the gauge
\eq{eq:backgauge} it now depends on the full gauge field $A$ and the
background field $\bar A$ separately, $\Gamma=\Gamma[\bar A;a,c,\bar
c]$, where $c,\bar{c}$ are the Faddeev-Popov ghosts. The path integral
representation is in terms of field-multiplet $\varphi=(a,c,\bar c)$
and their expectation values $\phi=\langle\varphi\rangle$,
\begin{equation}\label{eq:backPI}
  e^{-\Gamma[\bar A;\phi]}=\int \mathscr{D} \varphi\,\exp\left\{
    -S_A[\bar A;\varphi]+
    \int_x \0{\delta\Gamma}{\delta\phi}(\varphi-\phi)\right\}\,.
\end{equation}
The classical action is given by 
\beqa
S_A[\bar A;\phi]&=&\014 \int_x F_\mn^a F_\mn^a\nn  \\[1ex]
&& +\01{2 \xi}\int_x\left(\bar D^{ab}_\mu a_\mu^b\right)^2+ \int_x\bar
c^a\bar D_\mu D_\mu^{ab} c^b \,,
\label{eq:fixedaction} 
\eeqa 
where $F_{\mn}^a$ is the field strength tensor, $\xi$ the gauge fixing
parameter and the abbreviation $\int_x=\int_0^\beta d x_0\int d^3
x$. If we now identify the background $\bar A$ with the physical
background $A$, the expectation value of the field, we arrive at a
gauge invariant effective action
\begin{equation}\label{eq:gaugeinvG}
\Gamma[A,c,\bar c]=\Gamma[A; 0,c,\bar c]\,.
\end{equation}
The Polyakov loop potential is given by \eq{eq:gaugeinvG} evaluated on
a constant $A_0$-background, $A_0 = A_{\mu} \delta_{\mu 0}$,
\begin{equation}\label{eq:Vpol}
V[A_0^{\rm const}]:=\0{1}{\beta \rm \CV} \Gamma[A^{\rm const}_0;0]\,,
\end{equation}
where $\CV$ is the three-dimensional spatial volume. The Polyakov
loop, \eq{eq:Polloop}, is then evaluated at the minimum $\langle
A_0\rangle:=A_{0,\rm min}$. It has been proven in \cite{Braun:2007bx}
and \cite{Marhauser:2008fz} that \eq{eq:Polloop} evaluated on the
minimum of \eq{eq:Vpol} is an order parameter such as $\langle
L[A_0]\rangle \leq L[\langle A_0\rangle]$.

Functional equations for the effective action can be derived from the
FRG, DSEs and 2PI equations. All those equations depend on the
correlation functions $\Gamma^{(n)}$ of fluctuation fields $a$ only,
schematically given by
\begin{equation}\label{eq:Gamman} \Gamma^{(n)}[\bar A](p_1,...,p_n)
  =\left.\0{\delta^n
      \Gamma[\bar A;a]}{\delta a(p_1)\cdots 
      \delta a(p_n)}\right|_{a=0}\,,
\end{equation}
where we have suppressed the ghosts and the internal and Lorentz
indices. In \cite{Braun:2007bx} we have argued that the correlation
functions in the background Landau gauge, $\Gamma^{(n)}[\bar A]$, are
directly related to those in Landau gauge, $\Gamma^{(n)}[0]$.  This
allows us to use the latter correlation functions within the
computation of the effective potential. Here, we recall the argument
given in \cite{Braun:2007bx} for the ghost and gluon two-point
functions, $\Gamma^{(2)}_c, \Gamma^{(2)}_a$, which straightforwardly
extends to higher correlation functions:

Gauge covariance of the fluctuation field correlation functions which
constrains the difference between $\Gamma^{(n)}[0]$ and
$\Gamma^{(n)}[\bar A]$. At vanishing temperature the gluon two-point
function in Landau gauge splits into four-dimensionally transversal
and longitudinal parts with the projection operators
\begin{eqnarray}
  \Pi^{\bot}_{\mu \nu}(p)=\delta_{\mu \nu}-p_{\mu}p_{\nu}/p^2\,,\quad
  \Pi^{\parallel}_{\mu \nu}(p)=p_{\mu}p_{\nu}/p^2 \,.
  \label{eq:projections0} 
\end{eqnarray}
Hence, the gluon propagator is transversal for all cut-off scales $k$
even though the longitudinal part of the inverse gluon propagator,
$\Gamma^{(2)}_A$ receives finite corrections as well.

At non-vanishing temperature we have to take into account
chromomagnetic  and chromoelectric modes via the respective projection
operators $P^T$ and $P^L$,
\begin{eqnarray}
   P^{T}_{\mu \nu}(p_0, \vec{p}) &=& \left(1-\delta_{\mu 0} \right)
\left(1-\delta_{\nu 0} \right) \left( \delta_{\mu \nu}- p_{\mu}p_{\nu}/\vec{p}^{\ 2} \right),
  \nn\\
  P^{L}_{\mu \nu}(p_0, \vec{p}) &=& \Pi^{\bot}_{\mu \nu}(p)-P^{T}_{\mu \nu}(p_0, \vec{p})\,, 
  \label{eq:projections} 
\end{eqnarray}
where $\Pi^{\bot}_{\mu\nu}$ is the four-dimensional transversal
projection operator, see \eq{eq:projections0}. The parameterisation of
the gluon and ghost two-point functions in Landau gauge, i.e. the
chromoelectric/chromomagnetic gluon $\Gamma_{L/T}^{(2)}$ and the ghost
$\Gamma_{c}^{(2)}$, is then given by,
\cite{Fister:2011uw,Fister:2011um, Fister:Diss},
\begin{eqnarray} 
  \Gamma_{L}^{(2)}(p_0,\vec{p}) &=&
  Z_{L}(p_0^2, \vec p{\,}^2)\, p^2\,P^{L}(p_0,
  \vec{p})\,, \nn\\
  \Gamma_{T}^{(2)}(p_0,\vec{p}) &=& Z_{T}(p_0^2, \vec{p}{\,}^2)\, p^2\,P^{T}(p_0,
  \vec{p})\,,
  \nn \\
  \Gamma_{c}^{(2)}(p_0^2,\vec{p})&= & Z_c(p_0, \vec{p}{\,}^2)\,p^2\,, 
 \label{eq:parahatG}\end{eqnarray}
where the identity in colour space is suppressed and the wave function
renormalisations $Z$ are functions of $p_0$ and $\vec p$
separately. We now parameterise the background field correlation
functions in terms of the Landau gauge correlation functions in
\eq{eq:parahatG} evaluated at covariant momenta. For the gluon
$\Gamma^{(2)}_a$ this gives
\beqa
\nn 
\Gamma_a^{(2)}[A;\phi=0]&=&
\sum_{L/T}
P^{L/T}\  (-D^2) Z_{L/T}\, P^{L/T}
\\[1ex]&& 
+ F^{cd}_{\rho\sigma} f^{cd}_{\rho\sigma}(D)+ \Delta
m^2( D,A_0)\,,
\label{eq:genprop}
\eeqa 
with non-singular $f(0)$, and where the arguments with respect to the
covariant momentum of the projection operators onto the longitudinal
and transversal spaces $P_{\rm L/T}(-D_0,-\vec{D})$, respectively, and
of $Z_{L/T}(-D_0^2,-\vec{D}^2)$ have been omitted for clarity. Note
that the projection operators $P_{L/T}$ do not commute with $Z_{L/T}$
for general gauge fields. They do, however, for constant gauge fields
$A_0$. The $f$-term can not be obtained alone from the Landau-gauge
propagator, but is also related to higher Green functions.  However,
it does not play a r$\hat {\rm o}$le for our purpose. 

In the computations below we approximate the full inverse gluon
propagators by the first line in \eq{eq:genprop}. Similarly the inverse ghost 
propagator is approximated for constant temporal background $A_0$ as 
\begin{equation}\label{eq:approxghost}
\Gamma^{(2)}_c[A_0;\phi=0] \simeq (-D^2) Z_c(D^2) \,. 
\end{equation}
For these backgrounds no $f$-term as introduced in \eq{eq:genprop} is
present.  Note also that a mass term $\Delta m_c^2(0,A_0)$ is
kinematically forbidden for the ghost. Hence it can only contribute
for momenta larger than zero. As it is subject to the standard thermal
decay we neglect it as sub-leading.

\subsection{Thermal Corrections and Critical Scaling}\label{sec:critical}
Here we discuss in detail the impact of the neglected thermal
corrections $\Delta m^2$. They play a crucial r$\hat {\rm o}$le for the
correct critical scaling and the value of $T_c$ in the $SU(2)$-case
but is sub-leading in the $SU(3)$ case. This section might be skipped
in a first reading as the following results can be understood without
it.

We know that $\Delta m^2(D,A_0)$ vanishes at $A_0=0$ or $T=0$.
Moreover, the first two terms on the rhs in \eq{eq:genprop}
parameterise all terms in $\Gamma_a^{(2)}$ that only depend on the
covariant operator $D$. At finite temperature, however, the Polyakov
loop $L$ is a further invariant, i.e. the Polyakov line, $P(\vec x)$,
cf. \eq{eq:Polloop}, transforms covariantly under gauge
transformations. These terms are particularly important for the
chromoelectric $00$ component of the gluon two-point function
\eq{eq:genprop} as they depend on $A_0$. Moreover, these terms are not
covered by the Landau gauge term in \eq{eq:genprop} as the related
variable is not present for $T=0$. In addition, $\Delta m^2(D,0)$ has
the standard thermal decay for large momenta $p^2$. Hence we shall
only discuss it for the low momentum regime, that is $\Delta
m^2(0,A_0)$. Note also that \eq{eq:Polloop} is invariant under the
periodic gauge transformation $U(t)$ with
\begin{equation}\label{eq:gaugtrafotau} 
  U(t)=\exp{i2 \pi  N_c \tau_i t}\,,\qquad A_0\to A_0 + 
  \frac{2 \pi}{\beta g} N_c \tau^{i} \,,  
\end{equation}
with $\tau^{ i}$ being a generator of the Cartan subalgebra of the
respective gauge group, and $A_0$ in the Cartan subalgebra. This
entails that $\Delta m^2(0,A_0)$ is periodic under a shift of $A_0$ in
\eq{eq:gaugtrafotau} as is the Polyakov loop potential $V[A_0]$ in
\eq{eq:Vpol}. 

Moreover, the longitudinal correction $\Delta m_L^2(0,A_0)$ is derived
from the Polyakov loop potential directly. First we notice that
\begin{eqnarray}
  \Delta m_L^2(0,A_0) &=& \partial_{A_0}^2 V[A_0] -\partial_{A_0}^2 V[0]+ 
  \delta m_L^2\,,
\label{eq:Deltam}
\end{eqnarray}
where $A_0$ is the background field in a slight abuse of notation, and
$\delta m_L^2$ takes care of differences between derivatives with
respect to the background $A_0$ and the fluctuation field $a_0$ in
$\Delta m^2_L$. This term in \eq{eq:Deltam} follows from the Nielsen
identity in the background field formalism, \cite{Nielsen:1975fs}, in
the present context see \cite{Pawlowski:2001df,Litim:2002ce,%
  Pawlowski:2003sk,Pawlowski:2005xe}.  This identity reads
\begin{equation}\label{eq:Nielsen} 
(\partial_{\bar A_0}- \partial_{a_0}) \Gamma=
\012 \Tr\,  \01{\bar D^2}\, \partial_{\bar A_0}  \bar D^2  - \012 \Tr\,  G_c
\, \partial_{\bar A_0}  \bar D_\mu D_\mu  \,, 
\end{equation}
where both traces only sum over momenta and gauge group indices.  The
first term in \eq{eq:Nielsen} originates in the gauge fixing term, the
second one in the ghost term. Indeed the first term on the right hand
side is one loop exact, the second term is solely driven by the
ghost. Note also that evaluated at $a=0$, \eq{eq:Nielsen} is only
non-zero beyond one loop. Applying a further derivative
$\partial_{a_0}+ \partial_{\bar A_0}$ to \eq{eq:Nielsen} and
evaluating it at $a=0$ and $D=0$ relates it to the term $\delta m_L^2$
in \eq{eq:Deltam}.  The Nielsen identity \eq{eq:Nielsen} accounts for
the different RG-scaling of fluctuation propagators and background
propagators and higher correlation functions, as is well-known from
perturbative applications. We infer that the projection of
\eq{eq:Nielsen} on $\Delta m_L^2$ guarantees the correct RG-scaling
for the second derivative terms of the Polyakov loop potential in
\eq{eq:Deltam}. This is taken into account by applying the appropriate
RG-rescaling, $z_a/z_A$ to the first and second term on the rhs of
\eq{eq:Deltam}. Here, $z_a$ and $z_A$ are the renormalisation factors
of fluctuation field and background field, respectively. Higher order
corrections related to the momentum dependence of the RG-scaling which
we have neglected in the present discussion due to its thermal decay.

In summary we can estimate $\Delta
m_L^2$ on the basis of the Polyakov loop potential as
\begin{equation}\label{eq:00estimate}
  \Delta m^2_L\simeq \frac{z_a}{z_A}
  \left(\partial^2_{A_0} V[A_0]-
\partial^2_{A_0} V[0]\right)\,. 
\end{equation} 
\Eq{eq:00estimate} has the correct periodicity properties and the
correct limits. Moreover, it entails that the electric propagator,
$(1/\Gamma^{(2)})_L^{\ }$, carries critical scaling, see also
\cite{Maas:2011ez}. Note that the latter property does not depend on
the present approximation. This also indicates that the electric
propagator is enhanced for temperatures $T\lesssim T_c$ as found on
the lattice, \cite{Cucchieri:2007ta, Fischer:2010fx, Maas:2011ez,
  Bornyakov:2011jm,Cucchieri:2011di,%
  Maas:2011se, Aouane:2011fv,Aouane:2012bk}. A careful analysis of the
analytic consequences of \eq{eq:Nielsen} and \eq{eq:00estimate} will
be presented elsewhere. It is left to estimate $\Delta m_T^2$ which
also has to be proportional to ($A_0$-derivatives of) the Polyakov
loop potential.  In the transversal propagators such terms can only
occur together with (covariant) momentum dependencies or powers of the
field strength.  The latter vanishes for constant temporal backgrounds
while the former is thermally suppressed. We conclude that $\Delta
m^2_T\simeq 0$.

We are now in the position to estimate the impact of $\Delta m^2_L$
for the $SU(2)$ and $SU(3)$-computations: It is the electric
propagator which directly depends on the Polyakov loop potential. 
Moreover, this correction is the only term which is directly sensitive to
center symmetry. The standard approximations used in functional
methods are based on field expansions about vanishing fields and hence
are only sensitive to the $su(N_c)$-algebra.  In $SU(2)$ we expect a
second order phase transition. Then the critical temperature as well
as the correct critical scaling (Ising universality class) are
sensitive to the omission of the back-reaction. In the present context
this entails that we expect mean-field critical exponents as well as a
lowered critical temperature. In $SU(3)$ we expect a first order phase
transition and the back-reaction of the Polyakov loop potential is not
important for the value of the critical temperature. It should have a
an impact on the jump of the order parameter which should be increased
in the present approximation.

Both expectations are satisfied by the explicit results presented in
Section~\ref{sec:resultsPolPot}. It has been also confirmed within the
Polyakov gauge that the inclusion of the back-reaction leads to a
quantitatively correct critical temperature as well as the expected
critical scaling of the Ising universality class for $SU(2)$, see
\cite{Marhauser:2008fz}. This has been also confirmed within the
Landau gauge, see \cite{Spallek}.

\subsection{Flow Equation for the Polyakov Loop Potential}
\begin{figure}[t]
\includegraphics[width=.75\columnwidth]{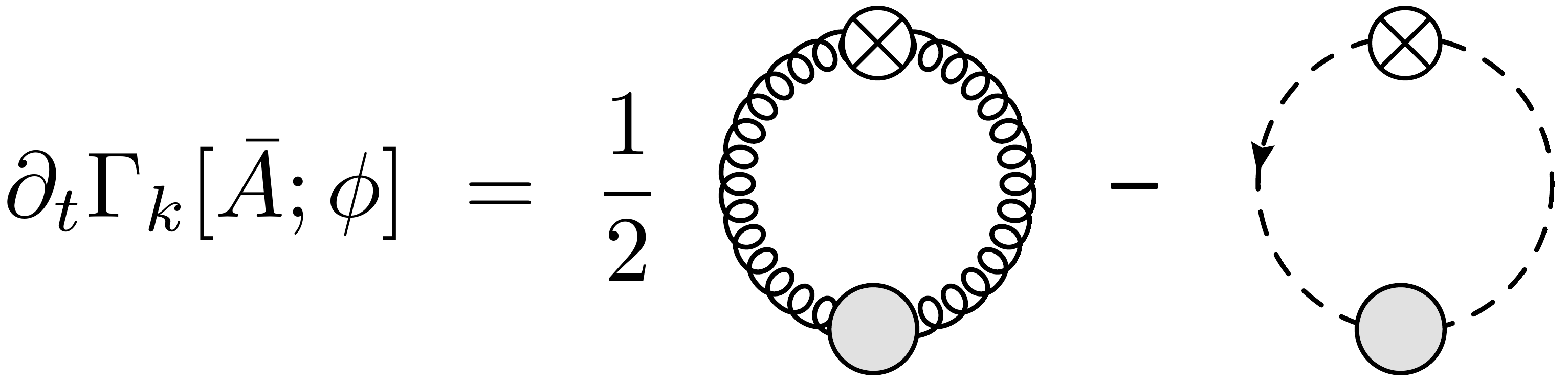}
\caption{Functional flow for the effective action. Lines with filled
  circles denote fully dressed field-dependent propagators
  \eq{eq:G}. Crossed circles denote the regulator insertion
  $\partial_t R_k$. }
\label{fig:funflow}
\end{figure}

We begin with the FRG representation, which has also been used in
\cite{Braun:2007bx, Marhauser:2008fz, Braun:2010cy, Eichhorn:2011gc},
for QCD-related reviews see e.g.Ê\cite{Litim:1998nf,Berges:2000ew,%
Polonyi:2001se, Pawlowski:2005xe, Gies:2006wv, Schaefer:2006sr,%
Igarashi:2009tj,Rosten:2010vm, Braun:2011pp}. We write the flow
equation for the Yang--Mills effective action, $\Gamma_k[\bar
A;\phi]$, at finite temperature $T$ as \beqa \nn
  \partial_t \Gamma_{k}[\bar A;\phi] &=\! & 
  \frac{1}{2} \sumint_p\ \left(G_a\right)^{ab}_{\mu\nu}[\bar A;\phi](p,p)
  \, {\partial_t} \left(R_a\right)_{\nu\mu}^{ba}(p)\\ &&-
  \sumint_p\ \left(G_{c}\right)^{ab}[\bar A;\phi](p,p)
  \, {\partial_t} \left(R_{c}\right)^{ba}(p)\,.\nn\\
\label{eq:funflow}
\eeqa 
where the integration involves the respective gluon and ghost
modes. Further, $t=\ln k$, with $k$ being the infrared cut-off
scale. The diagrammatic representation given in \Fig{fig:funflow}.
%
%

The momentum integration measure at finite temperature is given by
\begin{eqnarray}\label{eq:matsubaras}
\sumint_p=T \sum_{n\in \mathbb{Z}}\int \0{d^3 p}{(2 \pi)^3}\,,
\quad {\rm with}\quad  p_0=2 \pi T n\,,
\end{eqnarray} 
where the integration over $p_0$ turns into a sum over Matsubara
frequencies $n$. Both, gluons and ghosts have periodic boundary
conditions, $\phi(x_0+1/T,\vec x)=\phi(x_0,\vec x)$, which is
reflected in the Matsubara modes $2 \pi T n$ with a thermal zero mode
for $n=0$.  Naturally, at vanishing temperature we have $\sumint_p\to
(2\pi)^{-4}\!\!\int\!\! d^4p$.  The full field-dependent propagator
for a propagation from the fluctuation $\phi_1$ to $\phi_2$ is given
by
\begin{eqnarray}\label{eq:G}
  G_{\phi_1\phi_2}[\bar A;\phi](p,q)=\left(\0{1}{\Gamma_{k}^{(2)}[\phi]
      +R_k}\right)_{\phi_1\phi_2}(p,q)\,. 
\end{eqnarray} 
In short, we use
\beq
\left(G_a\right)_{\mn}^{bc}=G_{a_{\mu}^b a_{\nu}^c}\,, \   {\rm and} \  \left(G_{c}\right)^{ab}=G_{c^a \bar c^b}
\eeq
for the gluon and the ghost
propagator, respectively. In \eq{eq:G} we also introduced the regulator function in field
space, $R_{k,\phi_1\phi_2}$, with
\begin{eqnarray}\label{eq:Reg}
\left(R_{a}\right)_{\mu\nu}^{bc} = R_{k,a_\mu^b a_\nu^c}\,,  \quad  
\left(R_{c}\right)^{ab} = R_{k,\bar c^a
  c^b}=-R_{k, c^b
  \bar c^a}\,. 
\end{eqnarray} 
The above entails that \eq{eq:funflow} only depends on the
propagators of the fluctuations $\phi$ evaluated in a given background
$\bar A$. This also holds for the flow of the background effective
action $\Gamma_k[\bar A;a,c,\bar c]$. For more details we refer the reader to
\cite{Fister:2011uw,Fister:2011um, Fister:Diss} and 
Appendix~\ref{app:regs}.  

The effective Polyakov loop potential $V[A_0]$ is given by 
\begin{equation}\label{eq:flowV}
V[A_0]=V_\Lambda[A_0]+\0{1}{\beta \rm \CV} \int_\Lambda^0 dt\,
\partial_t \Gamma_k[A_0]\,,
\end{equation}
where $V'_\Lambda[A_0]\propto e^{-\Lambda/T} \to 0$ for sufficiently
large $\Lambda/T\gg 1$ and sufficiently smooth regulators, see
\cite{Fister:2011uw,Fister:2011um, Fister:Diss}.  We conclude that the
computation of $V[A_0]$ with FRG-flows only requires the knowledge of
the (scale-dependent) propagators $G_{a/c}$, see
\cite{Braun:2007bx,Braun:2010cy}.

\subsection{DSE for the Polyakov Loop Potential}\label{sec:DSEPolPot}
\begin{figure}[t]
\begin{center}
\includegraphics[width=.95\columnwidth]{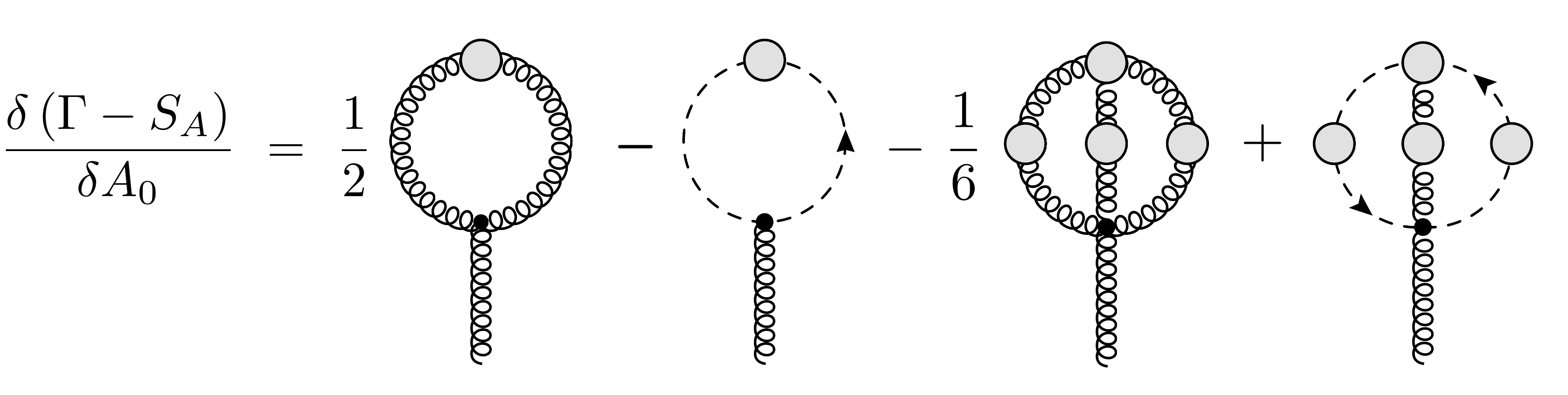}
\caption{DSE for the background gluon one-point function.}
\label{fig:DSE}
\end{center}
\end{figure}

The DSE for Yang--Mills theory relevant for the
Polyakov loop potential is that originating in a derivative of the
effective action \wrt the $A_0$-background at fixed fluctuation. It
can be written in terms of renormalised full propagators and vertices
and the renormalised classical action. The latter is written as 
\begin{equation}\label{eq:renS_A}
S_{A,\rm ren}[\bar A;\phi]=S_{A}[z_{ A}^{1/2}\bar A;z_\phi^{1/2}\phi; z_g g, 
1/z_\xi \xi]\,,
\end{equation}
with finite wave functions, coupling and gauge parameter
renormalisation $z_{A}, z_\phi, z_g, z_\xi$ and renormalised fields
$\bar A,\phi$, coupling $g$ and gauge fixing parameter $\xi$. We have
$z_\xi z_a=1$ due to the non-renormalisation of the gauge fixing
term. Moreover, background gauge invariance of the background field
effective action \eq{eq:gaugeinvG},
\begin{equation}\label{eq:backgaugetrafo} 
D_\mu^{ab}\0{\delta \Gamma[A;\phi=0]}{\delta A_\mu^b}=0\,,
\end{equation}
and non-renormalisation of the ghost-gluon vertex 
\cite{Taylor:1971ff, Lerche:2002ep} leads to
\begin{equation}\label{eq:ZgZbarA}
z_g z_{ A}^{1/2}=1\, \quad {\rm and} \quad z_g z_a^{1/2} z_c=1\,. 
\end{equation} 
The above arguments yield the DSE for the Polyakov loop potential, 
schematically written as
\beqa \nn
&&\hspace{-.3cm}   \0{\delta(\Gamma[A_0;0]-S_A[A_0;0])}{\delta A_0(x)}=
\012 S^{(3)}_{A_0 aa} G_{a} -
S^{(3)}_{A_0 c\bar c} G_{c}\\[1ex]
&& -\016 S^{(4)}_{A_0 aaa} G_{a}^3 \Gamma^{(3)}_{aaa} + 
S^{(4)}_{A_0  a c\bar c } G_{c}^2 G_{a} \Gamma^{(3)}_{ac \bar c}\,,
\label{eq:DSE}
\eeqa 
and the diagrammatic representation given in \Fig{fig:DSE}. 
In \eq{eq:DSE} and \Fig{fig:DSE} we have used the renormalised
classical vertices
\begin{equation}\label{eq:backvert1}
  S^{(3)}_{A_0 aa} =z_a \0{\delta^3 S_A}{\delta  A_0 \delta a^2}\,,
\quad S^{(3)}_{A_0 c\bar c} =z_c \0{\delta^3 S_A}{\delta A_0 \delta c\delta \bar c}\,, 
\end{equation}
at $\phi=0$ and 
\begin{equation}\label{eq:backvert2}
  S^{(4)}_{A_0 aaa} =z_g z_a^{3/2}\0{\delta^4 S_A}{\delta A_0 \delta a^3}\,, \quad  
S^{(4)}_{A_0 a c\bar c} =\0{\delta^4 S_A}{\delta A_0 \delta a\delta c\delta \bar c}\,, 
\end{equation}
due to \eq{eq:ZgZbarA}. The mixed three-point vertices in
\eq{eq:backvert1} have one background leg and two fluctuation legs and
differ from the standard vertices: they are $A_0$-derivatives of the
related fluctuation-field propagator. Using $z_g z_{A}^{1/2}=1$ from
\eq{eq:ZgZbarA} it follows that the vertices in \eq{eq:backvert1} have
renormalisation group properties of fluctuation two-point functions,
signaled by $z_a$ and $z_c$, respectively. The two-point function
property also entails that the gluonic vertex contains a piece from
the gauge fixing term proportional to $1/\xi$ and the ghost-gluon
vertex also involves the gauge field derivative of $\bar D$ and not
only $D$.

The mixed four-point vertices in \eq{eq:backvert2} have one background
leg and three fluctuation legs and have the renormalisation group
properties of the related fluctuation three-point functions multiplied
by $z_g$, signaled by $ z_g z_a^{3/2}$ and $1$, respectively. The
latter follows from $ z_g z_a^{1/2} z_c=1$ in \eq{eq:ZgZbarA}. The
ghost-gluon four-point function stems from the background field
dependence of the Faddeev-Popov operator $-\bar D_\mu D_\mu$, and causes 
an additional two-loop term in the DSE not present in the pure fluctuation DSE. 

At
asymptotically large momenta the finite wave function renormalisation
$z_\phi, z_A$ are related to the normalisation of the two point
functions with 
\begin{equation}\label{eq:renormZ} 
z_\phi=\0{\Gamma^{(2)}_{\phi\phi}(\mu^2)}{\mu^2} 
=Z_\phi(\mu^2)\,. 
\end{equation} 
Effectively this would amount to using propagators $z_\phi
G_{\phi\phi}(p^2)$ with
\begin{equation}\label{eq:asmyptprops}
p^2\, z_\phi  G_{\phi}(p^2\to \infty)\to 1\,. 
\end{equation}
Applying the DSE \eq{eq:DSE} to constant $A_0$ backgrounds we are
finally lead to the DSE equation for the effective Polyakov loop
potential with 
\begin{equation}\label{eq:DSEPolPot}
\0{\partial V[A_0]}{\partial A_0}=
\0{1}{\beta\CV}  \0{\partial \Gamma[A_0;0]}{\partial A_0}\,,
\end{equation}
with \eq{eq:DSE} or \Fig{fig:DSE} for the right hand side of \eq{eq:DSEPolPot}. This
equation is similar to the standard DSE for the fluctuation fields
with the exception of the two-loop ghost contribution, only the
vertices differ. In particular there is a contribution from the gauge
fixing vertex which gives a perturbative one-loop contribution to the
potential. The external field is a background gluon field with only a
temporal component $A_0$. This is reflected in the projections, see
Appendix~\ref{app:one-loop}. We conclude that the computation of
$V[A_0]$ only requires the knowledge of the propagators and of the
full three-point functions $\Gamma^{(3)}_{aaa}$ and
$\Gamma^{(3)}_{ac\bar c}$ in a constant background.

\subsection{2PI-Representation for the Polyakov Loop Potential}
\begin{figure}[t]
\includegraphics[width=.98\columnwidth]{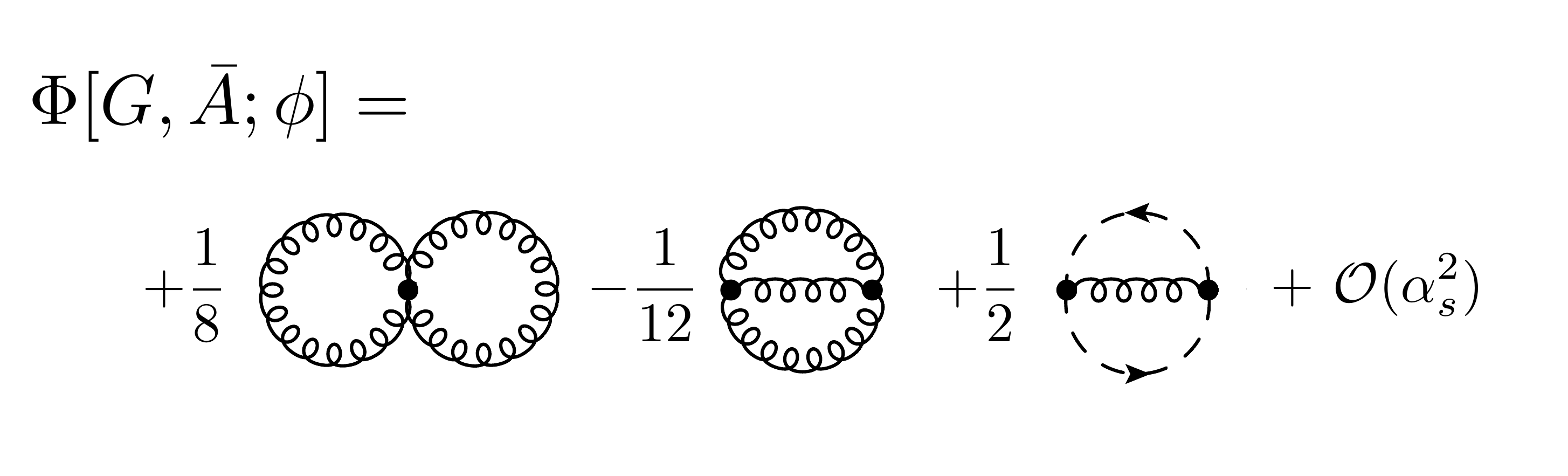}
\caption{Two-loop diagrams of the 2PI effective action. All
  propagators are dressed, the full circles denote bare
  vertices.}\label{fig:Phi}
\end{figure}

Now we extend our discussion to the 2PI approach, 
\cite{Luttinger:1960ua, Lee:1960zza, Baym:1962sx, deDominicis:1964zz, %
Cornwall:1974vz}, for applications to gauge theories see e.g.\
\cite{Blaizot:1999ip, Blaizot:1999ap, %
Blaizot:2000fc,Braaten:2001vr,Arrizabalaga:2002hn,Mottola:2003vx,%
Carrington:2003ut,Berges:2004pu,Andersen:2004re,Carrington:2007fp,%
Reinosa:2007vi, Reinosa:2008tv,Reinosa:2009tc}  
and the relation to the FRG-approach, see 
\cite{Blaizot:2010zx,Carrington:2012ea}.
Its application will simply lead to a
convenient resummation scheme for the DSE-equation for the Polyakov
loop potential \eq{eq:DSEPolPot} derived in the last chapter: The
functional DSE for the effective action displayed in \eq{eq:DSE} and
\Fig{fig:DSE} follows from the 2PI generating functional
\beqa\nn \Gamma_{\rm 2PI}[G,\bar A;\phi]&=& S_A[\bar A;\phi]-\012 \Tr
\log G_a + \Tr \log G_c\\[1ex]\nonumber &&-\012 \Tr \,\Pi_a G_a+
\Tr \,\Pi_c G_c \\[1ex]
&&+ \Phi[G,\bar A;\phi]\,,
\label{eq:2PI}
\eeqa 
where $\Phi$ contains only the 2PI pieces and
$\Pi_{a/c}=G_{a/c}^{-1}-S_{a/c}^{(2)}$ are the gluon vacuum
polarisation and ghost self-energy, respectively. Here, for the sake of
notational simplicity we have set the renormalisation factors
$z_{a/c}=1$. The two-loop diagrams of $\Phi$ are displayed
in \Fig{fig:Phi}. The 1PI effective action $\Gamma[\bar A;\phi]$ is
then given with
\begin{equation}\label{eq:1PI-2PI}
  \Gamma[\bar A;\phi]=\Gamma_{\rm 2PI}[G[\bar A;\phi],\bar A;\phi]\,
\end{equation}
with the stationarity condition
\begin{equation}\label{eq:gap}
  \left.\0{\delta\Gamma_{\rm 2PI}}{\delta G}\right|_{G=G[\bar A;\phi]}=0\,, 
\end{equation}
i.e. the effective action is the 2PI-effective action evaluated on the gap
equation.  Now we take the derivative w.r.t.\ $ A_0$ of
$\Gamma[\bar A;\phi]$ in its 2PI-representation given in
\eq{eq:1PI-2PI}. The derivative acts on the explicit $ A_0$-dependence
in the classical vertices as well as in the propagators. The
latter terms, however, vanish due to the gap equation displayed in
\eq{eq:1PI-2PI}, thus,
\begin{eqnarray}\label{eq:2PI-DSE}
  && \hspace{-.6cm} \0{\delta(\Gamma[A_0;0]-S_A[A_0;0])}{
    \delta A_0(x)}\\[1ex]\nonumber 
  &=&\left( \Tr \, \0{\delta \Gamma_{\rm 2PI}}{\delta G}
    \, \0{\delta G[\bar A;\phi] }{ \delta A_0}+ 
    \0{\delta \Gamma_{\rm 2PI}[G,\bar A;\phi] }{ 
      \delta  A_0}\right)_{G=G[ A_0;\phi]}
  \\[1ex]\nonumber 
  &=& \012 S^{(3)}_{A_0 aa} G_{a} -
  S^{(3)}_{A_0 c\bar c} G_{c} +
  \left.\0{\delta \Phi[G,\bar A;\phi] }{ 
      \delta  A_0}\right|_{G=G[ A_0;\phi]}\,,
\end{eqnarray}
where, by comparison with \eq{eq:DSE}, \Fig{fig:DSE}, the last term simply is
\begin{eqnarray}\label{eq:Phi-2loop}
  &&\hspace{-1cm}\left.\0{\delta \Phi[G,\bar A;\phi] }{ 
      \delta \bar A_0}\right|_{G=G[ A_0;\phi]}\\\nonumber 
  &=&-\016 S^{(4)}_{A_0 aaa} G_{a}^3 \Gamma^{(3)}_{aaa} + 
S^{(4)}_{A_0  a c\bar c } G_{c}^2 G_{a} \Gamma^{(3)}_{ac \bar c}  \,. 
\end{eqnarray}
\Eq{eq:Phi-2loop} can be proven in any order of a given 2PI-expansion
scheme such as 2PI perturbation theory or the $1/N$-expansion.  For
example, the two-loop terms in $\Phi$, depicted in \Fig{fig:Phi},
provide the right hand side in \eq{eq:Phi-2loop} with classical vertices only. In
the present work this is the approximation we shall use for the
explicit computations. In any case we conclude that for the present
purpose of studying the Polakov loop potential the 2PI-representation
and DSE-representation are quite close and we shall make use of the
similarities.  Note that this similarity does not hold for e.g.\
dynamics of a given system where conservation laws such as energy and
particle number conservation play a r$\hat {\rm o}$le. Then, using
self-consistent 2PI-schemes is mandatory.

\section{Confinement Criterion}\label{sec:confcrit}
The requirement of a confining potential at low temperatures restricts
the possible infrared behaviour of low-order correlation functions:
The FRG representation of $V[A_0]$ constrains the behaviour of the
propagators, and, furthermore, the DSEs and 2PI-equations constrain
the three-gluon vertex as well.

First we discuss the restriction of the infrared behaviour of the
propagators that follows from the FRG, \eq{eq:funflow}. The present
FRG-discussion extends and sharpens the criterion given in
\cite{Braun:2007bx}. There, the flow equation \eq{eq:funflow} was
rewritten in terms of a total derivative and a term proportional to 
$\partial_t \Gamma_k^{(2)}$. Schematically this reads 
\begin{eqnarray}\nonumber 
  \partial_t \Gamma_k[\bar A;\phi] &= & - 
  \sumint_p\ \left( \frac{1}{2} [\partial_t \log G_a ]_{\mu\mu}^{aa}
-\partial_t [\log G_c]^{aa}\right)\\ 
  & & \hspace{-1.5cm}-\sumint_p\ \left( \frac{1}{2} \left[G_a 
\, \partial_t \Gamma^{(2)}_a\right]_{\mu\mu}^{aa}  - 
\left[G_c   \, \partial_t \Gamma^{(2)}_c\right]^{aa}\right)\,.
\label{eq:total}\end{eqnarray}
The second line can be understood as a renormalisation group (RG)-improvement as it is
proportional to $\partial_t \Gamma_k^{(2)}$.  Moreover, the
$t$-integrals of both lines on the right hand side are independent of the choice
of the regulator as the integral of the first line caries this
property trivially. In \cite{Braun:2007bx} these arguments have been
used to drop the second line in \eq{eq:total} as a correction term. 
This leads to, see \cite{Braun:2007bx},
\begin{equation}
  \Gamma[\bar A;\phi] \simeq  - 
  \sumint_p\ \left( \frac{1}{2} [ \log G_a ]_{\mu\mu}^{aa}
    -[\log G_c]^{a}\right)+\Gamma_\Lambda[\bar A;\phi] \,.
\label{eq:inttotal}\end{equation}
which reads for the Polyakov loop potential 
\begin{equation}
  V[A_0] \simeq  - 
  \sumint_p\ \left( \frac{1}{2} [ \log G_a ]_{\mu\mu}^{aa}
    -[\log G_c]^{aa}\right)\,, 
\label{eq:Vinttotal}\end{equation}
as $V'_\Lambda[A_0]$ tends to zero exponentially for large initial
cut-off scales $\Lambda$. Note also that the approximation \eq{eq:Vinttotal} has been
used in \cite{Fukushima:2012qa,Kashiwa:2012td}. For confinement at
small temperatures, $T\to 0$, the potential is computed from the small
momentum regime with $p^2\to 0$ where the propagators are given by
$1/(p^{2(1+\kappa)})$ for gluon and ghost with the $T=0$ scaling
exponents $\kappa^{\ }_a$ and $\kappa_c$, respectively. It has been
shown in \cite{Braun:2007bx} that confinement enforces
\begin{equation}\label{eq:conf07} 
d-2+(d-1)\kappa^{\ }_a-2 \kappa_c <0\,, 
\end{equation}
in the approximation \eq{eq:Vinttotal}. 

The suppression argument concerning the RG-improvement term presented
above was tested in \cite{Braun:2007bx}. There, the $T\!=\!0$-propagators
computed in \cite{Fischer:2008uz} with an optimised regulator have
been used.  In the spirit of the suppression argument the potential
was computed with the $T\!=\!0$-propagators but with exponential regulator
functions which facilitates this specific computation. It has been
checked that the RG-improvement term indeed is small.  We have now
extended the analysis to a fully self-consistent computation with the
thermal propagators from \cite{Fister:2011uw,Fister:2011um,
  Fister:Diss}.  Surprisingly, the RG-improvement term turns out to be
large even though it has only a small impact on the phase transition
temperature. This relates to the fact that in the non-perturbative
regime RG-improvements are not parametrically suppressed by a small
coupling and suppression arguments have to be taken with care. This
mirrors the 20-30\% percent deviation of the standard DSE-results for
the $T\!=\!0$ ghost and gluon propagators at about 1GeV in comparison to
lattice results, being linked to the dynamics which drives the phase
transition. In the related DSE-approximations the two-loop term in
\Fig{fig:DSE} is dropped. This also implies that related approximation
schemes in the 2PI-approach have to be evaluated with care.

Here, we take into account the full flow and note that all loop
diagrams in \eq{eq:funflow} finally boil down to computing
\begin{equation}\label{eq:genloop}
  T \sum_{n\in \mathbb{Z}}\int_0^\infty\!\! \0{d |\vec{p\,}|}{(2 \pi)^3} \Omega_2\
 f_k(-D_0^2, \vec{p\,}^2,A_0)\,,
\end{equation}
where $\Omega_2 = 4 \pi$ is the two-dimensional spherical surface and
the dimensionless function $f_k$ is structurally given by
\begin{equation}\label{eq:defoff}
f_k=  \vec{p\,}^2 \frac{\partial_t R_k(x)}{ x\, Z(-D_0^2, \vec{p\,}^2)+
\Delta m^2( D,A_0)+R_k(x)}\,, 
\end{equation}
where $x=-D_0^2+\vec{p\,}^2$ and $Z=Z_L,Z_T,Z_c$. In \eq{eq:defoff} we do
not include the overall minus sign of the ghost contribution. The
function $R_k$ stands for the respective scalar parts of the regulator
functions. The choice of the regulators is at our convenience. For the
present discussion (not so much for numerics) it is most convenient to
choose $Z$-independent and $O(4)$-symmetric regulators,
\begin{equation}\label{eq:regs0}
 R_k(x)= x\, r(x/k^2)\,.
\end{equation} 
The terms  
\begin{equation}\label{eq:Deltamrep}
\Delta m^2(D, 0)=0\,,  
\end{equation}
with $\Delta m^2 =\Delta m^2_L,\Delta m^2_T,\Delta m^2_c$, are
temperature corrections related to the Polyakov loop. Their impact on
the phase transition has been discussed in detail in
Section~\ref{sec:critical}. The important property in the present
context is their decay with powers of the temperature for $T\to 0$.
At vanishing temperature we regain $O(4)$-symmetry and $Z(-D_0^2,
\vec{p\,}^2)\to Z(-D_0^2+\vec {p\,}^2)$. In terms of scaling
coefficients at vanishing temperature,
\begin{equation}\label{eq:scalingcoeffs}
Z(x)=\Cz\, x^\kappa\,, 
\end{equation}
the corresponding propagators exhibit non-thermal mass gaps $m_{\rm
  gap}$ for $\kappa<0$. Note that the prefactor $\Cz$ carries the
momentum dimension $-2 \kappa$. Hence, for $\kappa<0$ we can ignore
$\Delta m^2$ in the corresponding propagator for $T/m_{\rm gap}\to 0$
as it is suppressed with (potentially fractional) powers of $T/m_{\rm
  gap}$. In turn, for propagators with field modes with $\kappa\geq 0$
the temperature correction $\Delta m^2$ may play a r$\hat {\rm o}$le.
However, for the time being we treat all of them as sub-leading
corrections and discuss them later for the field modes with
$\kappa>0$. This amounts to
\begin{equation}\label{eq:Deltam0}
\Delta m^2(D, A_0)\equiv 0\,,   
\end{equation}
for the present purpose. The integrands in \eq{eq:defoff} have the
limits \begin{equation}\label{eq:asymptf} \lim_{x\to 0} f_k
  =0\,,\qquad \qquad \lim_{x\to\infty} x\, f_k\to 0\,,
\end{equation}
where we remark that $x=-D_0^2+\vec{p\,}^2\to 0$ implies both $-D_0^2\to
0$ and $\vec{p\,}^2\to 0$.  Computing the expression \eq{eq:genloop}
leads to a deconfining potential in $A_0$: Taking a derivative w.r.t.\
$A_0$ it vanishes at $A_0=0$ and the center-symmetric points, i.e. where
the Polyakov loop vanishes, $L=0$. The second derivative is positive
at $A_0=0$ while it is negative at the center-symmetric points.

Hence, with \eq{eq:Deltam0} the single loops in \eq{eq:funflow} give
deconfining potentials.  However, the ghost contribution has an
overall minus sign. Therefore, the ghost contribution provides a
confining potential. This already entails that the Polyakov loop potential in
covariant gauges requires the suppression of the gluonic
contributions.

In fact the same mechanism is at work in the DSE
representation. Restricting ourselves to the one-loop terms for the
moment, the gluonic modes give a deconfining potential whereas the
ghosts yield a confining contribution. A simple mode counting shows,
that the transverse gluons must be suppressed w.r.t.\ the ghosts for
confinement to be present. This is also seen in Appendix
\ref{app:one-loop}.  In other words, with the assumption
\eq{eq:Deltam0} the necessary condition for confinement is that
\begin{equation}\label{eq:IRlimits} 
  \lim_{x\to 0}\frac{1}{Z_{T/L}(x)}=0\,\quad {\rm and}\quad  
 \lim_{x\to 0}\frac{1}{Z_c(x)} > 0\,, 
\end{equation}
where the second condition guarantees for smooth $Z_c(p)$ that the
ghost contribution dominates the trivial contribution of the gauge
mode which is precisely $1/2 V_{SU(N_c)}$, where $V_{SU(N_c)}$ is the
one-loop perturbative potential, see \eq{eq:Weiss} in
Appendix~\ref{app:one-loop}. In terms of the scaling coefficients
introduced in \eq{eq:scalingcoeffs} this translates into 
\beq
\kappa_{T/L} < 0\,\qquad {\rm and} \qquad \kappa_c\geq 0\,,
\label{eq:confcrit}
\eeq 
where $\kappa_{T/L}$ are the anomalous dimensions of the
chromomagnetic and chromoelectric gluon, respectively.
\Eq{eq:IRlimits} is a sufficient condition for confinement in
Yang--Mills propagators (with the assumption \eq{eq:Deltam0}) as it
leads to a vanishing order parameter, the Polyakov loop expectation
value. Note also that it has been shown
\cite{Fischer:2006vf,Fischer:2009tn} within a scaling analysis that
$\kappa_c\geq 0$ has to hold in the infrared. With this additional
information we deduce that \eq{eq:IRlimits} encompasses the condition
\eq{eq:conf07} derived in \cite{Braun:2007bx}: For the minimal choice
$\kappa_c=0$ we get from \eq{eq:conf07} the condition
$\kappa_a<-(d-2)/(d-1)$ which satisfies \eq{eq:confcrit}.

These conditions can be tested in toy models, in which the gluon
propagator is suppressed very mildly, whereas the ghost propagator is
trivial. This choice is the minimal satisfaction of the confinement
criterion \eq{eq:IRlimits}. Anticipating the final expressions for the
FRG and DSE representations of the Polyakov loop potential, the
immanence of confinement at sufficiently low temperatures is shown
numerically in Appendix~\ref{app:conf_mildsupp}.

The infrared limits \eq{eq:IRlimits} leading to the scaling
coefficients \eq{eq:scalingcoeffs} are satisfied, as they must, by the
existing solutions in the literature, for a detailed overview over the
solutions found on the lattice and in the continuum see e.g.\
\cite{Fischer:2008uz}.  The different solutions vary in the deep
infrared. The decoupling solution, found on the lattice and in the
continuum, exhibits a trivial ghost and a finite, non-vanishing gluon
propagator, where the continuum allows for a scaling solution with a
vanishing gluon propagator and an enhanced ghost propagator. In terms
of the scaling exponents \eq{eq:scalingcoeffs} the decoupling solution
translates to $\kappa_a=-1$ and $\kappa_c=0$, the scaling solution
shows $\kappa_c=\kappa=-\kappa_a/2 $ with $\kappa \approx 0.595$
\cite{Zwanziger:2001kw,Lerche:2002ep,Pawlowski:2003hq}. We stress
again that both types meet the criterion \eq{eq:confcrit} and exhibit
confinement. In the computations presented below we have tested, that
the type of solutions does not affect the critical physics at the
confinement-deconfinement transition, neither qualitatively nor
quantitatively. This is expected, as the relevant range for the phase
transition is the mid-momentum region, where both decoupling and
scaling solutions agree quantitatively.

We close this section with a discussion of the consequences and
implications of \eq{eq:scalingcoeffs} and \eq{eq:IRlimits}: Note first
that confinement not only implies a vanishing order parameter but also
constrains other observables such as its correlations. The evaluation
of these observables in terms of the propagators (and higher vertices)
may lead to further, tighter constraints for the propagator. This
point of view is interesting for model computations as there the above
condition \eq{eq:IRlimits} is only a necessary one for deciding
whether the model is confining or not. For example, one can easily
construct models for Yang--Mills theory that have massive Yang--Mills
propagators and a trivial ghost propagator at vanishing
temperature. This can be parameterised as
\begin{equation}\label{eq:massive} 
 G_{a}(p^2)= \0{1}{p^2 +m^2}\,,\qquad G_c=\0{1}{p^2}\,.  
\end{equation}
Such a combination of propagators leads to a vanishing Polyakov loop
expectation value but lacks the necessary positivity violation in the
gluon system. It could model Yang--Mills theory in the Higgs phase but
not in the confining phase. Note that it is precisely this feature which
distinguishes the decoupling solution in Landau gauge Yang--Mills
theory with $0< G_{a}(0) <\infty$ from a massive solution put down
in \eq{eq:massive}.

\section{Confinement in Yang--Mills-Matter Systems}\label{sec:matter}
The confinement criterion discussed in the last section leads to a
simple counting scheme for general theories. In the present section we
put it to work in full dynamical QCD with fundamental or adjoint
quarks, see e.g.\ \cite{Braun:2009gm,Pawlowski:2010ht,Braun:2011fw,
  Braun:2012zq}, as well as in the Yang--Mills--Higgs system with
Higgs field in the fundamental or in the adjoint representation
coupled to Yang--Mills theory, formulated in the Landau gauge, e.g.\
\cite{Fister:2010ah, Fister:2010yw, Macher:2010di, Alkofer:2010tq,
  Macher:2011ys}. This also serves as a showcase for the general
simple counting schemes which has emerged by now. 

We concentrate on the one flavour case (Higgs or quark), the
generalisation to many flavours is straightforward.  Note also in this
context that the present confinement criterion constrains the physics
properties of a potentially confined phase, it does not prove or
disprove its dynamical existence in the theory at hand. For example,
it is well-known that for a sufficiently large number of quark
flavours in the fundamental or adjoint representation, QCD ceases to
be confining. Then the propagators simply do not satisfy the
confinement criterion.

\begin{figure}[t]
\includegraphics[width=.9\columnwidth]{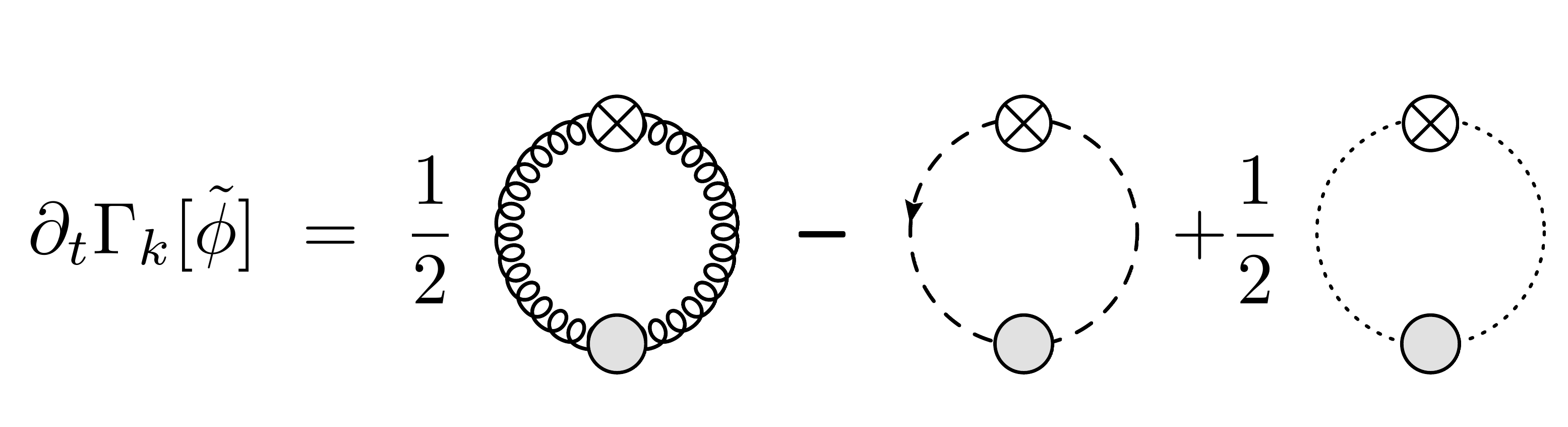}
\caption{Functional flow for the effective action of Yang--Mills
  theory with bosonic matter in the adjoint or fundamental representation,
  assembled in $\tilde\phi$. In either case, the matter fields are
  represented by the dotted lines adding the third loop.}
\label{fig:funflowhiggs}
\end{figure}

We start with a discussion of the adjoint Higgs $h$ (one flavour) with 
the action 
\begin{equation}\label{eq:ad}
  S_{\rm Higgs}[\bar A;\phi,h]=S_A+\int_x \tr_{\rm ad} h^\dagger D^\dagger_\mu D_\mu h+
  \int_x V(h^\dagger h)\,.
\end{equation}
The flow equation for the Polyakov loop potential is depicted in
\Fig{fig:funflowhiggs}.
We have already seen that in the Landau gauge the question of a
confining potential at low temperatures boils down to counting the
massless modes including the prefactors $1$ and $-2$ from the loops of
bosons and fermions, respectively, where we normalise the standard
boson loop with the prefactor $1/2$ to one. In the symmetric phase at
high temperatures, all modes are effectively massless for our
counting: All vacuum masses are suppressed with the temperature. Our
counting is normalised at the (one-loop) Polyakov loop potential in
Yang--Mills theory which is computed from 2 boson loops, that are
related to the physical transversal gauge modes; the third transversal
contribution and that of the longitudinal gauge mode are canceled by
the ghost contribution,
\begin{equation}\label{eq:YMadlargeT}
\underbrace{\underbrace{2}_{\rm phys.\ pol.} +\underbrace{1}_{\rm transversal}
+\underbrace{1}_{\rm longitudinal}}_{\rm gauge\ bosons}-\underbrace{2}_{\rm ghost} = 2\,.
\end{equation}
Hence, in the present case we have $4$ contributions from the gauge
field (4 vector modes), $ -2$ ghost contributions (relative factor
$-2$ for the loop), and $1$ Higgs contribution. This amounts to an
overall factor of $3$ for the symmetric phase of the adjoint
Yang--Mills--Higgs system,
\begin{equation}\label{eq:higgsadlargeT}
\underbrace{4}_{\rm gauge\ bosons} -\underbrace{2}_{\rm ghost} +  
\underbrace{1}_{\rm Higgs}= 3\,.
\end{equation}
This has to be compared with the counting factor $2$ for the pure
Yang--Mills system, leading to the one-loop Polyakov loop potential
$V_{SU(N_c)}$ \cite{Weiss:1980rj, Gross:1980br}, given in 
Appendix~\eq{eq:Weiss}. The one-loop potential for the
Yang--Mills-adjoint Higgs system, $V_{\rm ad-h}$, is hence given by
\begin{equation}\label{eq:ad-Higgs-Weiss}
V_{\rm ad-h}(\varphi)=\032 V_{SU(N_c)}(\varphi)\,.
\end{equation}
In turn, in the fully-broken phase of the Yang--Mills--Higgs system,
we have one radial massive (away from the phase transition) Higgs mode
with expectation value $h_0$ and $N_c^2-2$ Goldstone bosons. In the
glue sector, $N_c^2-2$ gauge bosons acquire an effective mass due to
the non-vanishing expectation value of the Higgs field. In the Landau
gauge this leads to massive propagators for the $3 (N_c^2-2)$
transversal modes, if the theory is evaluated at the expectation value
of the Higgs field. The gauge mode, even though it also acquires a
mass, is effectively massless: Its propagator, evaluated at the
expectation value of the Higgs, reads
\begin{equation}\label{eq:gaugeprop} 
\0{\xi }{ p^2(1+\xi) + \xi (g h_0)^2 }\delta^{ab}\Pi_{\mn}^L(p)\,,
\end{equation} 
and, for Landau gauge, $\xi\to 0$, it only contributes via the
longitudinal part of the regulator proportional to $1/\xi$ in the flow
equation or via the gauge-fixing part proportional to $1/\xi$ in the
DSE and 2PI formulation. We conclude that the gauge mode in the Landau
gauge stays effectively massless in any phase. Together with the
remaining massless gauge boson we have $N_c^2+2$ massless modes in the
gauge boson sector. 

For example, in $SU(2)$ we have $h^1,h^2,h^3$ and without loss of
generality we take
\begin{equation}\label{eq:hvev}
\langle h^i\rangle =\delta^{i3} h_0\,, \quad {\rm with}\quad V'(h_0^2)=V(h_0^2)=0\,,
\end{equation}
leading to 
\begin{eqnarray}\nonumber 
  S_{\rm Higgs}[\bar A;\phi,h] &=&S_A+
  g^2 h_0^2\epsilon^{3ab} A_\mu^b \epsilon^{3ac} A_\mu^c\\[1ex] 
  &=&S_A+(g h_0)^2 
  \left[ (A^1_\mu)^2+ (A^2_\mu)^2\right]\,.
\label{eq:massad}
\end{eqnarray}
Hence we have $4 (N_c^2-2) = 8$ massive modes. This includes the gauge
mode which, however, is effectively massless in the Landau gauge due
to \eq{eq:gaugeprop}. Note that the latter fact is special to Landau
gauge. In the unitary gauge also the gauge mode gets massive as does the 
Faddeev--Popov operator. 

We proceed to the Polyakov loop potential. First we note that the
massive radial Higgs mode effectively blocks one colour direction with
projection operator $\CP_{\rm radial}$. For the transversal gauge
field propagators, the situation is exactly the opposite. All the
modes in the subspace $\id -\CP_{\rm radial}$ are massive and the
massless transversal colour direction is that in direction $\CP_{\rm
  radial}$. Since we are finally only interested in the sign of the
sum of the contribution we simply remark that adding one of the three
transversal massless gauge boson conbributions to the Higgs
effectively restores the Higgs boson contribution in the symmetric
regime. We also still have the trivial contribution from the gauge
mode. Moreover, the ghost is essentially unchanged as in the pure
Yang--Mills case, it still has a massless dispersion. We conclude that
due to the relative sign of the ghost contribution the Higgs, the
effectively 'restored' Higgs contribution and the gauge contribution
are canceled by the ghost contribution. This is the manifestation of
the Higgs--Kibble mechanism for the Polyakov loop potential in the
Landau gauge: The apparently massless contributions cancel each other
including those of the Goldstones. In the unitary gauge the Goldstone
modes are included in the longitudinal gauge field modes, which would
be massive and would have dropped out due to their masses.

What is left is the contribution from the remaining two massless
transversal gauge bosons with the colour direction $\CP_{\rm
  radial}$. For gauge fields $A_0$ with $\hat A_0 \CP_{\rm radial}\neq
0$ or $\CP_{\rm radial} \hat A_0\neq 0$ this provides a deconfining
contribution to the Polyakov loop potential which counts as $1/N_c$ in
the present counting. Again, all this is apparent in the
$SU(2)$-example. Adding all the massless contributions we arrive at
\begin{equation}\label{eq:higgsadlowT}
  \underbrace{2+1/N_c}_{\rm gauge\ bosons\ \&\ Higgs} -
  \underbrace{2}_{\rm ghost}= \0{1}{N_c}\,.
\end{equation}
Hence, the Polyakov loop potential for the Yang--Mills-adjoint Higgs
system in the Higgs phase is deconfining, albeit suppressed with order
$1/N_c$. 

We close the analysis on the Yang--Mills-adjoint Higgs system with a
remark on a confining phase in this theory: In this phase all gluon
modes are expected to be gapped, except for the gauge mode. Keeping
the other modes unchanged, the sum in \eq{eq:higgsadlowT} turns
negative implying confinement,
\begin{equation}\label{eq:higgsconf}
  \underbrace{1+1/N_c}_{\rm gauge\ bosons\ \&\ Higgs} -
  \underbrace{2}_{\rm ghost}= -1+\0{1}{N_c}\,.
\end{equation}
Note however, that \eq{eq:higgsadlowT} includes gapped Higgs modes. If
these modes became massless they would counter-balance the
gapping of the confined gluons. We conclude that gapped Higgs modes
are required in the confined phase. This is a reminiscent of the
expectation in Yang--Mills-quark systems where one usually expects
chiral symmetry breaking in the confined phase.

\begin{figure}[t]
\includegraphics[width=.9\columnwidth]{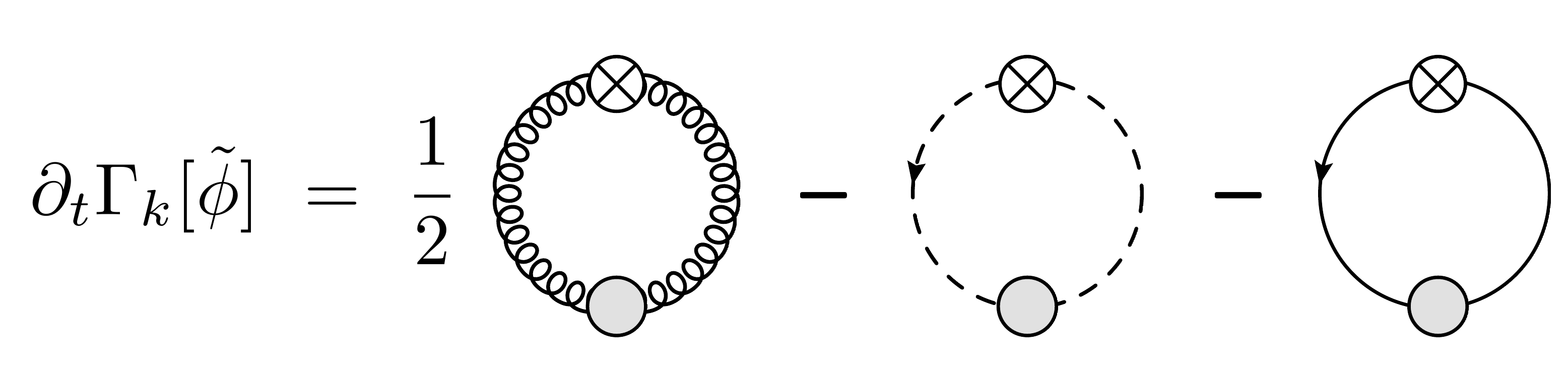}
\caption{Functional flow for the effective action of Yang--Mills
  theory with fermionic matter in the adjoint or fundamental representation,
  assembled in $\tilde\phi$. In either case, the matter fields are
  represented by the straight lines adding the third loop.}
\label{fig:funflowferm}
\end{figure}

Let us compare this with the situation in QCD with adjoint
fermions. There, the action reads
\begin{equation}\label{eq:adQCD}
  S_{\rm QCD_{\rm ad}}[\bar A;\phi,\psi,\bar\psi]=S_A+\int_x 
  \tr_{\rm ad} \bar\psi  \dr \psi\,.
\end{equation}
The flow equation for the Polyakov loop potential is depicted in
\Fig{fig:funflowferm}. 
The fermionic one-loop contribution to
the Polyakov loop potential reads
\begin{equation}\label{eq:fermoneloop}
  \Tr_{\rm ad+spinor} \ln \dr =\012 \Tr_{\rm ad+spinor} \ln D_\mu^2 
  =4 \, \012 \Tr_{\rm ad} \ln D_\mu^2 \,,
\end{equation} 
where we have used that $\dr^2=( D_\mu^2 -1/2 g
[\gamma_\mu,\gamma_\nu] F_{\mu\nu})\id =D_\mu^2 \id $ for constant
fields $A_0$, and $\id$ is the identity in spinor space with $\tr_{\rm
  spinor} \id =4$. The fermionic contributions are deconfining
as the quarks have anti-periodic boundary conditions,
$\psi(t+\beta)=-\psi(t)$, and, hence, the Matsubara frequencies are
shifted: $2\pi T(n+1/2+\varphi)$. This entails that the Polyakov loop
potential is shifted by a factor $1/2$:
\begin{equation}\label{eq:Poladjointq}
V_{\rm ad-q}(\varphi)=-V_{\rm SU(N_c)}(\varphi+\012)\,, 
\end{equation}
which is deconfining with the same strength as the Yang--Mills
potential. This argument stays valid beyond one loop. For large
temperature we conclude that the adjoint quarks for large temperatures
lead to $4 (N_c^2-1)$ contributions, 
\begin{equation}\label{eq:quarkadlargeT}
\underbrace{4}_{\rm gauge\ bosons} -\underbrace{2}_{\rm ghost} +  
\underbrace{4}_{\rm quark}= 6\,.
\end{equation}
This has to be compared with $2 (N_c^2-1)$ for the pure Yang--Mills
system, leading to the one-loop Polyakov loop potential $V_{SU(N_c)}$,
\eq{eq:Weiss}. The one-loop potential for the Yang--Mills-adjoint Higgs
system, $V_{\rm ad-h}$, is hence given by
\begin{equation}\label{eq:ad-quark-Weiss}
V_{\rm ad-h}(\varphi)=3 V_{SU(N_c)}(\varphi)\,.
\end{equation}
In turn, in the chirally broken phase all quarks are massive and their
contributions are removed from the Polyakov loop potential. The glue
sector is qualitatively the same as in Yang--Mills theory, the transversal
modes are gapped, while the gauge mode is effectively massless.  Hence,
we are lead to
\begin{equation}\label{eq:quarkadlowT}
  \underbrace{1}_{\rm gauge\ bosons} -
\underbrace{2}_{\rm ghost}= -1\,, 
\end{equation}
the theory is confining in clear contradistinction to the Higgs phase
in the Yang--Mills-adjoint Higgs system. Note also that this directly
implies {\it no confinement without chiral symmetry breaking} for QCD
with adjoint quarks if the gapping in the quark sector relates to
chiral symmetry breaking. Strictly speaking, however,
\eq{eq:quarkadlowT} implies {\it no confinement without gapped
  quarks}. In turn, in \cite{Braun:2012zq} it has been shown that
confinement, that is a vanishing expectation value of the Polyakov
loop, leads to chiral symmetry breaking. These two results tightly
link chiral symmetry breaking and confinement for QCD with adjoint
quarks.

In a potential chirally symmetric low energy phase, or better massless
low energy phase, we have
\begin{equation}\label{eq:quarkadlowTchiral}
  \underbrace{1}_{\rm gauge\ bosons} -
\underbrace{2}_{\rm ghost}+  
\underbrace{4}_{\rm quark}= 3\,, 
\end{equation}
which relates to a deconfining Polyakov loop potential. 

We close the section with a discussion of matter in the fundamental
representation coupled to Yang--Mills theory. There, center symmetry
is explicitly broken. The Polyakov loop is not an order parameter
anymore and the question of the confining or Higgs phase has to be
answered differently. Indeed, we know that on the lattice these two
phases are not well-separated \cite{Osterwalder:1977pc,
  Fradkin:1978dv, Kertesz:1989}. A full discussion of this goes far
beyond the aims of the present contribution. Here we simply discuss
our counting scheme and compare its outcome for the
Yang--Mills-fundamental Higgs system with physical QCD.  The action of
the Yang--Mills-fundamental Higgs system reads
\begin{equation}\label{eq:fun}
  S_{\rm Higgs}[\bar A;\phi,h]=S_A+\int_x \tr_{\rm f}\, h^\dagger
 D^\dagger_\mu D_\mu h+
  \int_x V(h^\dagger h)\,, 
\end{equation}
similarly to the action for an adjoint Higgs, but with the group
trace in the fundamental representation, the Higgs living in the
$U(N_c)$. For $SU(2)$, the Higgs is a complex 2$\times$2-matrix,
\begin{equation}\label{eq:Higgs}
h=\left(\begin{array}{cc} h_0+h_3 &\quad  h_1- i h_2 \\ h_1+i h_2 
&\quad  h_0- h_3\end{array}\right)\,.
\end{equation}
The diagrammatics of the flow equation for the Polyakov loop
potential does not change in comparison to the adjoint Higgs and is
depicted in \Fig{fig:funflowhiggs}, the last loop now involving a
trace in the gauge group in the fundamental representation.

The fundamental Higgs has $N_c^2$ modes due to its
$U(N_c)$-representation. In the symmetric phase at high temperatures
all of them are effectively massless. The group traces, however, are
in the fundamental representation of $SU(N_c)$, and, hence, they are
suppressed by a factor $1/ N_c$ (in the large $N_c$ limit) in
comparison to the traces in the adjoint representation. More
precisely, the multiplicity of the non-vanishing eigenvalues of $A_0$
is relatively suppressed with $1/N_c$. The fundamental representation
also leads to different eigenvalues of $A_0$, see in particular
\cite{Braun:2009gm}. For example, for $SU(2)$ we have for
contributions in the adjoint and fundamental representation
\begin{equation}\label{eq:fundpol}
V_{\rm fund}(\varphi)= V_{\rm ad}(\varphi/2)\,, 
\end{equation}
which can be nicely tested for the one-loop contributions. Note also,
that the $U(1)$-mode $h_0$ does not couple to the $SU(N_c)$-gauge
field and $V_{\rm fund}$ is a pure $SU(2)$-potential. In
\eq{eq:fundpol} the factor $1/2$ on the right hand side signals the
explicit breaking of center symmetry. This also means that one should
not simply add the counting factors of contributions in different
representations. We merely remark here that as long as the
contributions from the center-symmetry breaking sector is small, one
may still have a phase transition of the order determined by
$Z_N$-symmetry. The larger the contributions of the center-symmetry
breaking sector are, the weaker the transition or finally the
cross-over gets. The interpretation of the latter as distinguishing
different phases is not clear.

Restricting ourselves to the center-symmetric field modes we get for
large temperatures
\begin{equation}\label{eq:higgsfunlargeT}
\underbrace{4}_{\rm gauge\ bosons} -\underbrace{2}_{\rm ghost}=2\,, 
\end{equation}
the counting of pure Yang--Mills theory. This entails that the Polyakov
loop potential of the glue sector is deconfining. At one-loop we have a 
deconfining total potential,
\begin{equation}\label{eq:fun-Higgs-Weiss}
V_{\rm YM-h_{\rm fund}}(\varphi)= V_{\rm SU(N_c)}(\varphi)+V_{\rm h_{\rm fund}}(\varphi)\,.
\end{equation}
In turn, in the fully-broken phase of the Yang--Mills-fundamental
Higgs system we have one radial massive (away from the phase
transition) Higgs mode with expectation value $h_0$ which does not
couple to the gauge field. We also have $N_c^2-1$ Goldstone bosons. In
the glue sector, all $N_c^2-1$ gauge bosons acquire an effective mass
due to the non-vanishing expectation value of the Higgs field.
In the Landau gauge this leads to massive propagators for
$3(N_c^2-1)$ transversal gauge modes and $N_c^2-1$ massless gauge
modes. The ghost is massless, and in summary this leads to
\begin{equation}\label{eq:higgsfunlowT}
\underbrace{1}_{\rm gauge\ bosons} -\underbrace{2}_{\rm ghost} 
= -1\,, 
\end{equation}
a confining potential for the glue sector. The $N_c^2-1$ massless
$SU(N_c)$ modes give a center-symmetry breaking potential, which,
however, is suppressed by a factor $1/N_c$ for large $N_c$.  Whether
for finite $N_c$ one has a cross-over or still a first order,
($N_c>2$), or second order, $(N_c=2)$, phase transition is decided
dynamically and is not accessible by the present counting.  Evidently,
however, is the existence of a confining potential in the large
$N_c$-limit. In this limit the pure glue sector dominates the Polyakov
loop potential due to the $1/N_c$-suppression of the Higgs
contributions.

Finally we compare the situation in the Yang--Mills-fundamental Higgs
system with physical QCD with fundamental quarks, the most interesting
and most studied system. Its action is given by
\begin{equation}\label{eq:funQCD}
  S_{\rm QCD}[\bar A;\phi,\psi,\bar\psi]=S_A+\int_x  \bar\psi  \dr \psi\,, 
\end{equation}
and the flow equation for the Polyakov loop potential is depicted in
\Fig{fig:funflowferm}. The Higgs in the fundamental representation
breaks center symmetry in the same way a fundamental quark does, and
the Polyakov loop potential shows no phase transtion. The order
parameter $L[\langle A_0\rangle]$ is a smooth function of temperature
similarily to two-flavour QCD studied in \cite{Braun:2009gm}. One may
be tempted to interpret the behaviour in terms of a cross-over;
however, the Wilson loop shows no area law which is in accord with the
behaviour of the Polyakov loop. As for the Higgs the pure glue sector
shows the counting \eq{eq:higgsfunlowT}. The part of the potential
computed from the massless fermionic modes satisfies
\begin{equation}\label{eq:fermoneloopfund}
  \Tr_{\rm f+spinor} \ln \dr =\012 \Tr_{\rm f+spinor} \ln D_\mu^2 
  =2 \, \Tr_{\rm f} \ln D_\mu^2 \,,  
\end{equation} 
and with anti-periodic boundary conditions for the fermion we get 
\begin{equation}\label{eq:Polfundq}
V_{\rm fund-q}(\varphi)=-2 V_{\rm fund}(\varphi+\012)\,, 
\end{equation}
where $V_{\rm fund}$ is the potential of one bosonic mode in the
fundamental representation. In the large $N_c$-limit the fermionic
contributions are suppressed (unless the number of flavours increases
with the number of colours) and yield a confining theory with a first
order phase transition. At finite $N_c$, and in particular $N_c=2,3$,
we expect a significant influence of the fermions at temperatures
above the chiral symmetry breaking scale. Below this scale the
fermionic contributions are more and more suppressed and we are left
with the pure glue counting.

\section{Results for the Polyakov Loop Potential and
  $T_c$}\label{sec:resultsPolPot}
Here, we compute the Polyakov loop potential within the FRG, DSE and
2PI representation with the propagators at vanishing and finite
temperatures obtained from the FRG in
\cite{Fister:2011uw,Fister:2011um, Fister:Diss}. There, the finite
temperature propagators for all cut-off scales $k$ have been computed
on the basis of a given set of propagators at vanishing temperature
and vanishing cut-off scale $k\!=\!0$. This minimises the systematic
error of a given approximation: Only the difference of the $k\!=\!0$-
to the $k\!\neq\!  0$-propagator and the thermal to the
$T\!=\!0$-propagator, respectively, is sensitive to the
approximation. The results for the Matsubara zero mode for the
longitudinal and transversal propagator, $G_{L/T}$, are displayed in
\Fig{fig:thermalLong} and \Fig{fig:thermalTrans}, respectively, for
some temperatures below and above the phase transition. They are
compared to lattice results \cite{Maas:2011ez,Maas:2011se}.
\begin{figure}[t]\begin{center}
\includegraphics[width=.95\columnwidth]{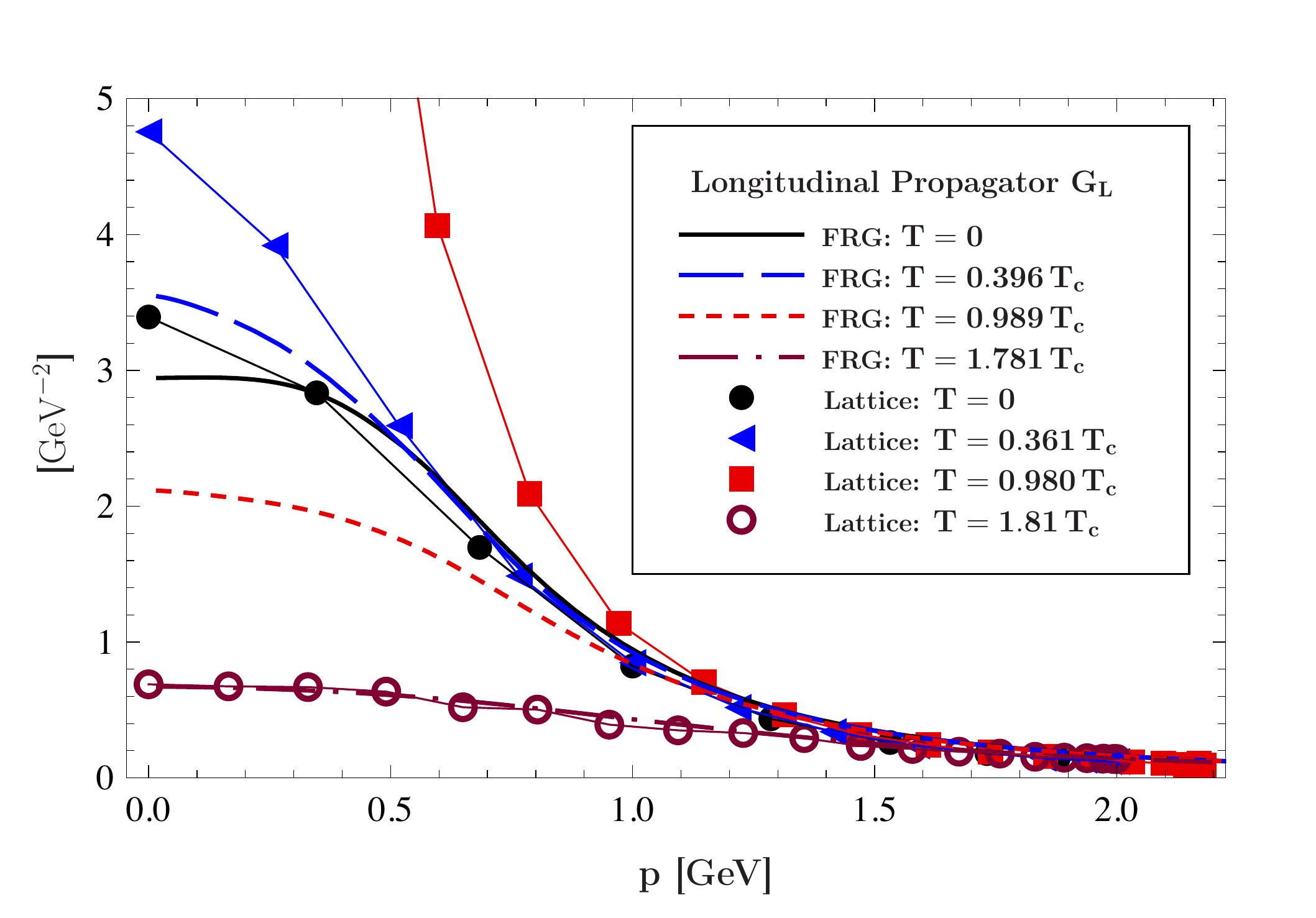}
\caption{Zeroth Matsubara mode of the thermal longitudinal gluon
  propagator from the FRG \cite{Fister:2011uw,Fister:2011um,
    Fister:Diss} in comparison with lattice results
  \cite{Maas:2011ez,Maas:2011se}.}
\label{fig:thermalLong}
\efc
\begin{figure}[t]\begin{center}
\includegraphics[width=.95\columnwidth]{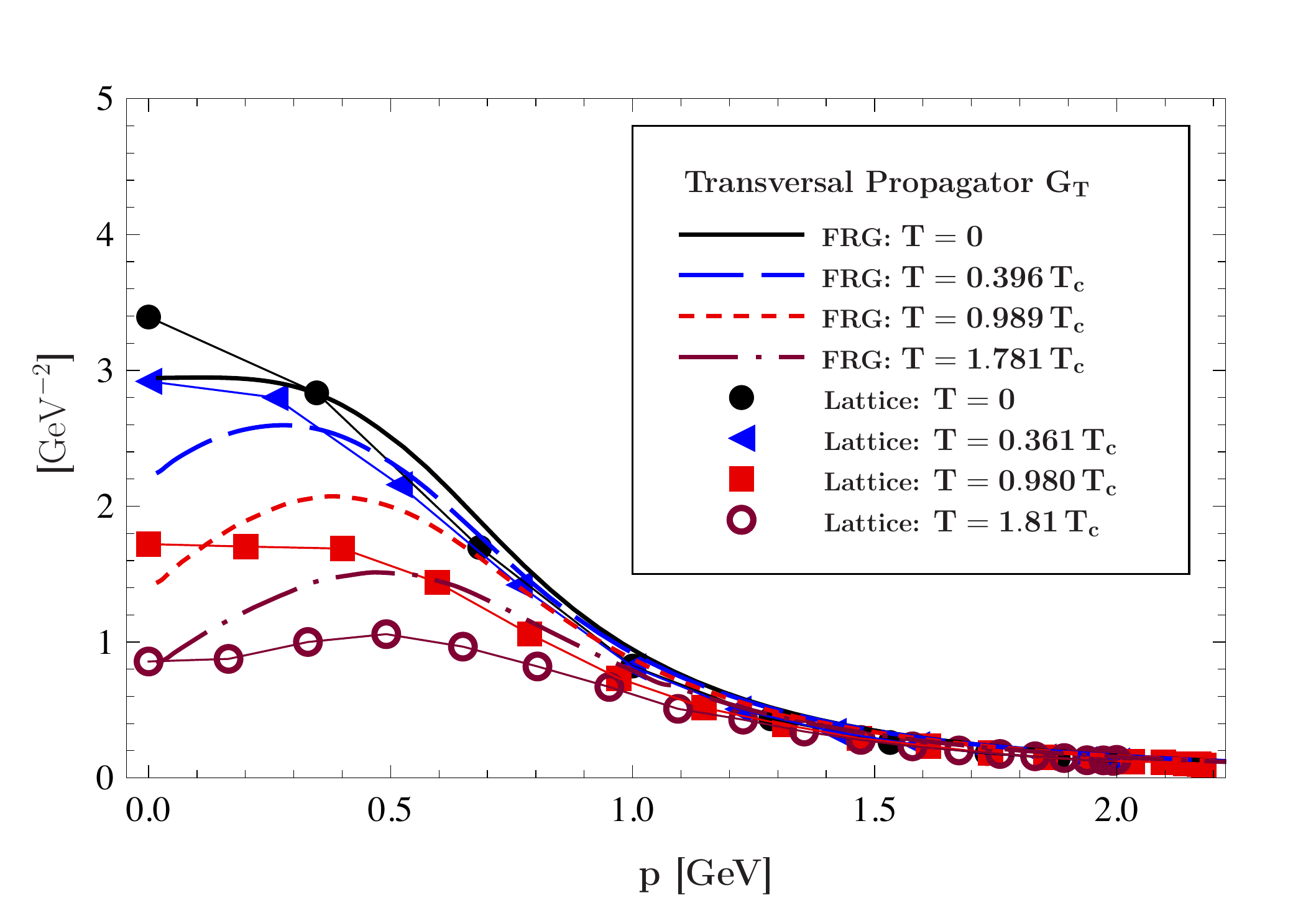}
\caption{Zeroth Matsubara mode of the thermal transversal gluon
  propagator from the FRG \cite{Fister:2011uw,Fister:2011um,
    Fister:Diss} in comparison with lattice results
  \cite{Maas:2011ez,Maas:2011se}.}
\label{fig:thermalTrans}
\efc
The FRG results have been obtained in a slightly modified
approximation in comparison to \cite{Fister:2011uw,Fister:2011um}, see
\cite{Fister:Diss}. We resort to the additional approximation
\begin{equation}\label{eq:zeroall}
Z(p_0^2,\left.\vec p\right.^2) \to Z(0,p_0^2+\left.\vec p\right.^2)\,, \quad 
\end{equation}
which has proven to be an accurate approximation in former studies
\cite{Maas:2005hs, Cucchieri:2007ta, Fischer:2010fx, Fister:Diss}.
For the present purpose of computing the Polyakov loop potential this
is a quantitatively reliable approximation as the finite temperature
effects in the propagators decay already rapidly for momenta with
$p^2/k^2\gtrsim (2 \pi T)^2$ in the FRG computations.  We conclude
that already the first Matsubara mode is close to the zero mode
evaluated as in \eq{eq:zeroall}. Note however, that this does not
apply to thermodynamical quantities as already pointed out in
\cite{Fister:2011uw,Fister:2011um, Fister:Diss}, for non-relativistic
analogues see \cite{Boettcher:2012cm, Boettcher:2012dh}.

\subsection{Results from the FRG}\label{sec:resFRG}
As discussed before the flow equation has the minimal representation
for the Polyakov loop potential as it only requires the propagators in
a given background. The present computations improves upon that in
\cite{Braun:2007bx,Braun:2010cy} with the use of the thermal
propagators.

In addition, further emphasis is put on the
regulator-independence, see Appendix \ref{app:regs}.  The potential is
depicted for temperatures above and below the critical temperature for
$SU(2)$ in \Fig{fig:SU2_FRG_T}, and for $SU(3)$ in
\Fig{fig:SU3_FRG_T}. Without loss of generality we shift the potential
trivially such, that $V(\varphi=0)=0$, and divide by the canonical
dimension $T^4$ to simplify the comparison, as the Weiss potential in
this normalisation is independent of the temperature.
\begin{figure}[t]
\begin{center}
\includegraphics[width=.95\columnwidth]{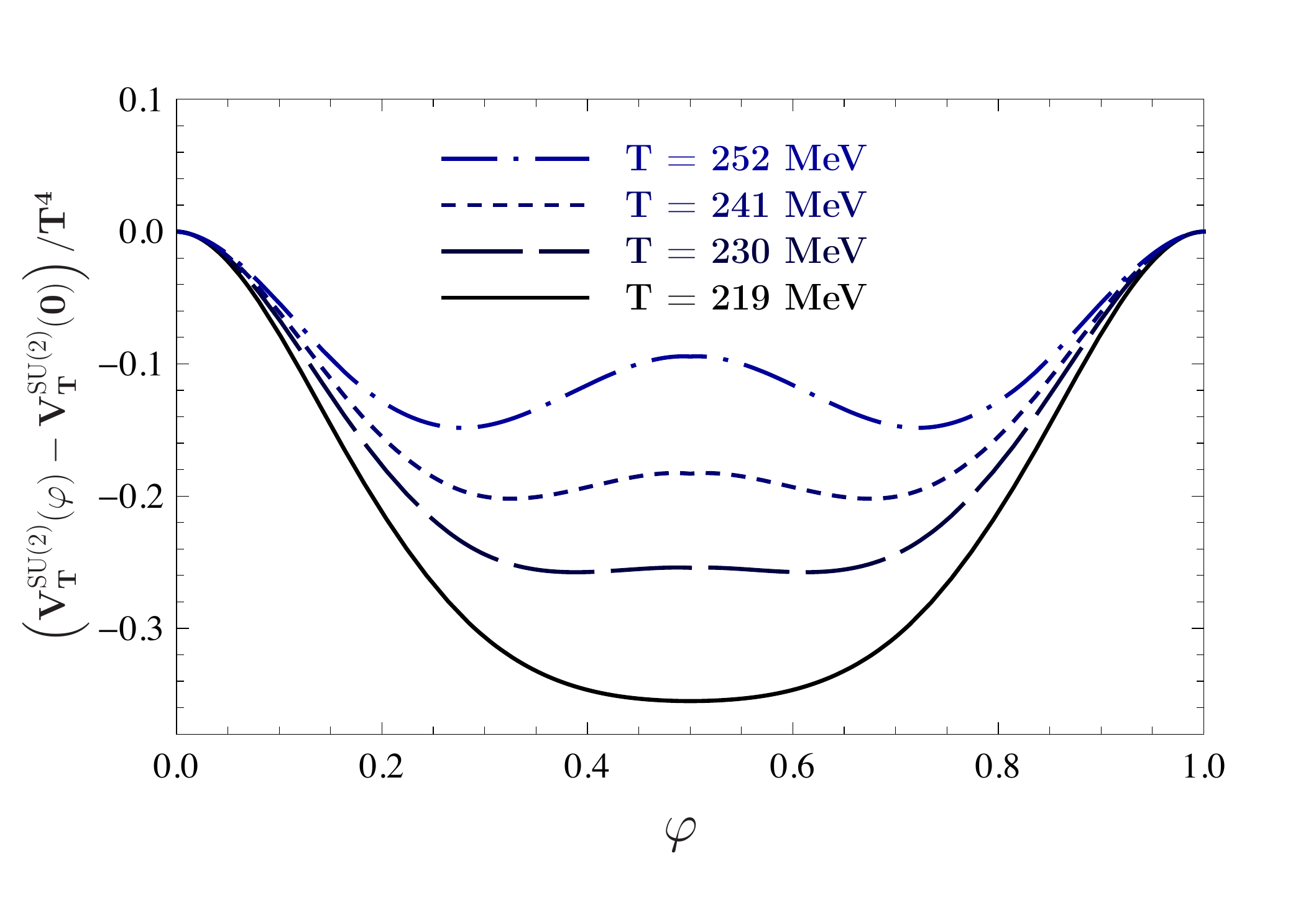}
\caption{Polyakov loop potential for $SU(2)$ obtained from the FRG.}
\label{fig:SU2_FRG_T}
\end{center}
\end{figure}
\begin{figure}[t]
\begin{center}
\includegraphics[width=.95\columnwidth]{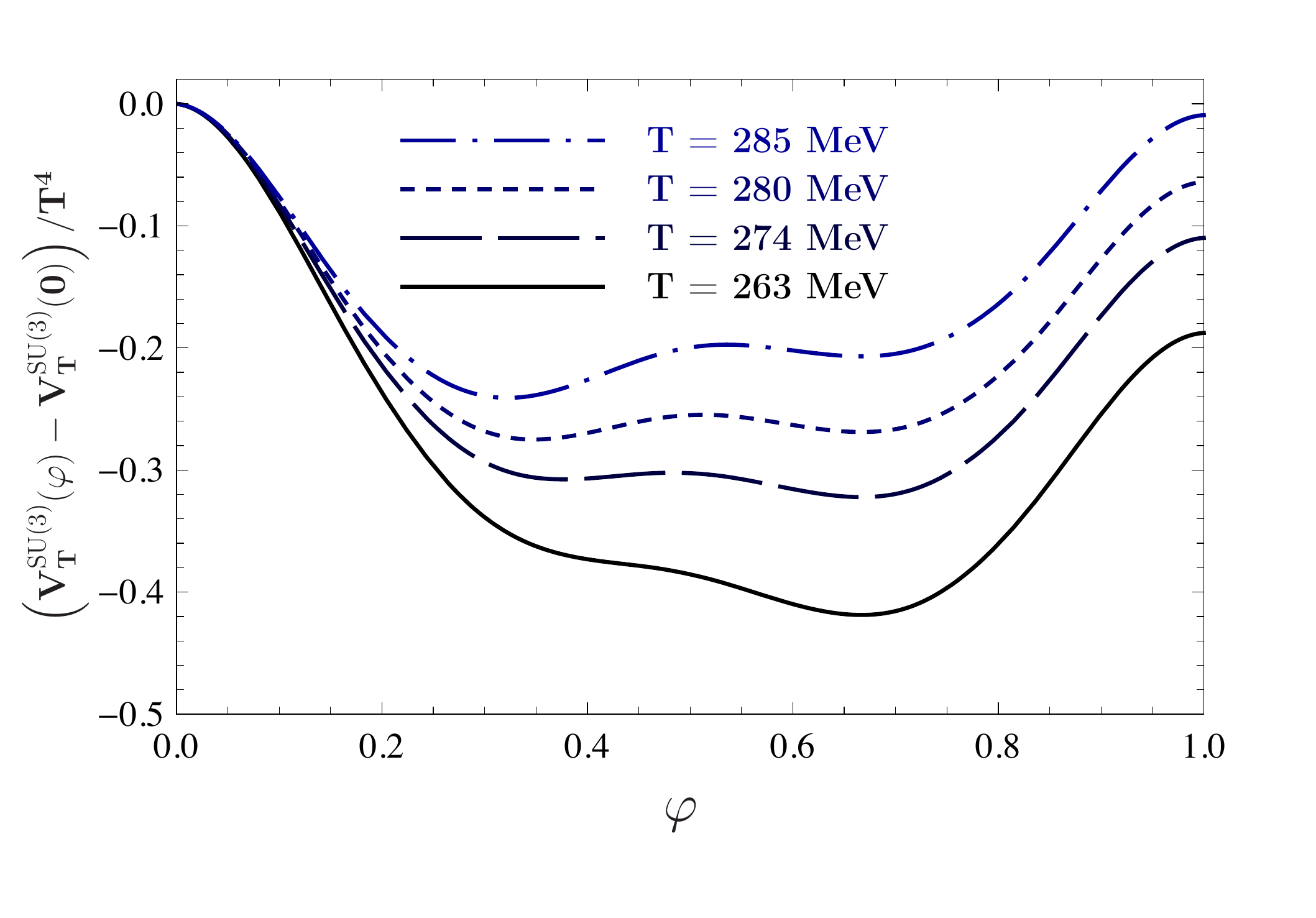}
\caption{Polyakov loop potential for $SU(3)$ obtained from the FRG.}
\label{fig:SU3_FRG_T}
\end{center}
\end{figure}
From these potentials the confinement-deconfinement critical
temperatures $T_c$ are determined at
\begin{equation}\label{eq:FRG-Tc}
  T_{c}^{SU(2)}= 230\pm 23\, {\rm MeV}\ \ {\rm and}  \ 
  \ T_{c}^{SU(3)}= 275\pm 27\, {\rm MeV}\,.  
\end{equation} 
The absolute temperatures in \eq{eq:FRG-Tc} are set in comparison to
absolute scales of the lattice propagators with a string tension
\begin{equation}
\sigma= (420\,  {\rm MeV})^2\,,
\end{equation}
which leads to the dimensionless ratios of 
\begin{equation}
T_c^{SU(2)}/\sqrt{\sigma}
\approx 0.548 \quad {\rm and}\quad T_c^{SU(3)}/\sqrt{\sigma} \approx 0.655\,.
\end{equation} 
For $SU(3)$ our result is in quantitative agreement
with the lattice results,
\begin{equation}\label{eq:lattice-Tc}
  T_{c}^{SU(2)}= 295\, {\rm MeV}\ \ {\rm and}  \ 
  \ T_{c}^{SU(3)}=270 \, {\rm MeV}\,,
\end{equation}
which yield the ratios 
\beq\label{eq:lattice-Tc_ratios}
T_c^{SU(2)}/\sqrt{\sigma} \approx0.709 \quad {\rm and} \quad
T_c^{SU(3)}/\sqrt{\sigma} \approx 0.646\,.
\eeq
For a review on lattice results see e.g.\ \cite{Lucini:2012gg}.  Our
relative normalisation is taken from the peak position of the lattice
propagators in \cite{Maas:2011ez,Maas:2011se,Fischer:2010fx}. Note
however, that this normalisation has an error of approximately $10\%$
which is reflected in the errors in \eq{eq:FRG-Tc}.

In turn, for $SU(2)$, the critical temperature $T_c^{SU(2)}$ is
significantly lower than the lattice temperature in
\eq{eq:lattice-Tc}. This is linked to the missing back-reaction of the
fluctuations of the Polyakov loop potential, see
Section~\ref{sec:funPolPot}. Including the back-reaction in the Landau
gauge flow leads to a transition temperature of
\begin{equation}\label{eq:FRG-Tcsu2}
  T_c^{SU(2)}=300\pm 30\, {\rm MeV} \,,
\end{equation}
see \cite{Spallek}. More details will be presented elsewhere. This
result agrees quantitatively with the lattice temperature as well as
with the result obtained with flows in the Polyakov gauge
\cite{Marhauser:2008fz} which also reproduced the correct critical
scaling of the Ising universality class.

\subsection{Results from DSE \& 2PI}\label{sec:resDSE}
\begin{figure}[t]
\begin{center}
\includegraphics[width=.95\columnwidth]{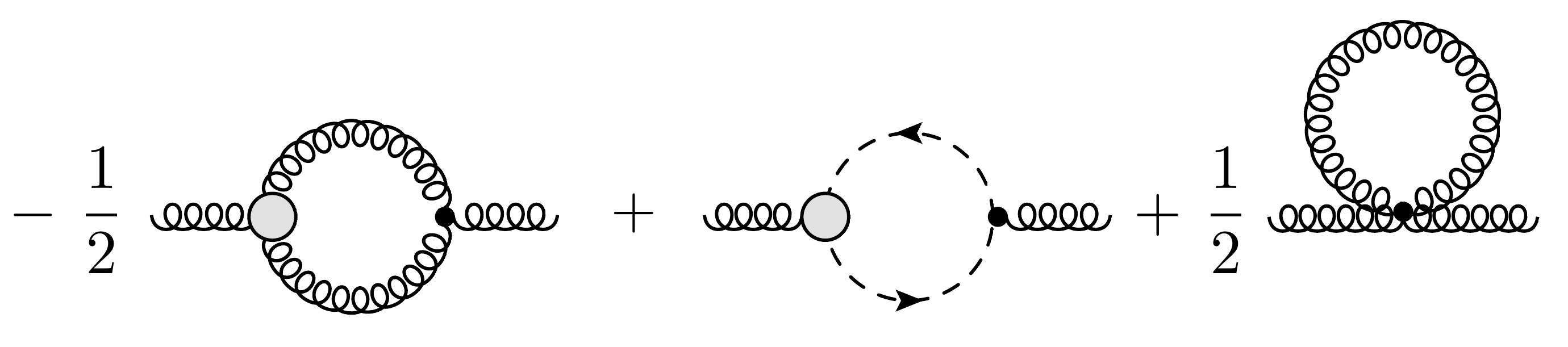}
\caption{Gluon vacuum polarisation.}
\label{fig:gluonvacpol}
\efc
\begin{figure}[t]
\begin{center}
\includegraphics[width=.45\columnwidth]{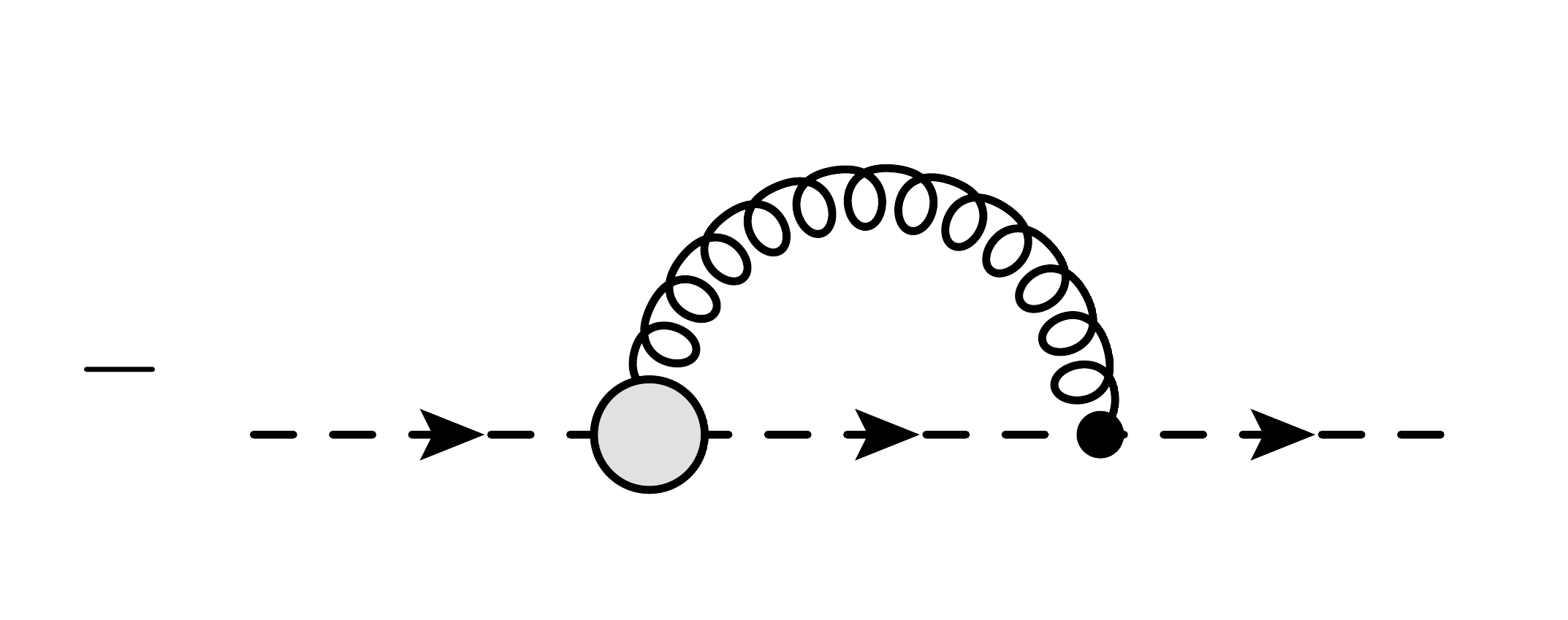}
\caption{Ghost self-energy.}
\label{fig:ghostself}
\efc
%
\begin{figure}[t]\begin{center} 
\includegraphics[width=.9\columnwidth]{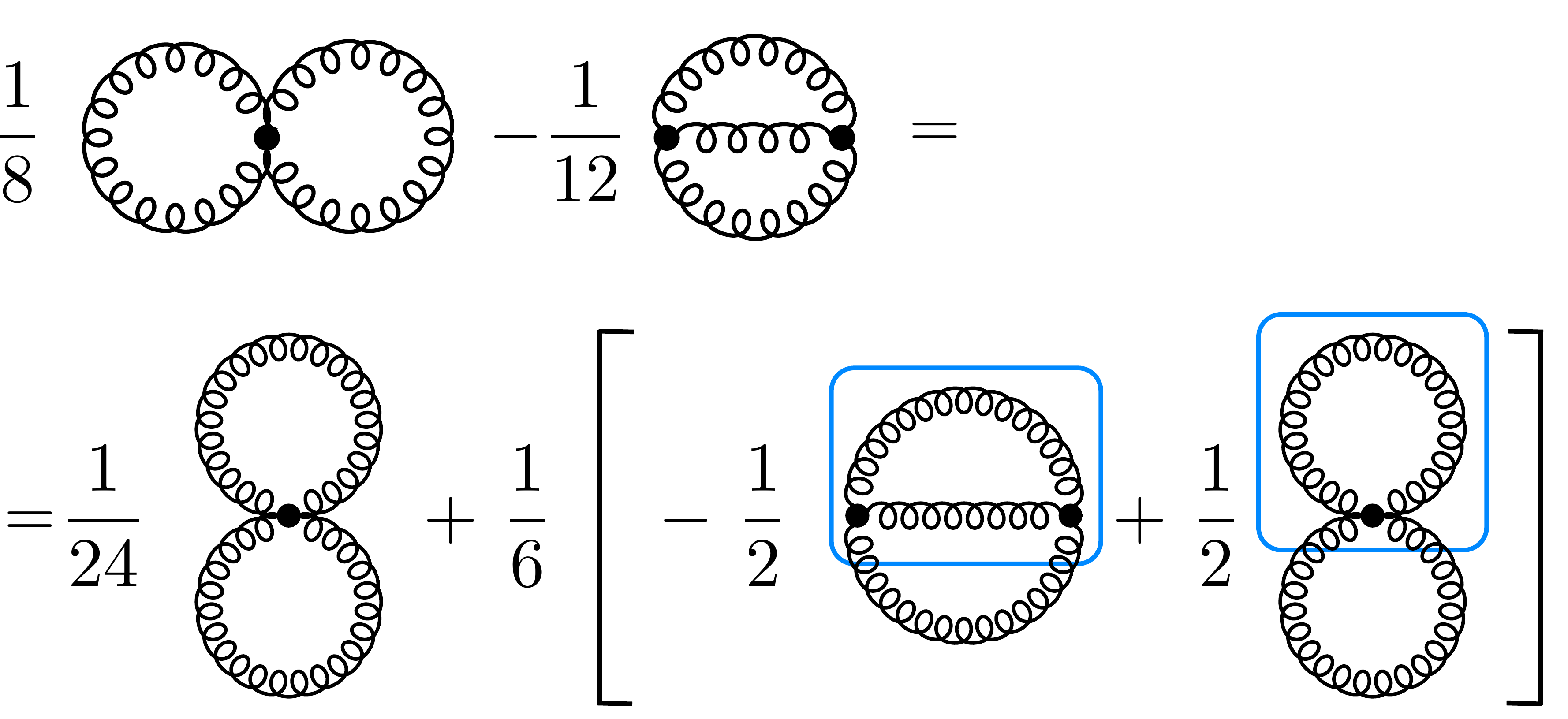}
\caption{Identification of gluonic part of the vacuum polarisation in
  the gluonic two-loop diagrams in the 2PI effective action depicted in
  \Fig{fig:Phi}.}
\label{fig:gluon_2_loop_split}
\efc
%
\begin{figure}[t]\begin{center} 
\includegraphics[width=.95\columnwidth]{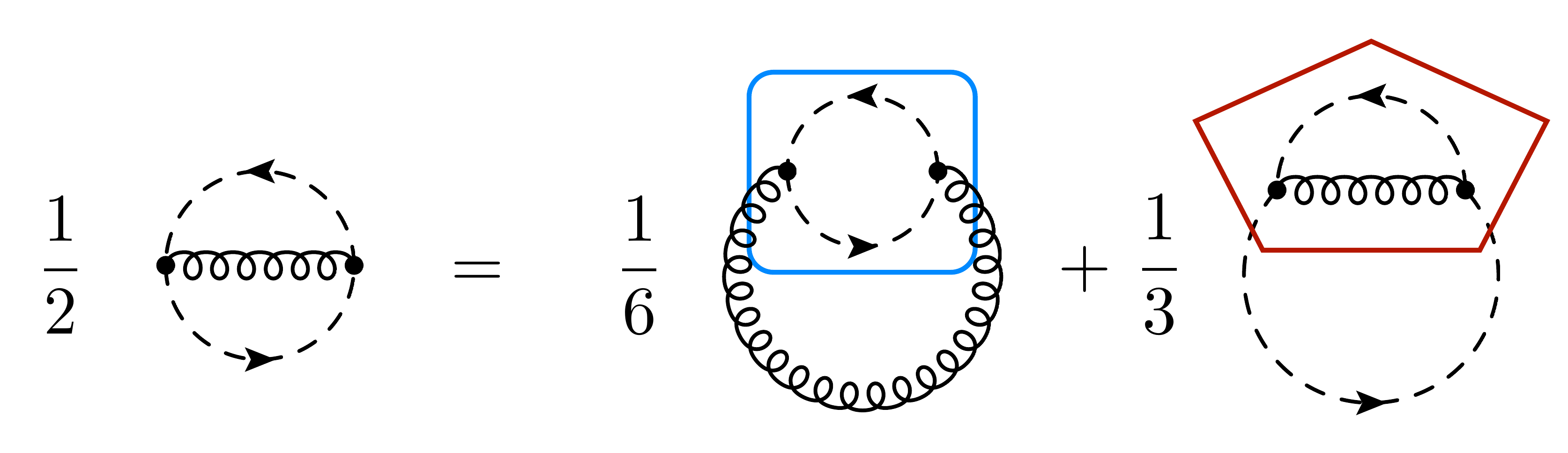}
\caption{Identification of the ghost part of the vacuum polarisation
  and the ghost self-energy in the ghost-gluon two-loop diagrams in the
  2PI effective action depicted in \Fig{fig:Phi}.}
\label{fig:ghost_2_loop_split}
\efc
%

The full computation of the Polyakov loop potential \eq{eq:DSEPolPot}
from the DSE including the two-loop terms requires the knowledge of the
full three-gluon vertex and the full ghost-gluon vertex in the
presence of an $A_0$-background, see \eq{eq:DSE} and
\Fig{fig:DSE}. Moreover, one has to compute two-loop diagrams in the
presence of such a background. Here, we present a simple resummation
scheme for effectively computing these terms in an expansion of the
Polyakov loop potential in full propagators and full vertices. The
simplest way to do this is to consider the 2PI-hierarchy in
\Fig{fig:Phi} up to two-loop. As it also provides some additional
information about the resummation scheme we also present a derivation 
solely within the DSE-approach in Appendix~\ref{app:2loop-DSE-rep}. 

The two-loop terms in $\Phi$ can be rewritten in terms of one-loop
diagrams with full propagators and gluon vacuum polarisation and ghost
self-energy, see \Fig{fig:gluonvacpol} and \Fig{fig:ghostself},
respectively. We easily identify the corresponding diagrams in
$\Phi$. The first gluonic diagrams in $\Phi$ as in
\Fig{fig:Phi} can be rewritten in terms of (-1/6 of) the gluonic part
of the vacuum polarisation $\Pi_a$ contracted with the full gluon propagator,
which is depicted in \Fig{fig:gluon_2_loop_split}. 

Then, the last two-loop diagram in $\Phi$, \Fig{fig:Phi}, is rewritten
in terms of the missing ghost contribution to the gluonic vacuum
polarisation $\Pi_a$ in \Fig{fig:gluon_2_loop_split}, and (-1/3) of the
ghost self-energy $\Pi_c$ contracted with the full ghost propagator,
see \Fig{fig:ghost_2_loop_split}. Both terms are subtractions to the
1PI-terms $\Tr ( -1/2 \Pi_a G_a + \Pi_c G_c)$ in the 2PI effective
action, \eq{eq:2PI}.  This leads to the final form of the effective
action,
\begin{eqnarray}\label{eq:2PI-2loop}
 \hspace{-.4cm}\Gamma[ A_0;\phi]&=& S_A[A_0;\phi]-\012 \Tr
\log G_a + \Tr \log G_c\\[1ex]
&& -\023 \left(\012 \Tr \,\Pi_a G_a- 
 \Tr \,\Pi_c G_c\right)+\Delta\Gamma_3[A_0;\phi]\,, 
\nonumber  \end{eqnarray}
with 
\begin{equation}\label{eq:DeltaG}
  \Delta\Gamma_3[A_0;\phi]= \01{24} G_a S^{(4)}_{aaaa} G_a 
  +O(\alpha_s^2)\,,  
\end{equation}
comprising the contributions of order $\alpha_s^2$ in a perturbative
2PI ordering as well as a tadpole contribution. As mentioned before, a
derivation of \eq{eq:2PI-2loop} within the DSE-approach can be found in 
Appendix~\ref{app:2loop-DSE-rep}. In the explicit
computations presented below we will drop $\Delta\Gamma_3$. The
tadpole term is set to zero at $T=0$ in standard renormalisation
schemes as it is a mass-contribution. Its thermal $A_0$-dependent
contribution is suppressed by roughly one order of magnitude, $1/8$,
relative to the contribution of the full tadpole in the final
expression, see \eq{eq:2PI-2loop}. Moreover, its $A_0$-dependent part
solely contributes to the $\Delta m^2(0;A_0)$ defined in
\eq{eq:Deltam} which we have consistently dropped in the explicit
computations presented here. 

In conclusion, with $\Delta\Gamma_3\approx 0$, we have arrived at a
representation \eq{eq:2PI-2loop} of the effective action which is
manifestly RG-invariant. Indeed, both the
first and the second line are separately RG-invariant. One is tempted
to even drop the second line in \eq{eq:2PI-2loop} in a first
computation. However, even though this should not have a qualitative
impact on the phase transition temperature, it turns out to have a big
impact on the amplitude of the potential. We have checked this within
the flow equation for the Polyakov loop potential: \eq{eq:2PI-2loop}
resembles \eq{eq:total}. The first lines agree up to the
normalisation at $k=\Lambda$ which is trivial. Thus, we deduce that up to the
above normalisation the
second line of \eq{eq:2PI-2loop} is identical to
\begin{eqnarray}\label{eq:agree}
  &&\hspace{-.6cm}-\023 \left(\012 \Tr \,\Pi_a G_a- 
    \Tr \,\Pi_c G_c\right)+\Delta\Gamma_3[\bar A;\phi] \\
  &=&-\int_\Lambda^0 \0{d k}{k}\sumint_p\ \left( \frac{1}{2} \left[G_a 
      \, \partial_t \Gamma^{(2)}_a\right]_{\mu\mu}^{aa}  - 
    \left[G_c   \, \partial_t \Gamma^{(2)}_c\right]^{aa}\right)\,,
  \nonumber \end{eqnarray}
where the right hand side has the manifest RG-invariance due to being an integrated flow. 
We have already discussed in Section~\ref{sec:confcrit} below \eq{eq:conf07} that 
the integrated flow in \eq{eq:agree} is not negligible for the potential even though 
it does not play a major r${\rm \hat o}$le for the phase transition temperature. 
We emphasise that a detailed
comparison of the different approximations is of great interest for
phenomenological applications in QCD, and in particular also for 
approaches to QCD within Polyakov-loop enhanced low energy models, see e.g.\  
\cite{Meisinger:1995ih,Pisarski:2000eq,Fukushima:2003fw,%
  Mocsy:2003qw,Dumitru:2003hp,Ratti:2005jh,Ghosh:2006qh,%
  Schaefer:2007pw,Sakai:2008py,Skokov:2010wb,Herbst:2010rf}. 
This analysis is beyond the scope of the present work and will be presented elsewhere. 

Hence, we also consider the second line in \eq{eq:2PI-2loop} and use
the scheme-independence of the result in \eq{eq:2PI-2loop} to further
simplify the computation: With the Dyson equation,
\begin{equation}\label{eq:Dyson}
G_\phi^{-1}= z_\phi S^{(2)}_\phi+ \Pi_\phi\,,
\end{equation} 
we rewrite the DSE for the effective potential, \eq{eq:DSEPolPot}, in
terms of the one-loop diagrams in \Fig{fig:DSE} and the rest,
\beqa \nn \hspace{-.3cm}\beta \CV\0{\partial V[A_0]}{\partial A_0}&=&
\012 S^{(3)}_{A_0 aa} \,G_{a} - S^{(3)}_{A_0 c\bar c}\, G_{c}\\[1ex]
\nonumber &&
+ \013\left( \012 \Tr\, G_a\0{\partial
 \, \Pi_a}{\partial A_0} - \Tr\, G_c\0{\partial
 \, \Pi_c}{\partial A_0} \right)\\[1ex]
& &
-\023\left( \012 \Tr \,\0{\partial G_a}{\partial A_0}  \Pi_a
-\Tr \,\0{\partial G_c}{\partial A_0}  \Pi_c\right)\,.
\label{eq:DSERGinv}
\eeqa
After performing the Lorenz traces in the second line of \eq{eq:DSERGinv}, 
the remaining operator traces are simply of the form 
\begin{equation}\label{eq:optrace} 
\Tr \,\0{\partial x}{\partial A_0} \0{Z+x\,Z' -z}{x\, Z} \,,\qquad 
\Tr\, \0{\partial x}{\partial A_0} \0{(Z-z)(Z + x\, Z')}{x \, Z^2} 
\end{equation}
where $x=-D^2(A_0)$, $Z=Z(-D_0^2,-\left.\vec D\right.^2)$ are the respective wave
functions, see \eq{eq:parahatG}. The abbreviation $Z'$ stands for the
the $p_0^2$-derivative of $Z$, 
\begin{equation*}
Z'(-D_0^2,-\left.\vec
D\right.^2)=\left.\partial_{p_0}^2 Z(p_0^2,-\left.\vec D\right.^2)\right|_{p_0^2=- D_0^2}\,.
\end{equation*} 
Now we minimise the size of the contributions from the second and
third line in \eq{eq:DSERGinv} close to the transition temperature by
using an appropriate RG-scheme, that is an appropriate renormalisation
scale $\mu^2$ with $z=Z(\mu^2)$ at $T=0$. Since we are interested in
the physics at temperatures at about $\Lambda_{\rm QCD}$ this already
implies $\mu\approx 1$ GeV. We can further restrict this choice
by demanding that for $p_0=0$ and $A_0=0$ we have
\begin{eqnarray}  \nonumber 
&& \hspace{-.4cm}\Biggl[ \012 \0{Z_a+x\,Z_a' -z_a}{
    Z_a}
  +\0{(Z_a+x\,Z_a')(Z_a-z_a) }{ Z_a^2}\nonumber\\[1ex]
  &&\hspace{-.4cm}- \0{Z_c+x\,Z_c' -z_c}{ Z_c}- 2
  \0{(Z_c+x\,Z_c')(Z_c-z_c)}{ Z_c^2}\Biggr]_{x=\mu} =0\,.\nn\\
 \label{eq:zero} \end{eqnarray} 
\Eq{eq:zero} minimises the integrand in the momentum integrals in the
second and third line in \eq{eq:DSERGinv} at the momenta with the
largest contributions. In \eq{eq:zero} a sum over the chromoelectric
and the two chromomagnetic polarisations is understood. 
For temperatures $T \lesssim 200-300$ MeV \eq{eq:zero} is solved 
for 
\begin{equation}\label{eq:muopt} 
\mu_{\rm opt} \approx 1.08\ {\rm GeV} \,.
\end{equation} 
We also can investigate the $\mu$-dependence of the splitting related
to the normalisation with $z$. To that end we apply the related variation to 
\eq{eq:zero}, that is the operator $\mu\partial_\mu z_{\phi_i} \partial_{\phi_i}$ and   
obtain from \eq{eq:zero}
\begin{equation} \label{eq:zerotest}
\left[\012 \eta_{a,0}
  (3-\eta_{a,T})-\eta_{c,0} (3-\eta_{c,T})\right]_{x=\mu}
  =0\,, 
\end{equation} 
with 
\begin{equation}\label{eq:eta}
\eta_{\phi,T}(p^2)=-\0{p\partial_p Z_{\phi}(p^2)}{Z_\phi(p^2)} \,.
\end{equation}
Again this yields $\mu\approx 1$ GeV which entails that the
contribution of the second and third line in \eq{eq:DSERGinv} can be
minimised for temperatures close to $T_c$ along \eq{eq:zero}.  Note
also that, since the differences are marginal, it can also be
evaluated at $T=0$ as a first good approximation. There we have
$z=Z(\mu^2)$ and \eq{eq:zero} simplifies to
\begin{eqnarray}\label{eq:zeroT0}
 \left[\012 \0{Z_a+x\,Z_a' -z_a}{x\, Z_a} 
- \0{Z_c+x\,Z_c' -z_c}{x\, Z_c}\right]_{x=\mu} =0\,,
\end{eqnarray}
which is satisfied for $\mu\approx 1$ GeV as well. 
The constraint \eq{eq:zero} fixes the RG-scale with $\mu_{\rm
  opt}=1.08$ GeV close to $\Lambda_{\rm QCD}$ and $\pi T_c$ in the
temperature regime between $T=0 - 300$ MeV.  Moreover, it can be shown
that a variation of $\mu$ in this regime by 100\% only leads to
changes of the results on the percent level, in particular $T_c$ does
not change at all. For the numerical computations we have chosen
\begin{equation}\label{eq:muchoice} 
z_a=Z_a(\mu^2)\,,\quad z_c=Z_c(\mu^2)\,,\quad {\rm at}\quad 
\mu=1.08\ {\rm GeV}\,. 
\end{equation}
Note that this RG-scheme effectively normalises the propagators used
in the DSE and 2PI hierachy to one at the RG-scale, $z/(
\Gamma^{(2)}(\mu^2)/\mu^2) = 1$. At finite temperature this only holds
approximately. 

In summary we have derived that for appropriately
chosen RG-schemes with $z=Z(\mu)$ the DSE for the Polyakov loop
potential effectively boils down to a simple one-loop form, depicted
in \Fig{fig:DSE_1_loop}.
\begin{figure}[t]
\begin{center}
\includegraphics[width=.7\columnwidth]{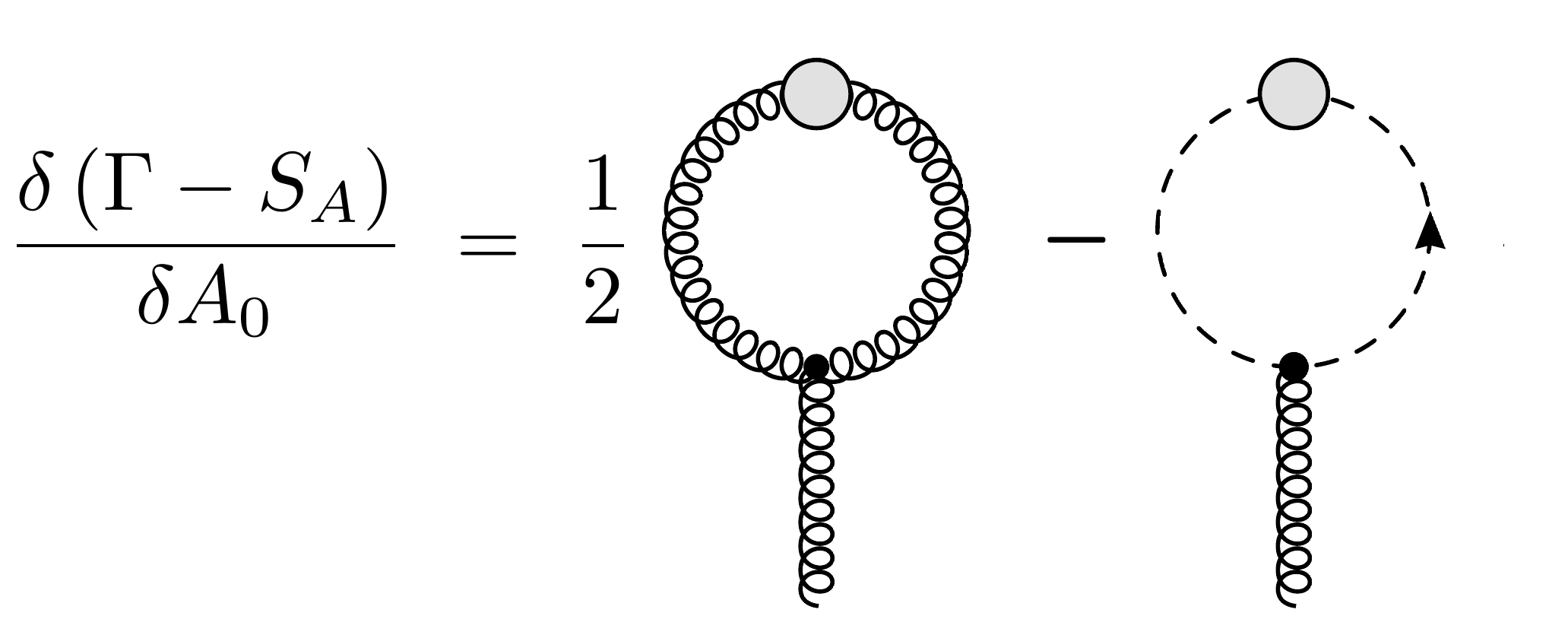}
\caption{One-loop truncated DSE for the effective action.}
\label{fig:DSE_1_loop}
\end{center}
\end{figure}
\begin{figure}[t]
\begin{center}
\includegraphics[width=.95\columnwidth]{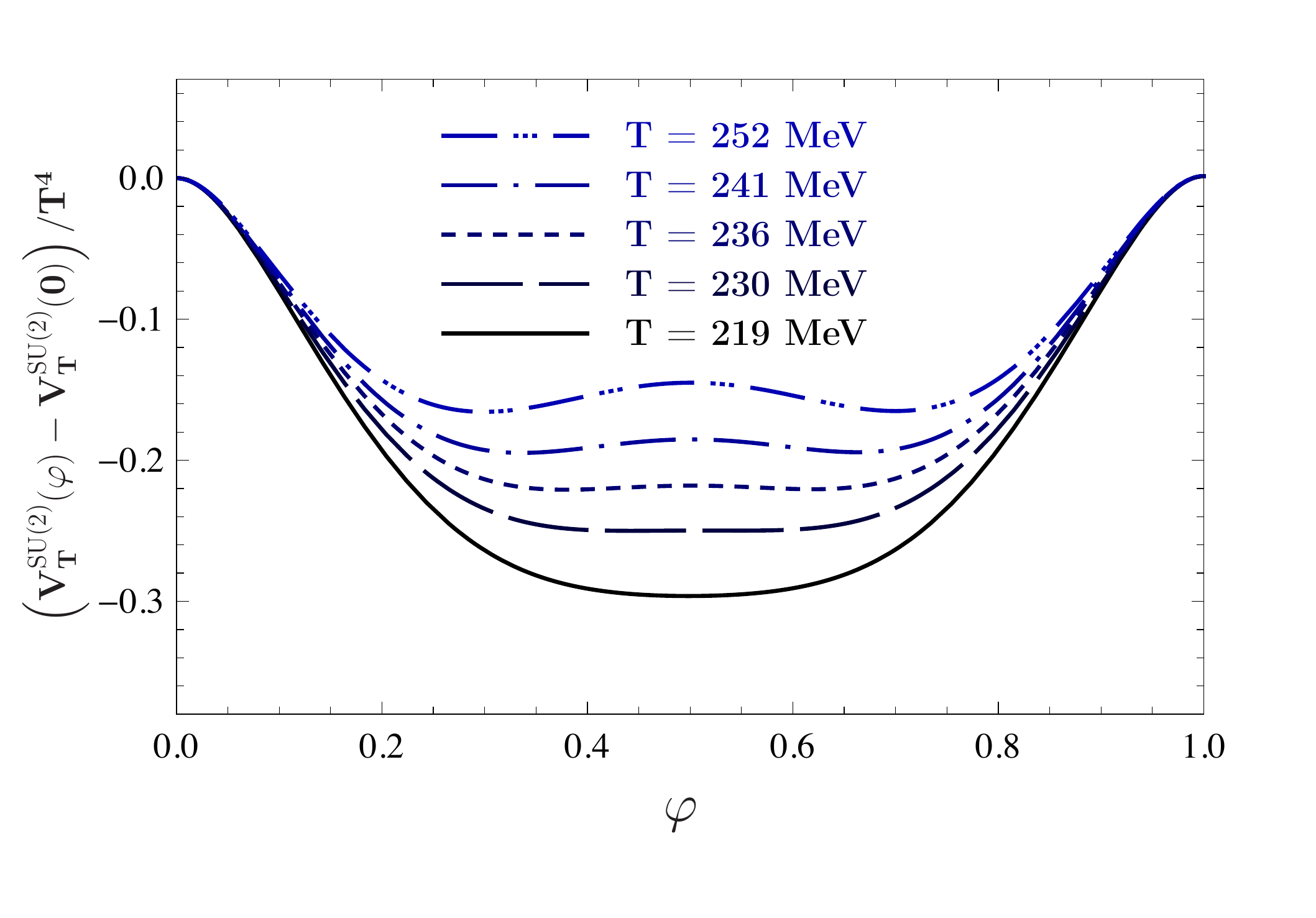}
\caption{Polyakov loop potential for
  $SU(2)$ from its DSE/2PI-representation within an optimised RG-scheme.}
\label{fig:SU2_DSE_T}
\end{center}
\end{figure}

In Appendix~\ref{app:one-loop} the one-loop diagrams in
\Fig{fig:DSE_1_loop} are reduced to 
\begin{eqnarray} \nonumber V'(\varphi)& =& \012 V'_{\rm
    Weiss}(\varphi)+\sumint_q \,2 \pi T\left(n+\varphi
  \right)\\[1ex]
  && \hspace{-.5cm} \times \left[ z_a G_L\left(q+2 \pi T\varphi
    \right)+ 2\, z_a G_T\left(q
      +2 \pi T\varphi \right)\right.\nonumber\\[1ex]
  &&\left.-2 z_c G_c\left(q+2 \pi T \varphi\right)\right] \,,
\label{eq:DSEpolpot}\end{eqnarray}
where $G_{L},\,G_{T},\,G_{c}$ are the propagators for the
chromoelectric gluon, the chromomagnetic gluon and ghost propagators
normalised $z_{a/c}\!=\!Z_{a/c}(\mu^2)$ at vanishing temperature at
the point $p^2=\mu^2$, and the first term originates in the gauge
mode. The evaluation of \eq{eq:DSEpolpot} leads to the potential
depicted in \Fig{fig:SU2_DSE_T} for $SU(2)$ and \Fig{fig:SU3_DSE_T}
for $SU(3)$.  From those the confinement-deconfinement critical
temperature is determined at
\begin{equation}\label{eq:DSE+2PI-Tc}
  T_{c}^{SU(2)}= 235\pm 25\, {\rm MeV}\ \ {\rm and}  \ \
  T_{c}^{SU(3)}= 275\pm 25\, {\rm MeV}\,, 
\end{equation} 
which gives 
\begin{equation}
T_{c}^{SU(2)}/\sqrt{\sigma}\approx 0.56 \quad {\rm  and
}  \quad T_{c}^{SU(3)}/\sqrt{\sigma}\approx 0.655\,.
\end{equation}  
These results are in quantitative agreement with the FRG results,
\eq{eq:FRG-Tc}, and can be compared with lattice results in
\eq{eq:lattice-Tc} and \eq{eq:lattice-Tc_ratios}.

We close this chapter with a comparison of the temperature-dependence
of the FRG and DSE/2PI potentials. Note that the DSE/2PI computation
presented here approximately also takes into account the integrated
RG-improvement in \eq{eq:agree} for temperatures about the phase
transition temperature. Hence, not only $T_c$ but also the potentials
themselves should agree quantitatively for temperatures close to
$T_c$. We also remark that for $T\to 0$ the present approximation may
lose its quantitative character as the normalisation of the ghost and
gluon propagators changes rapidly for momenta $p^2 \approx (2 \pi
T)^2$ important for the Polyakov loop potential. In \Fig{fig:SU2comp}
we only show the comparison of the potentials for $SU(2)$, as that for
$SU(3)$ simply follows from an weighted sum over $SU(2)$-potentials,
cf. Appendix~\ref{app:one-loop}. The potentials agree quantitatively
to a surprising level of accuracy which confirms the quantitative
nature of both, FRG and DSE/2PI-computations. For temperatures
$T/T_c\to 0$ there is a significant deviation as the
DSE/2PI-approximations used here loses its quantitative nature. Note
that this also holds for the approximation \eq{eq:Vinttotal}. In turn
for $T/T_c\to \infty $ both approaches converge towards the
perturbative potential. The present approximation in the
DSE/2PI-approach is easily improved by taking into account the
two-loop terms in \eq{eq:DSERGinv}. The discussion of the respective
results goes beyond the scope of the present work and will be
presented elsewhere.
\begin{figure}[t]
\begin{center}
\includegraphics[width=.95\columnwidth]{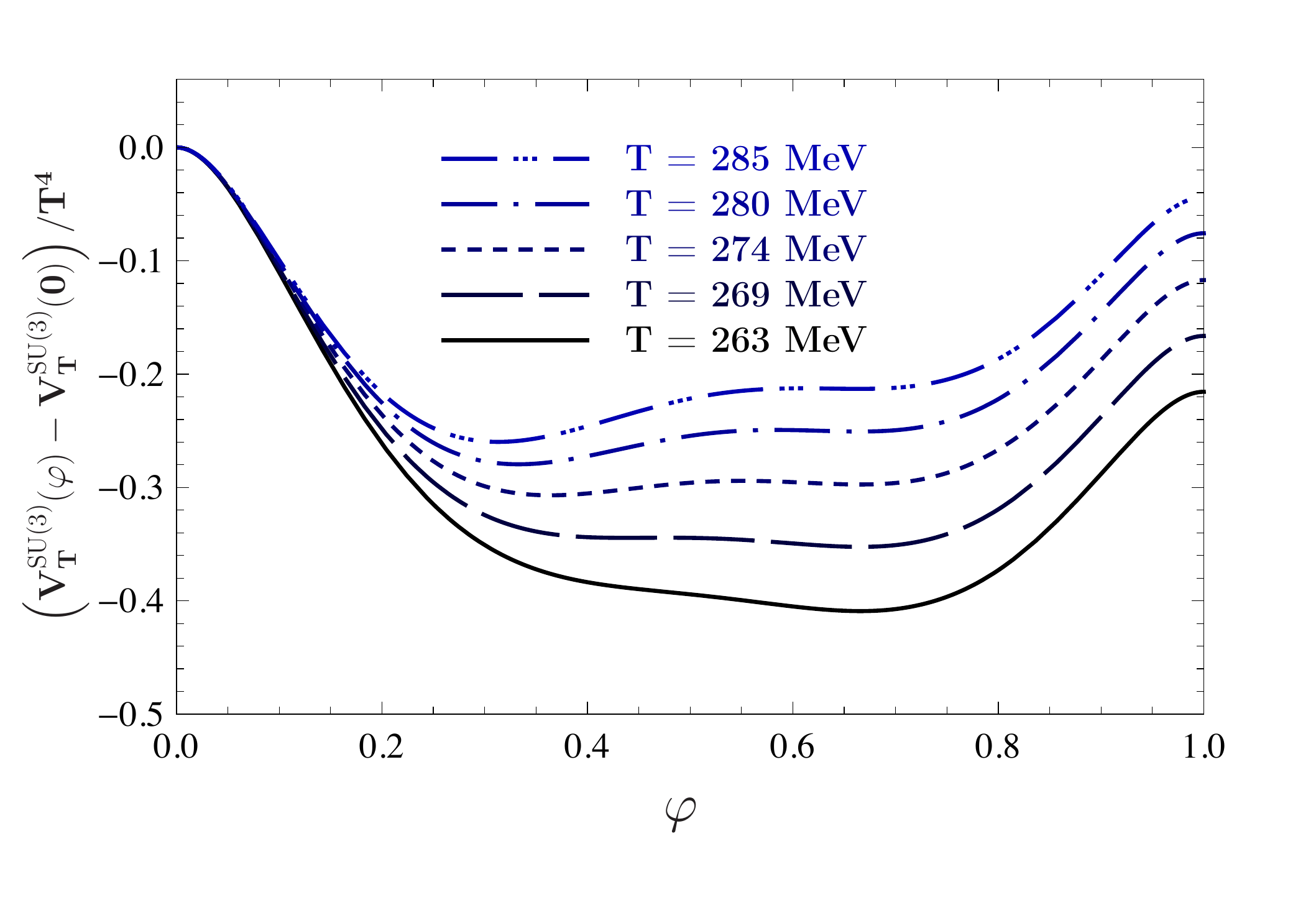}
\caption{Polyakov loop potential for
  $SU(3)$ from its DSE/2PI-approach within an optimised RG-scheme.}
\label{fig:SU3_DSE_T}
\end{center}
\end{figure}
\begin{figure}[b]
\begin{center}
\includegraphics[width=.95\columnwidth]{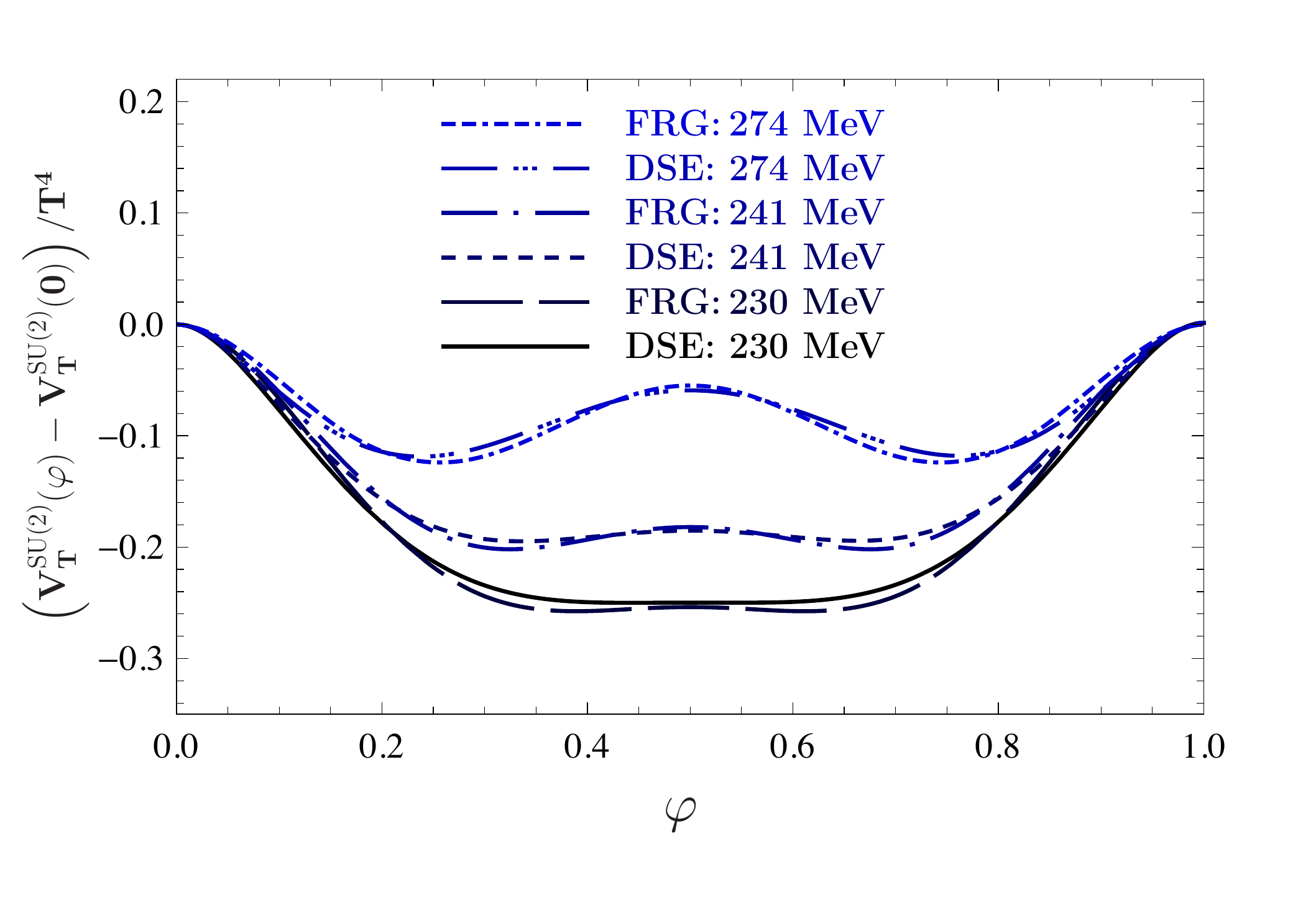}
\caption{Comparison of the Polyakov loop potential for
  $SU(2)$ from the DSE/2PI- with the FRG-approach.}
\label{fig:SU2comp}
\end{center}
\end{figure}

\section{Conclusions}\label{sec:conclusions} 
We have studied the confinement-deconfinement phase transition of
static quarks with the effective potential of the Polyakov. The
position of the minima signals the presence or absence of confinement,
as they can be related to the free energy of a single quark. The
potential itself was computed in the background field formalism in the
Landau--deWitt gauge. This allows for the use of Landau gauge
propagators at vanishing background, which had been obtained
previously in the framework of the FRG
\cite{Fister:2011um,Fister:2011uw, Fister:Diss}. The study here was
done within different non-perturbative functional continuum
approaches, i.e. using FRG-, DSE- and 2PI-representations of the
effective action.

In all variants we find a second order phase transition for $SU(2)$
and a first order transition for $SU(3)$, as it is expected from
lattice QCD. The corresponding temperatures agree quantitatively for
all functional methods within the estimated error of approximately
10\%, originating in the normalisation of the total momentum scale.
The results including the lattice temperatures are summarised and
discussed in Section~\ref{sec:resFRG} and Section~\ref{sec:resDSE}.
These results are stable w.r.t.\ the choice of scaling or decoupling
type solutions for the propagators and, furthermore, to the variation
of the regulator in the FRG approach. 
The critical temperature for $SU(3)$ from functional methods agrees
quantitatively with the corresponding lattice temperature. For $SU(2)$
the critical temperature from functional methods is significantly
lower than the lattice temperature. This is linked to the missing
back-reaction of the fluctuations of the Polyakov loop potential, see
Section~\ref{sec:critical}. Its inclusion leads to critical
temperatures in quantitative agreement with the lattice results and
the correct Ising-type critical scaling, see
\cite{Marhauser:2008fz,Spallek}.

The dependence of the Polyakov loop potential on the individual gluon
and ghost modes is easily tractable. This was exploited to derive a
simple criterion for static quark confinement. For infrared suppressed
gluon but not suppressed ghost propagators confinement is immanent at
sufficiently low temperatures. Independent of the functional approach
chosen above, this is due to the fact, that the two non-suppressed
confining ghost modes always overcompensate the deconfining trivial
gluon gauge mode. As a result of their suppression, the remaining
transversal gluon modes decouple at temperature scales far lower than
the suppression scale and confinement is realised.

This confinement criterion has also been applied to general
Yang--Mills-matter systems with matter in the adjoint and the
fundamental representation. For matter in the adjoint representation
again an infrared suppressed gluon is mandatory. Moreover, the
criterion is sensitive to the difference between Higgs phase and
confined phase even though both phases have infrared suppressed
gluons. For matter in the fundamental representation the matter part
of the potential breaks center symmetry. Hence it is only possible to
evaluate the relative importance of center symmetry modes and center
breaking modes.

In summary the work establishes both on the qualitative and
quantitative level the functional approach to confinement by means
of the Polyakov loop potential. It only requires the computation and
discussion of background-gauge propagators and hence allows a very
simple access to structural questions such as the confinement
mechanism as well to explicit analytic and numerical computations of
the phase structure.

{\it Acknowledgments} --- We thank Jens Braun, Astrid Eichhorn,
Christian S.~Fischer, Holger Gies, Jan~L\"ucker and Axel Maas for
discussions.  This work is supported by the Helmholtz Alliance
HA216/EMMI and by ERC- AdG-290623. LF is supported by the Science
Foundation Ireland in respect of the Research Project 11-RFP.1-PHY3193
and the Helmholtz Young Investigator Grant VH-NG-322.

\appendix

\section{Regulators}\label{app:regs}
The regulator function introduced in the FRG approach can be chosen
freely as long as it fulfills certain limits. Firstly, it must have
the same tensor structures as the corresponding two-point
function. Neglecting those tensors for the moment, $R_k(p)$ must (i)
regulate the propagator in the infrared, (ii) leave the full quantum
theory for vanishing renormalisation group scale $k$ and (iii)
suppress fluctuations at large scales to reduce to the classical
action. Formally, these requirements are met by 
\beqa
&{\rm (i)}& \lim_{p^2/k^2 \rightarrow 0}R_k(p) > 0\,,\\
&{\rm (ii)}& \lim_{k^2/p^2 \rightarrow 0}R_k(p) = 0\,,\\
&{\rm (iii)}& \lim_{k^2/p^2 \rightarrow \infty}R_k(p) \rightarrow
\infty\,.  
\eeqa 
The full theory is independent of the choice of
$R_k(p)$, however, in truncations this may not be satisfied. This
rules out certain forms of the regulator. Another restricting factor
is the applicability in the computations, as some regulators are not
feasible numerically.

For the thermal gluon and ghost propagators of Yang--Mills theory,
four-dimensional exponential regulators, 
\beq R^{(0)}_{k,m}(p)=p^2
r_m(p^2/k^2)=p^2 \frac{\left(p^2/k^2\right)^{m-1}}{e^{\left(p^2/k^2
    \right)^m}-1}, 
\eeq 
with $m=2$, have proven to be well suited for
numerical purposes \cite{Fister:2011um, Fister:2011uw,
  Fister:Diss}. Note however, that for the numerical computation of
the Polyakov loop potential the regulator must not be too sharp, thus,
$m\lesssim 2$, cf.  \cite{Fister:Diss}.

The regulation of the propagator in the infrared demands that
$R_k(p)$ is of approximately the same amplitude than the
$\Gamma^{(2)}_k$ at each $k$. This is ensured by a prefactor such,
that 
\beq R_{L/T/c,k,m}(p)=\bar{Z}_{L/T/c,k} R^{(0)}_{k,m}(p)\,.
\label{eq:regs}
\eeq 
For the purpose presented here the precise form of $\bar Z$ is
not crucial but lengthy, thus, for the exact definitions we refer the
interested reader to \cite{Fister:2011uw,Fister:2011um,Fister:Diss}.
The results presented in section~\ref{sec:resFRG} are obtained with
regulators \eq{eq:regs}.

In order to test the (in)dependence of the Polyakov loop potential on the
choice of the regulator, we have studied different functions
$R^{(0)}_k(p)$ and prefactors in the full regulator function. This was
done for temperature-independent propagators, since not all choices of
$R_k(p)$ are viable for thermal effects.

The scale-dependency of the propagators, of both scaling and
decoupling type, was obtained with \eq{eq:regs} for $m=1,2$
\cite{Fister:2011uw,Fister:2011um, Fister:Diss}. These results were,
firstly, combined with self-consistent choices in the determination of
the Polyakov loop potentials. Secondly, the Polyakov loop potential
was computed with these propagators but $R_k(p)=R^{(0)}_{k,m}(p)$ with
$m=1,2$. Measured on the level of the confinement-deconfinement phase
transition temperature of $SU(2)$, we find good stability w.r.t.\ the
variation of the regulators and, furthermore, w.r.t.\ scaling of
decoupling type propagators, as all choices are within 5\%
deviation. Furthermore, we found that for different choices of the
regulator in propagators and Polyakov loop potential, namely those
which have relative shifts in the renormalisation group scale, the
truncation effects may grow larger.

\section{One-Loop Diagrams in the DSE for the Polyakov Loop
  Potential}\label{app:one-loop}
The one-loop diagrams are given in \Fig{fig:DSE_1_loop}. Both couple
the external temporal gluonic background to the fluctuating loop
fields. For clarity we consider temperature-independent propagators
first, before we decompose the tensor structures for the case of
non-trivial temperature dependence of the propagators.

The gluon loop, depicted in \Fig{fig:gluon_1_loop}, involves the
three-gluon vertex with one background leg. At vanishing fields it is
given by 
\beq\label{eq:aaA-vertex} {\left(S^{(3)}_{A_0 aa}\right)}_{\rho\mn} = {\left(S^{(3)}_{a
    aa}\right)}_{\rho\mn} +\0{1}{\xi} \0{\delta (\bar D_\mu\bar
  D_\nu)}{\delta A_0} \,, 
\eeq 
where the second term originates from the background field dependence
of the gauge fixing term, see \eq{eq:fixedaction}. The classical
vertex is given in \Fig{fig:A3}.
\begin{figure}[t]\begin{center}\subfigure[Gluon one-loop
diagram.]{\includegraphics[width=0.3\columnwidth]{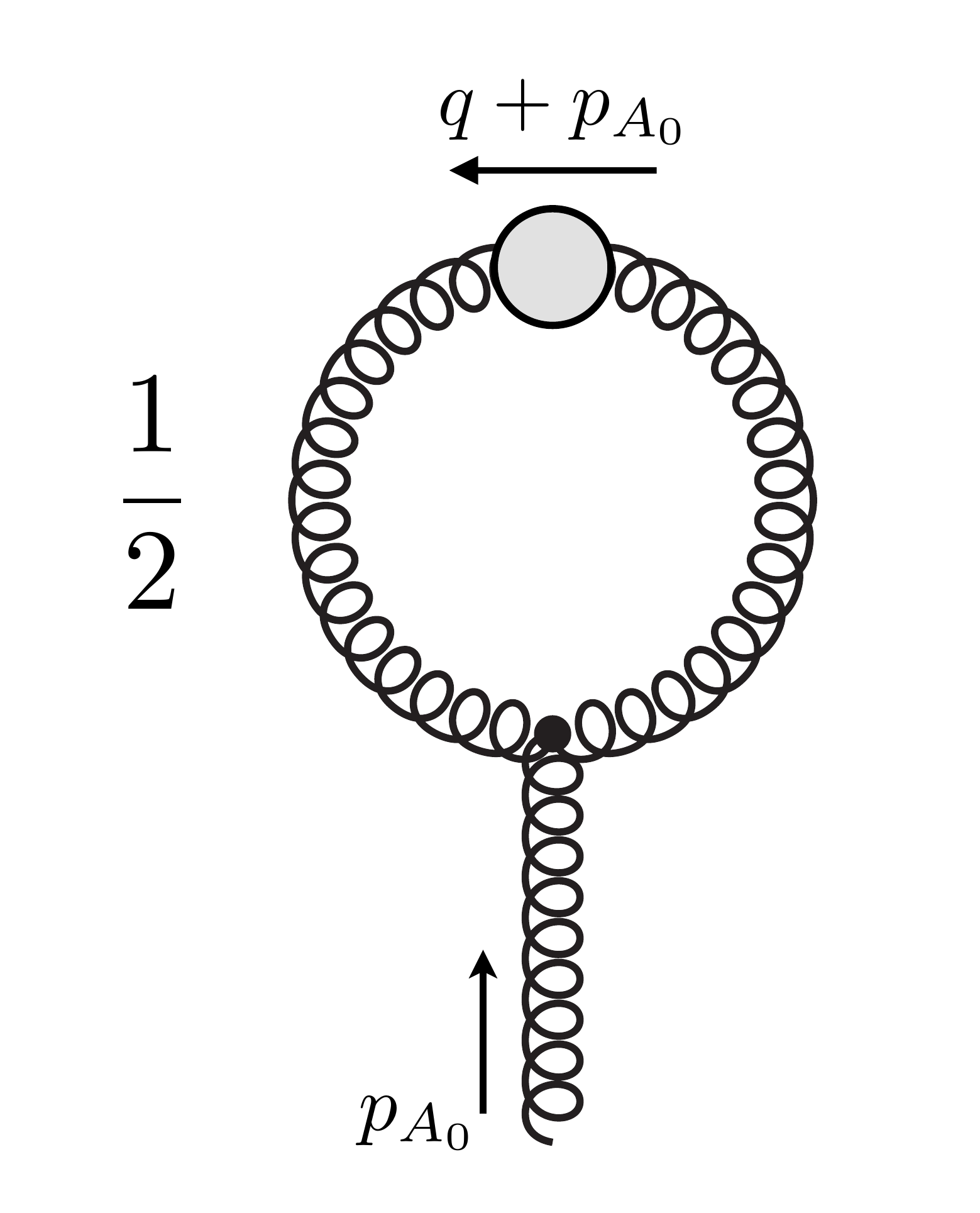}
  \label{fig:gluon_1_loop}} \hspace{.1\columnwidth}
\subfigure[Three-gluon
vertex.]{\includegraphics[width=0.4\columnwidth]{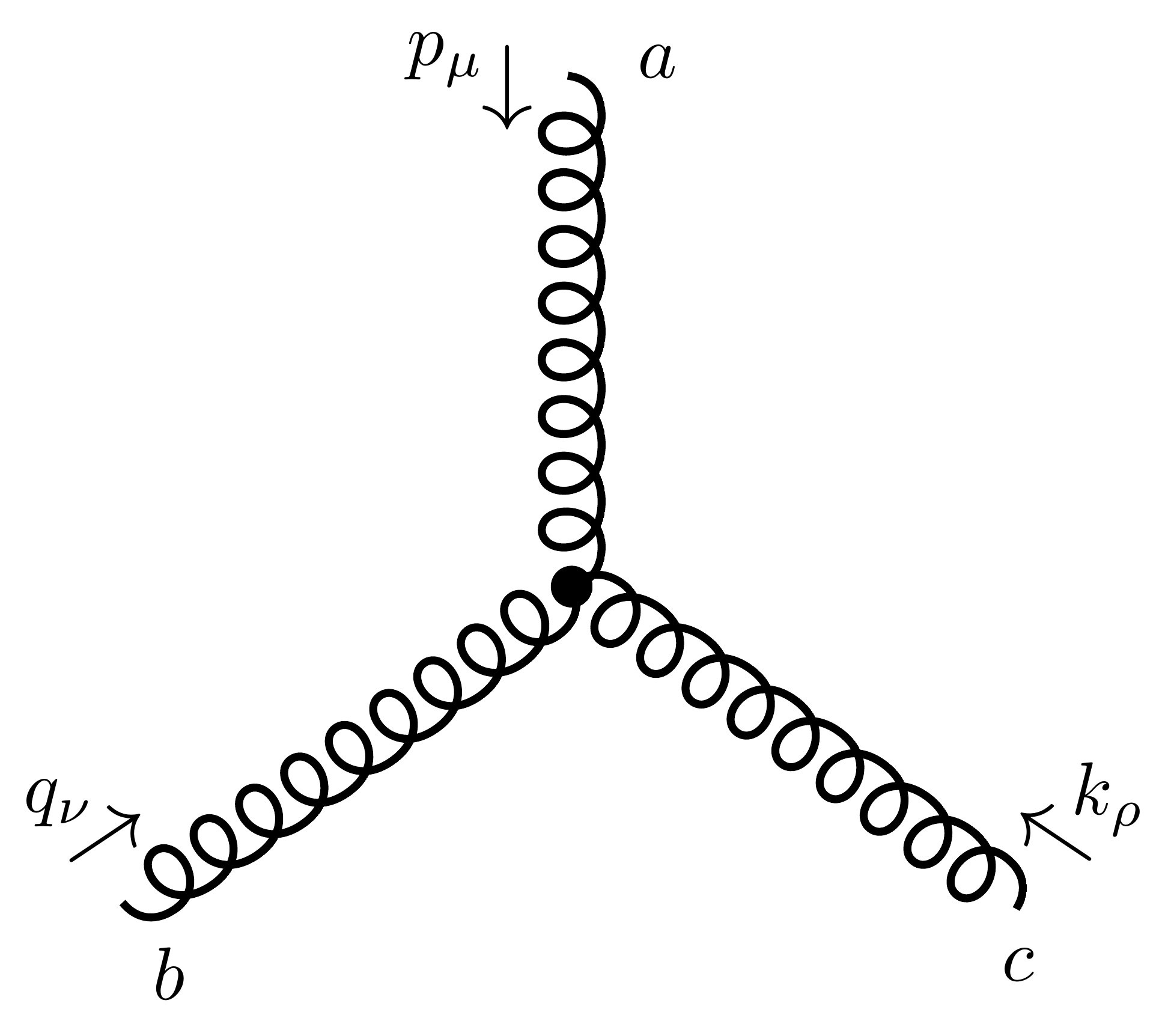}\label{fig:A3}}
\caption{Gluon one-loop diagram in the DSE for the Polyakov loop potential.}
\efc
For this very definition it is given by
\beqa\label{eq:Saaa} &&\left(S^{(3)}_{aaa}\right)^{abc}_{\mu \nu
  \rho}(p,q,k)\\[1ex]\nonumber &=& i g f^{abc} \left( \delta_{\mu
    \nu} \left(p-q \right)_{\rho} + \delta_{\nu \rho} \left( q -k
  \right)_{\mu} + \delta_{\rho \mu} \left(k-p \right)_{\nu} \right)\,.
\eeqa

At $T=0$ the gluon propagator has four degrees of freedom, three
transversal modes and the longitudinal gauge mode, thus, at vanishing
fluctuation fields and a constant background we can split the gluon
propagator according to
\begin{equation}\label{eq:gaugesplit} 
  {\left(G_{A}\right)}_{\mn} = {\left(G_{A}^{\bot}\right)}_{\mn} + \xi \bar D_\mu\0{1}{\bar D^2} \bar D_\nu\,,
\end{equation}
where the first term is the non-trivial transversal part $ \bar D_\mu
{\left(G_{A}^{\bot}\right)}_{\mn}=0$, whereas the second term represents the gauge
mode. This is a consequence of the corresponding Slavnov--Taylor
identitites: The quantum corrections to the propagator are transversal
and, hence, the gauge sector is unchanged. Together with the
background vertex \eq{eq:aaA-vertex} this leaves us with a
contribution of the gauge sector
\begin{equation}
  \012 \Tr \left\{\0{1}{\xi} \0{\delta(\bar D_\mu\bar D_\nu)}{\delta \bar A_\rho} 
  \ \xi \bar D_\nu\0{1}{\bar D^2} \bar D_\mu\right\} = \0{\delta}{\delta \bar A_\rho} \012 
  \Tr \log (-\bar D^2)\,, 
\end{equation}
where the trace on the right hand side does not include a trace over
Lorentz indices and gives $1/2 V_{\rm Weiss}$, the perturbative
one-loop potential. The latter is given by 
\beq V^{\rm
  Weiss}_{SU(N_c)} = - \frac{d-2}{\pi^{d/2}}\Gamma(d/2) \, T^d \,
\sum_{l}^{N_c^2-1}\sum_{n=1}^{\infty} \frac{\cos \left\{ 2 \pi n \nu_l
    \, |\varphi \, |\right\}}{n^d}\,,
\label{eq:Weiss}
\eeq
where $d$ is the number of spacetime dimensions and $\nu_l$ are the
eigenvalues of the generators of the gauge group spanned via the
Cartan subalgebra. For $SU(2)$ the eigenvalues are
$\nu_l={0,1,-1}$. For $SU(3)$ and the background-independent
propagators one can construct the potential according to the
eigenvalues,
\beqa V_{\rm SU(3)} (\varphi^3, \varphi^8)&=& V_{\rm SU(2)}
(\varphi^3)
+ V_{\rm SU(2)} \left(\frac{\varphi^3+3 \sqrt{\varphi^8}}{2}\right) \nn\\
&&+ V_{\rm SU(2)} \left(\frac{\varphi^3-3 \sqrt{\varphi^8}}{2}\right)
\,.  \eeqa
The non-trivial part of the computation concerns the transversal part
${\left(G_{a}^{\bot}\right)}_{\mn}(p)=G_{a}^{\bot}(p) \Pi_{\mn}^{\bot}(p)$, which
couples to the standard part of the three-gluon vertex, $S^{(3)}_{aaa}$,
given in \eq{eq:Saaa}. Note that at vanishing temperature the
propagator is $O(4)$-symmetric, thus, it only depends on the absolute
value of the momentum, not on the spatial and temporal components
separately like at non-vanishing temperature, see \eq{eq:parahatG}.

For a purely temporal momentum $p_{A_0}$ the loop diagram in
\Fig{fig:gluon_1_loop} translates to
\begin{eqnarray}
&&\hspace{-.8cm}\frac{1}{2}\sumint_q  z_a{\left(G_{a}^{\bot}\right)}_{\mn}(p_{A_0}\!\!+q)
{\left(S^{(3)}_{aaa}\right)}_{0\mu \nu }(p_{A_0},p_{A_0}\!\!+\!q,-p_{A_0}\!-\!\!q)\nn\\
&&\hspace{-.75cm}=ig\sumint_q\, z_a G_{a}^{\bot} (p_{A_0}+q) \left(p_{A_0}
  +q\right)_{\rho} \left( \delta_{\mu\mu}-1\right),
\label{eq:compaaa}\end{eqnarray}
where due to symmetry only the term $\propto p_{A_0}+q_0 = 2 \pi
T(n_q+\varphi)$ survives, with $q_0=2\pi T n_q$ and $z_a$ according to
\eq{eq:renormZ}.

In \eq{eq:compaaa} the factor $\left(\delta_{\mu\mu}-1\right)$ counts
the three coinciding transversal modes. For thermal propagators the
projection is such, that it accounts for the chromoelectric and
chromomagnetic modes. Thus, without repeating the computation along
the lines of \eq{eq:compaaa}, we get the final expression for the loop
diagram \Fig{fig:gluon_1_loop} as
\beqa
&&\hspace{-1.5cm}ig \sumint_q\, 2 \pi T (n_q\!+\!\varphi)\left(z_a G_{L}
\left(2\pi T(n_q+\varphi),\vec{q}\right)\right.\nn\\
&&\hspace{1.3cm} \left.+ 2 z_a G_{T}(2\pi T(n_q+\varphi),\vec{q})
\right), 
\eeqa 
with $G_L$ and $G_T$ being the chromoelectric and chromomagnetic gluon
propagators with the normalisation at vanishing temperature given in
\eq{eq:renormZ}.

Having the similiar structure as the gluon loop, the ghost loop is
given in \Fig{fig:ghost_1_loop}.
\begin{figure}[t]\begin{center}\subfigure[Ghost one-loop
    diagram.]{\includegraphics[width=0.28\columnwidth]
{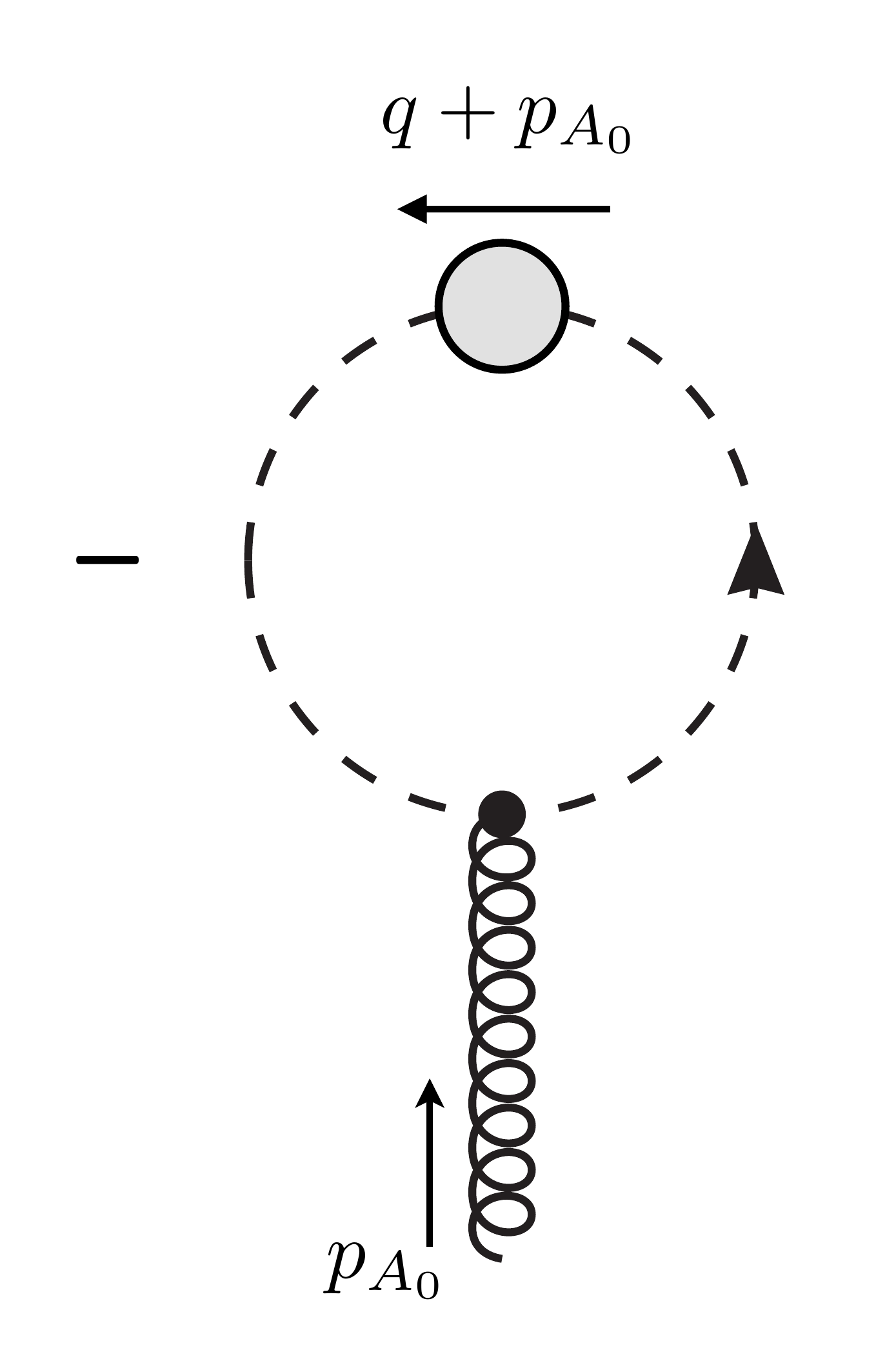}\label{fig:ghost_1_loop}}
\hspace{.1\columnwidth} \subfigure[Ghost-gluon
vertex.]{\includegraphics[width=0.4\columnwidth]{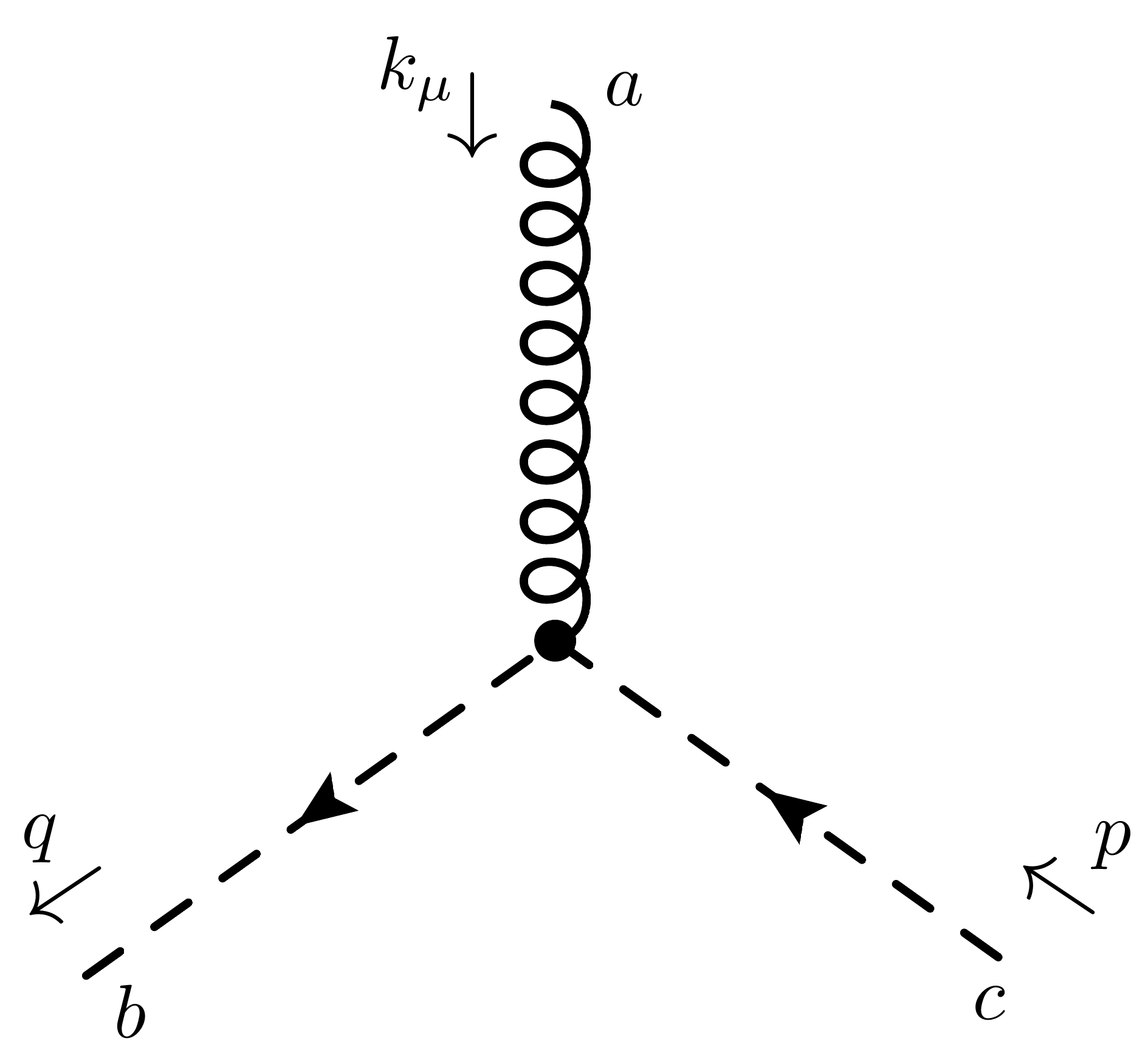}
  \label{fig:cAcb}}
\caption{Ghost one-loop diagram in the DSE for the Polyakov loop
  potential.}
\efc
The first ingredient of this diagram is the classical ghost-gluon
vertex as defined in \Fig{fig:cAcb}. It is given by
\beq \left(S^{(3)}_{A_0
    c\bar c}\right)^{abc}_{\mu}(0,q,p)=2 \left(S^{(3)}_{a c\bar
    c}\right)^{abc}_{\mu}(0,q,p)=2i g f^{abc} q_{\mu} \,.  
\eeq 
Note that the ghost-gluon vertex involves an additional factor of $2$
compared to the vertex in standard Landau gauge, due to the appearance
of the covariant derivative $\bar D_\mu$ (instead of a plain
derivative $\partial_\mu$) in the last term of \eq{eq:fixedaction}.

The second component in the ghost loop is the full ghost
propagator. It is a Lorentz scalar, thus, dropping the colour
structure it is simply given by a scalar function
$G_c(p_0,\vec{p})$. Thus, for trivial and non-trivial implicit
temperature dependence of the propagator, the diagram attached to a
purely temporal gluon field in \Fig{fig:ghost_1_loop} reduces to \beqa
&&\hspace{-1cm}-\sumint_q\!\! z_c G_c(\left(p_{A_0}\right)_0\!\!+
\!q_0,\vec{q})\! \left(S^{(3)}_{Ac\bar c}\right)_{\rho}(p_{A_0}\!+
\!\!q,\!p_{A_0},\!-\!p_{A_0}\!-\!q)\nn\\
&=& -2 i g \sumint_q z_c G_c(\left(p_{A_0}\right)_0+q_0,\vec{q}) \,
(p_{A_0}+q)_\mu\,, \eeqa where $z_c=z_c(\mu)$. Due to symmetry, this
only leaves the term \beq -2 i g\sumint_q z_cG_c(2\pi
T(n_q+\varphi),\vec{q}) \ 2\pi T (n_q+\varphi)\,.  \eeq In combination
with the gauge mode, the sum of the individual contributions yields
the final expression for the DSE of the Polyakov loop potential
\eq{eq:DSEpolpot}.

Due to the different sign in the ghost loop and the gluon suppression
the potential shows confinement for small temperatures. The additive
contributions of the different modes, the chromoelectric,
chromomagnetic, gauge modes from the gluon and the ghost modes, are
illustrated in \Fig{fig:contribs} at a temperature around the phase
transition. For lower temperatures, the ghost contributions grow
stronger and the transverse gluons decrease further. In contrast, for
higher temperatures the propagator approach the perturbative forms,
thus, the potential approaches the (deconfining) Weiss potential
\eq{eq:Weiss}.
\begin{figure}[t]\begin{center}
\includegraphics[width=.9\columnwidth]{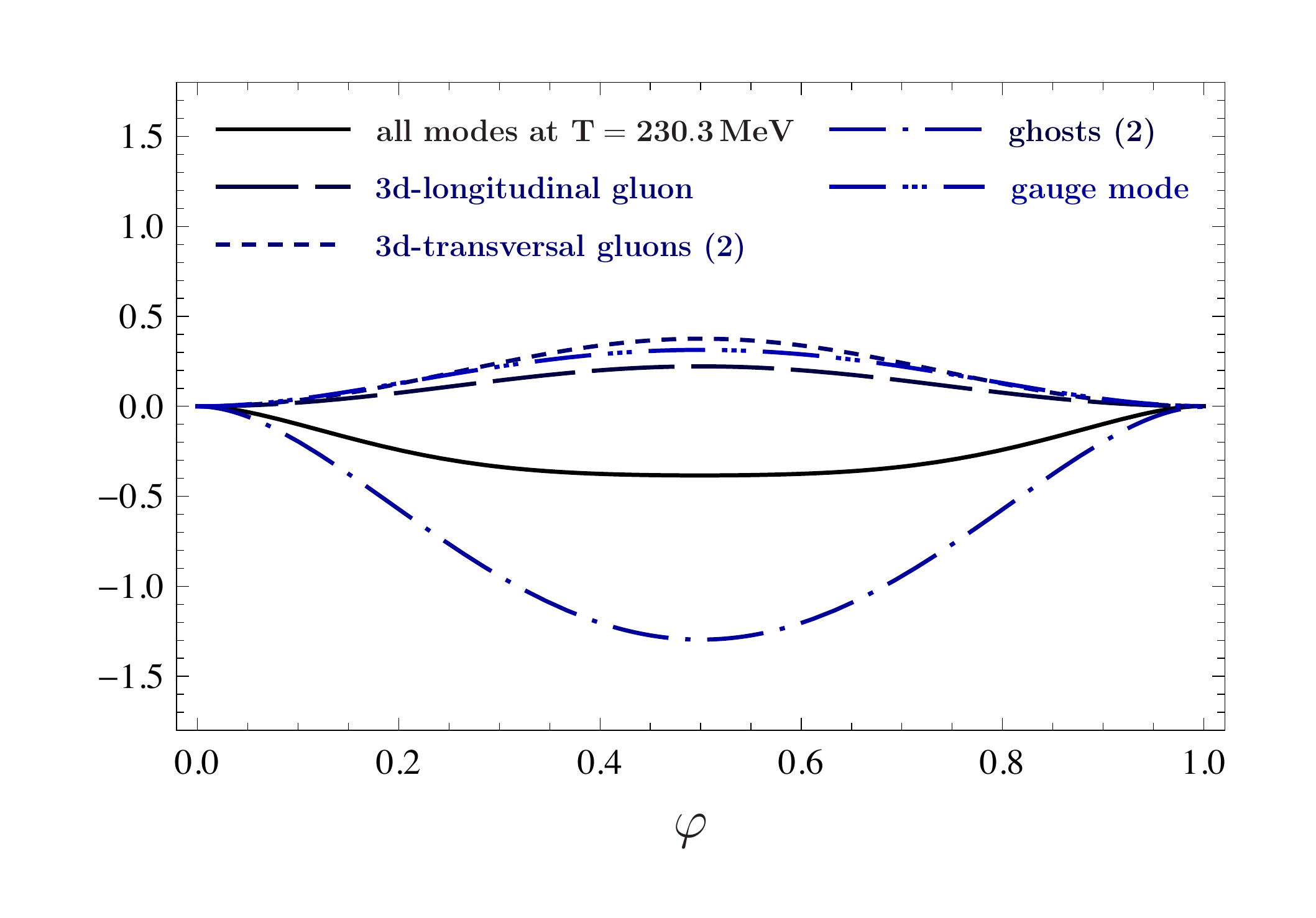}
\caption{Contributions from individual modes to the Polyakov loop
  potential.}
\label{fig:contribs}
\efc

\section{Confinement for Mildly Suppressed Gluons}\label{app:conf_mildsupp}
\begin{figure}[h!]\begin{center}
\includegraphics[width=.95\columnwidth]{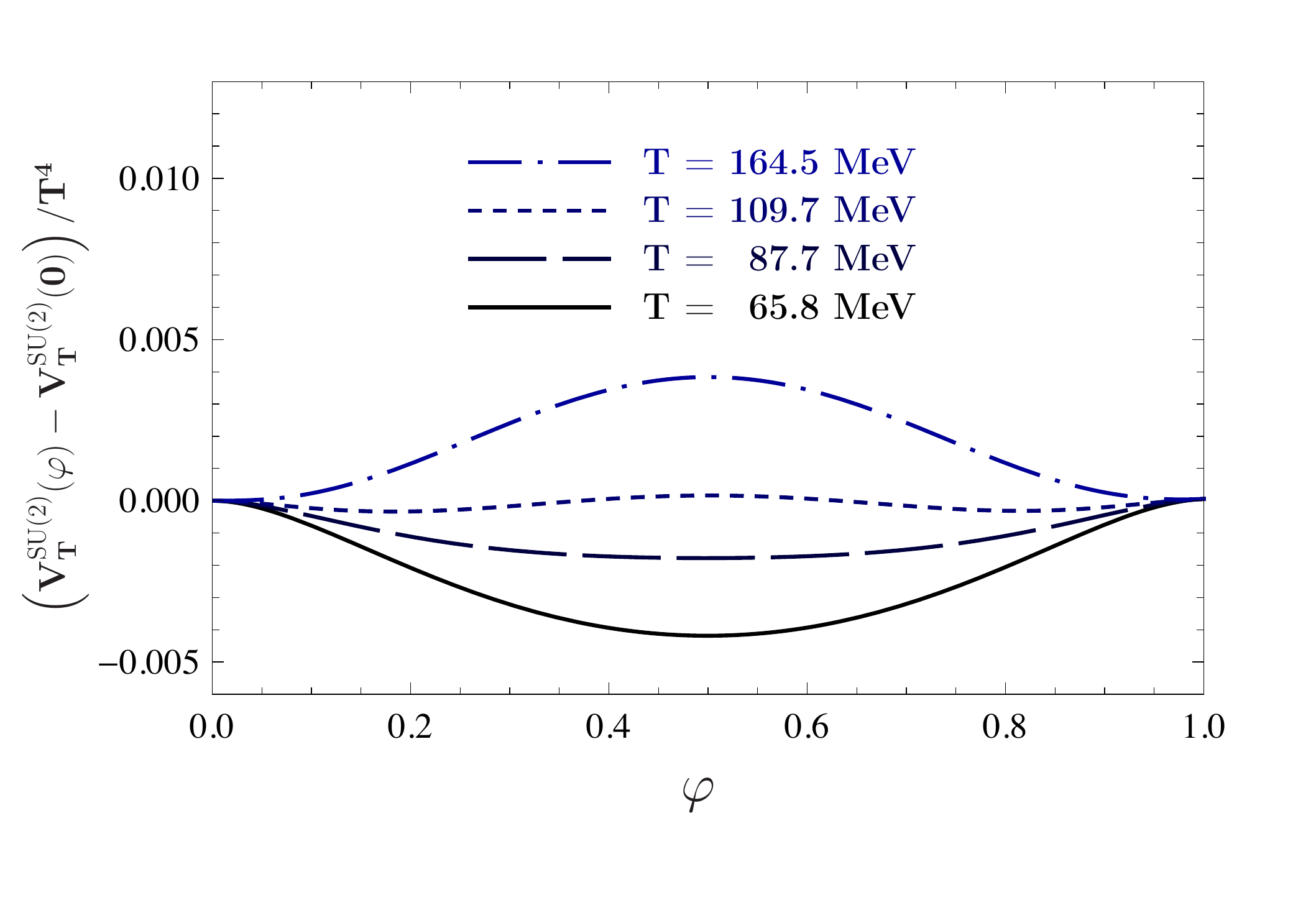}
\caption{The Polyakov loop potential from its DSE for mildly
  suppressed transversal gluons and trivial ghosts shows confinement
  for sufficiently low temperatures.}
\label{fig:DSE_mild_gluon_supp}
\efc
In this section we confirm the hypothesis made in
section~\ref{sec:confcrit} in a numerical application for both
approaches DSEs and the FRG. In these computations the ghosts are left
trivial, thus, they give a confining potential which overcompensates
the gauge mode, and both terms sum up to $-1/2 V_{\rm Weiss}$. We
show, that already a soft suppression of transversal gluons yields a
decrease of the deconfining effect such, that confinement is immanent
at sufficiently low temperatures.

For the DSE we choose an ansatz, in which transversal gluons are
suppressed for infrared momenta, thus, 
\beq
 \Gamma^{(2)}_a(p^2)= p^2 \left(
  1+\left(p^2\right)^{\kappa_a}\right)\,, {\rm with} \quad \kappa_a
=-0.2\,.  
\eeq
The full result is given in \Fig{fig:DSE_mild_gluon_supp}, where
clearly a confining potential is realised at low temperatures.
\begin{figure}[h!]\begin{center}
\includegraphics[width=.95\columnwidth]{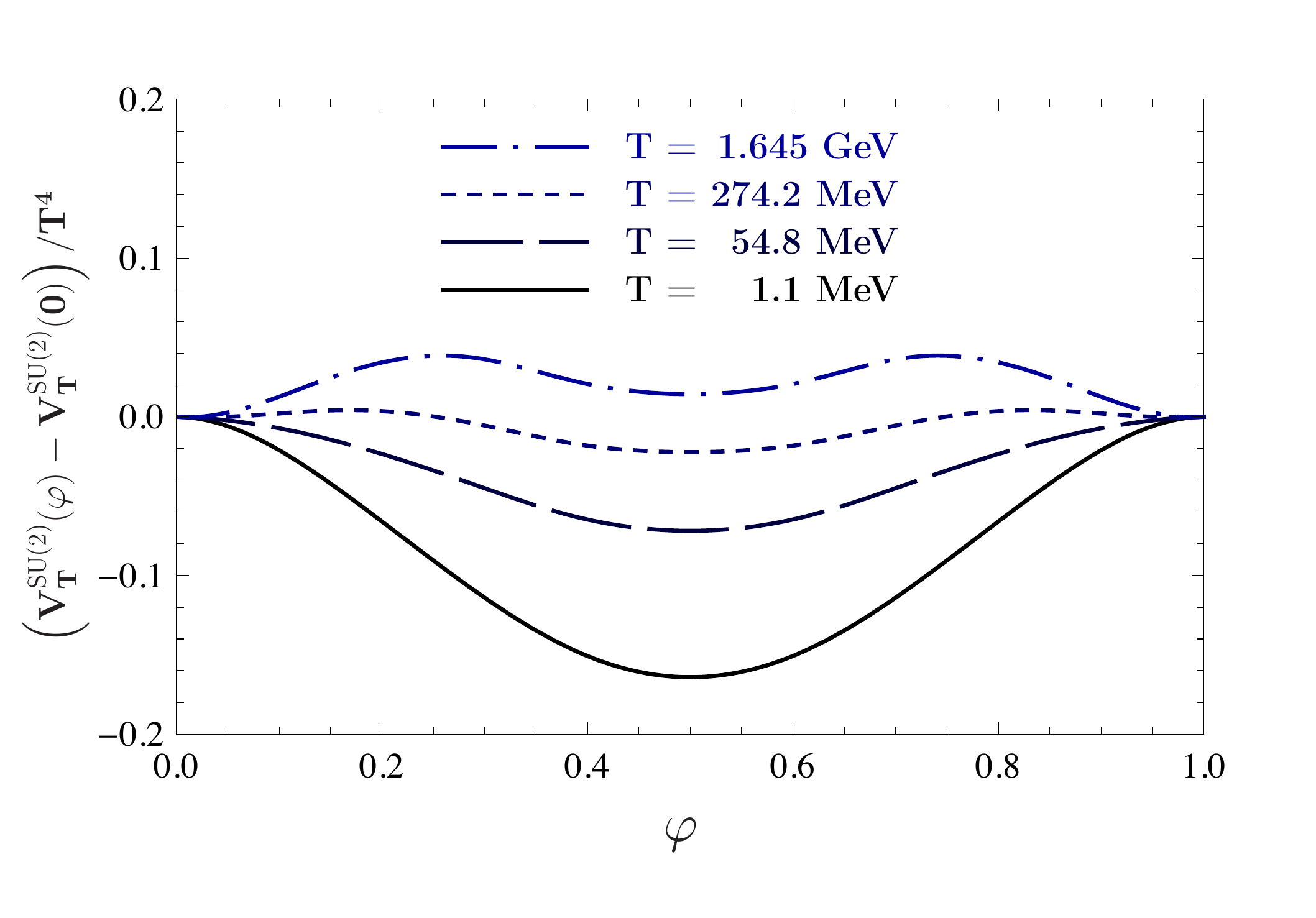}
\caption{The Polyakov loop potential from its FRG equation for mildly
  suppressed transversal gluons and trivial ghosts shows confinement
  for sufficiently low temperatures.}
\label{fig:FRG_mild_gluon_supp}
\efc

In the case of the FRG the suppression must also be present in the
renormalisation group flow along the scale $k$. Thus, a qualified
ansatz is given by 
\beq {\Gamma^{(2)}_{a,k}}(p^2)= \left(p^2 +
  k^2\right)^{1+\kappa_a} \,, {\rm with} \quad \kappa_a =-0.2\,.
\eeq 
The result for this ansatz is given in \Fig{fig:FRG_mild_gluon_supp}.

From these two examples we can infer, that indeed a mild gluon
suppression is sufficient to yield a confining potential at low
temperatures, as we claimed in
section~\ref{sec:confcrit}. Nevertheless, the sum of gluonic, ghost
and gauge modes is non-trivial, thus, we refrain from giving phase
transition temperatures here, as these examples only serve as a proof
of principle and are of no quantitative physical relevance.

\section{Direct resummation of the DSE-hierachy}
\label{app:2loop-DSE-rep}
\begin{figure}[t]
\begin{center}
\includegraphics[width=.22\columnwidth]{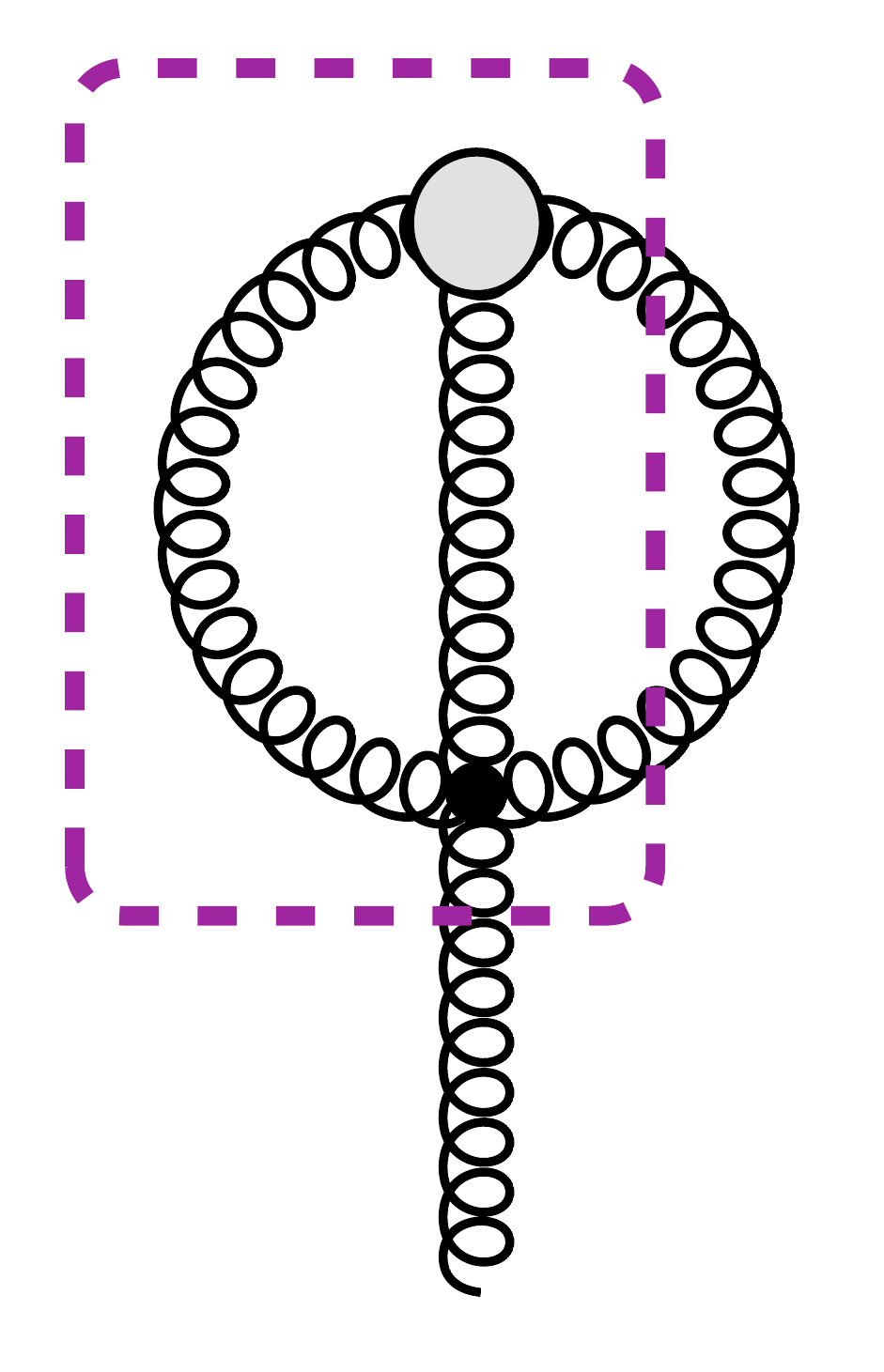}
\caption{Three gluon vertex correction from the two-loop diagram}
\label{fig:3glue-vertexcorrection}
\end{center}
\end{figure}
\begin{figure}[b]
\begin{center}
\includegraphics[width=1\columnwidth]{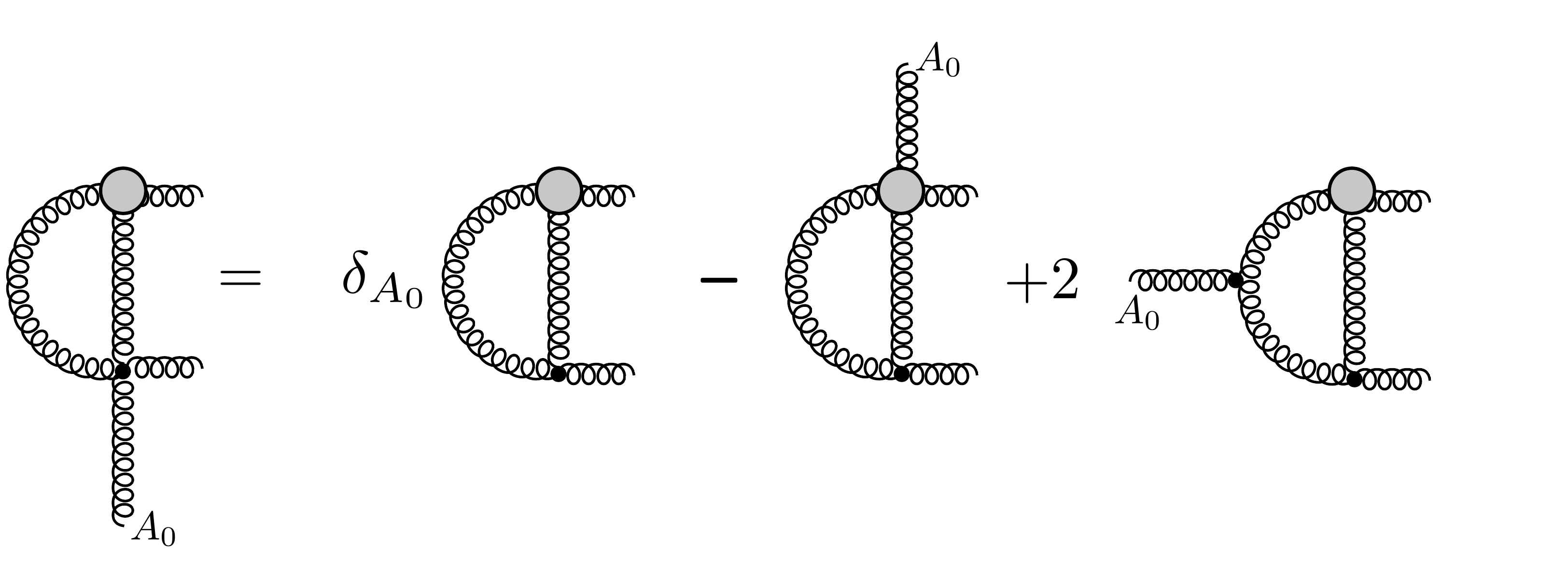}
\caption{Approximation of the substructure of the dashed box in
  \Fig{fig:3glue-vertexcorrection}.}
\label{fig:three_gluon_trafo}
\efc
\begin{figure}[t]
\begin{center}
\includegraphics[width=.82\columnwidth]{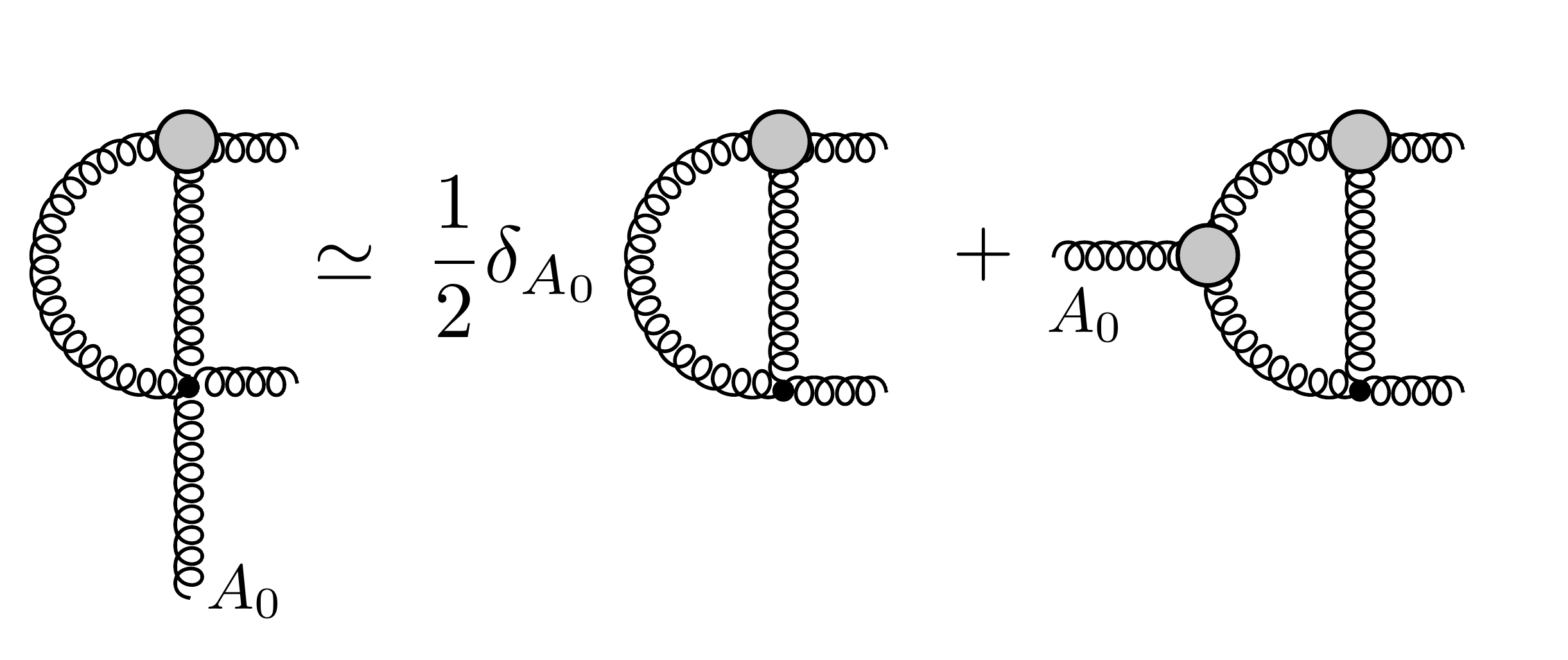}
\caption{Approximation for the three-gluon vertex of \Fig{fig:three_gluon_trafo}.}
\label{fig:three_gluon_trafo_approx}
\efc
%
In this appendix we provide the necessary steps for deriving
\eq{eq:2PI-2loop} solely within the DSE-approach. Note that this
simply amounts to using the ghost and gluon gap equations in the DSE
hierarchy which is trivially implemented in the 2PI-approach. To that
end we have to rewrite the two-loop diagrams in \eq{eq:DSE},
\Fig{fig:DSE}. Similarly to the 2PI argument leading to
\eq{eq:2PI-2loop} this diagram can be turned into a one-loop form
within a resummation scheme which neglects some order
$\alpha_s^2$-corrections to full vertices but takes into account full
propagators. In the following we explicitly perform this analysis for
the gluon two-loop diagram, the same steps can be trivially taken over
for the ghost two-loop diagram: First, we note that the two-loop
diagram has no contributions from the gauge fixing vertices
proportional to $1/\xi$ which contributes in the one-loop diagram, see
Appendix~\ref{app:one-loop}. The reason is that the full and
renormalised classical vertices involve more than 2 fluctuating fields
$a$. Moreover, the two-loop contribution can be seen as a
vertex-correction diagram to the gluonic one-loop diagram, see
\Fig{fig:3glue-vertexcorrection}.
%
\begin{figure}[t]
\begin{center}
\includegraphics[width=.85\columnwidth]{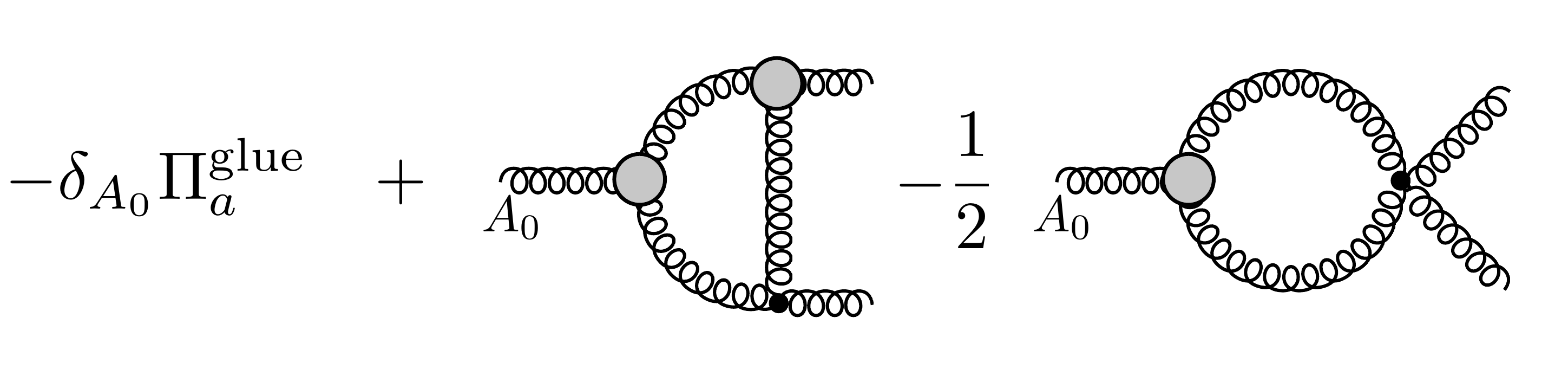}
\caption{Identification of the gluon vacuum polarisation in the two-loop diagram of the DSE.}
\label{fig:three_gluon_trafo_approx_self_energy}
\efc
%

%
\begin{figure}[b]
\begin{center}
\includegraphics[width=.95\columnwidth]{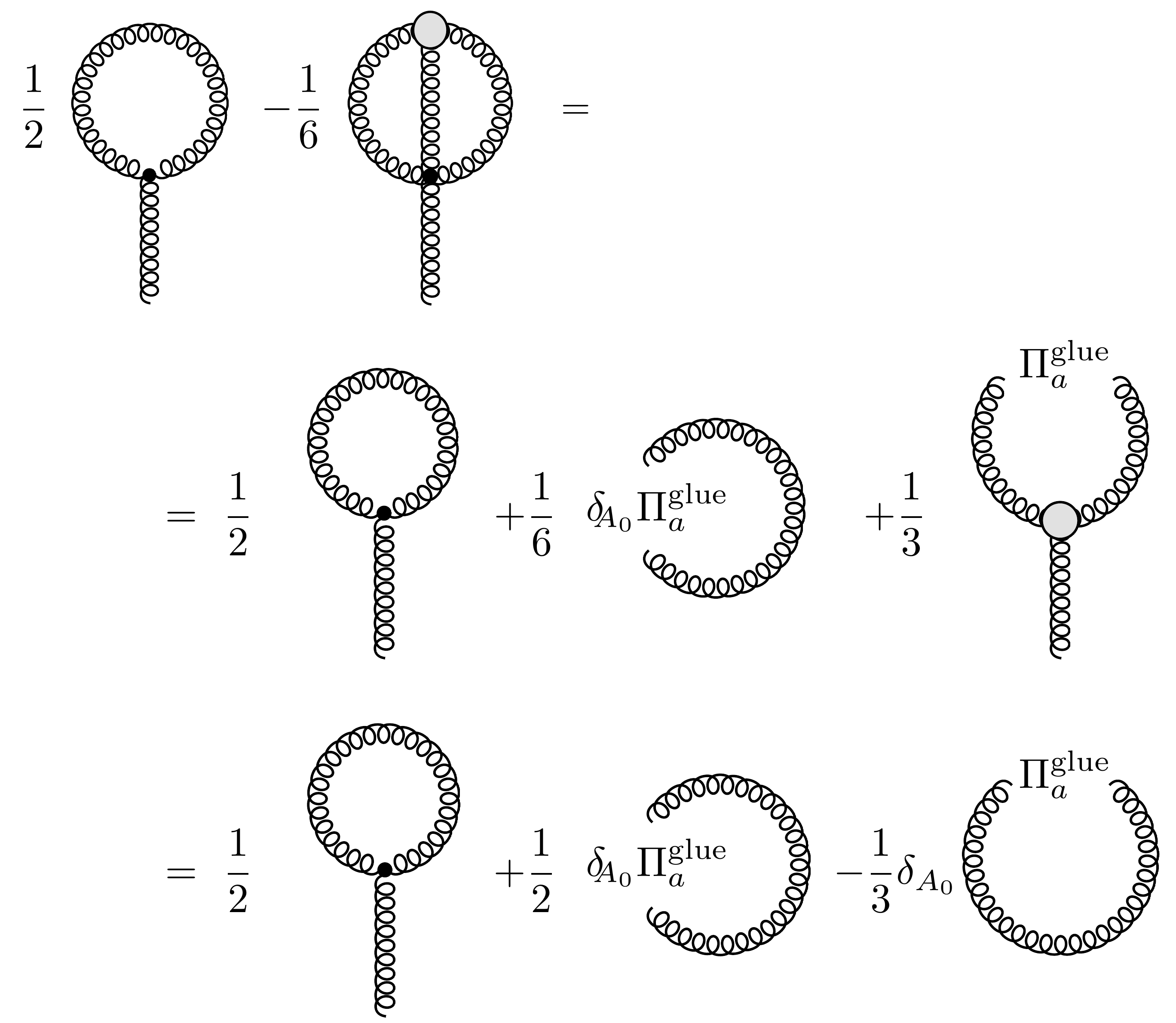}
\caption{Approximation of glue part of the DSE for the Polyakov loop potential.}
\label{fig:DSE_corr_full}
\efc
%

The vertex correction can be rewritten in terms of a total
$A_0$-derivative, a three-gluon vertex diagram and a term which
resembles (minus) the vertex correction, see
\Fig{fig:three_gluon_trafo}. The only difference is that the $A_0$ leg
is attached to the full vertex. Since the two fluctuating fields are
attached to the additional propagator in the two-loop diagram, the
original diagram and this one differ in order $\alpha_s^2$ in the
vertex. Hence, in the present approximation we identify it with (minus)
the vertex correction. We also note that strictly speaking the last
diagram in \Fig{fig:three_gluon_trafo} only involves the $A_0$-vertex
originating in $S_A$. However, the gauge fixing part does not
contribute anyway as the vertex is contracted with two transversal
propagators. Hence, for the sake of simplicity we will always use the
full $A_0$-vertices, but the gauge-part is projected out in the diagrams.

Moving the vertex correction to the left hand side side leads to the form
\Fig{fig:three_gluon_trafo_approx} of the vertex correction, up to the 
vertex correction-terms of order $\alpha_s^2$. The
first term on the right hand side of \Fig{fig:three_gluon_trafo_approx} is (minus)
the three-gluon vertex part of the gluonic part $\Pi_a^{\rm glue}$ of the gluon vacuum
polarisation  $\Pi_a$, see \Fig{fig:gluonvacpol}. We
rewrite it accordingly and the right hand side takes the form displayed in 
\Fig{fig:three_gluon_trafo_approx_self_energy},
\begin{equation}\label{eq:vertcorrect} 
 -\013 \partial_{A_0} \Pi_a^{\rm glue} + {\rm diagrams}\,.
\end{equation}
Upon closing the open $a$-lines in
\Fig{fig:three_gluon_trafo_approx_self_energy} with the remaining
gluon propagator the corresponding one-loop sub-diagrams constitute
$\Pi_a^{\rm glue}$ up to a factor $1/2$ in the tadpole diagram. As for
the 2PI-derivation we drop this diagram. Inserting
\Fig{fig:three_gluon_trafo_approx_self_energy} into
\Fig{fig:3glue-vertexcorrection} we are led to
\Fig{fig:DSE_corr_full}. 
%
\begin{figure}[b]
\begin{center}
\includegraphics[width=.95\columnwidth]{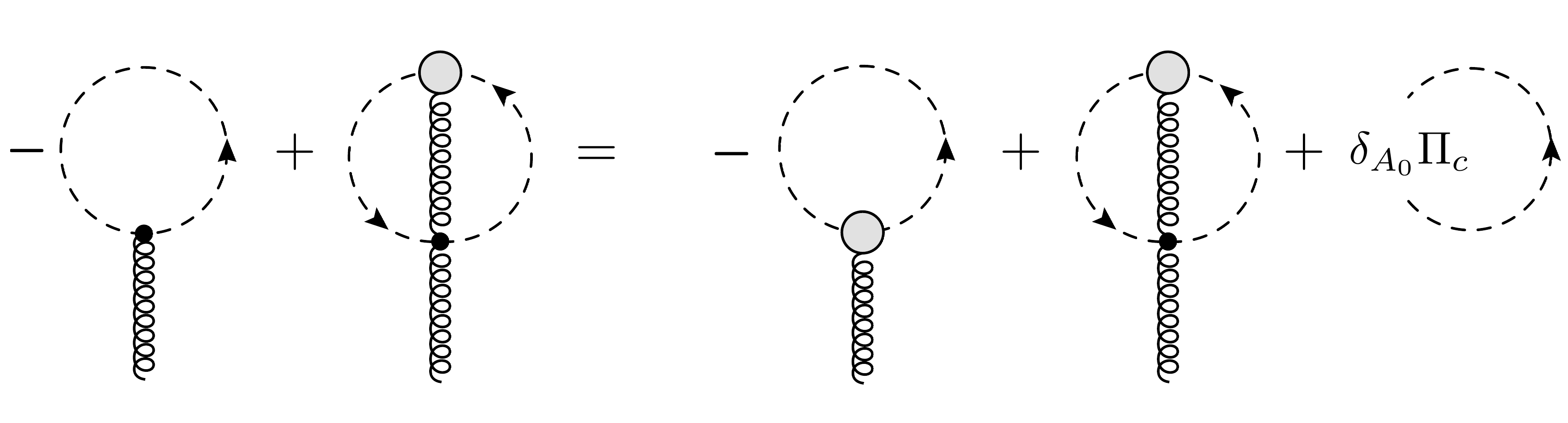}
\caption{Approximation of ghost part of the DSE for the Polyakov loop potential.}
\label{fig:DSE_ghost_corr}
\efc
%

Similar steps as above can be done for the ghost two-loop
diagram. Here we simply note that the two ghost terms in the DSE
\Fig{fig:DSE} can be rewritten as displayed in \Fig{fig:DSE_ghost_corr}. 
The first term on the rhs in \Fig{fig:DSE_ghost_corr} is the
$A_0$-derivative of the $\Tr \log G_c$-term in \eq{eq:2PI-2loop}.  In turn, the two
remaining terms equal 
\begin{equation}
\0{\delta}{\delta \bar A_0} \left[-\023 \left(\012 \Tr \,\Pi_a^{\rm ghost} G_a- 
 \Tr \,\Pi_c G_c\right)+O(\alpha_s^2)\right]\,.  
\end{equation}
This is most easily seen by just comparing the perturbative 
two-loop diagrams in both expressions.  

In summary the gluonic and ghost terms in the DSE add up to the total
$A_0$-derivative \eq{eq:2PI-2loop}. 

\bibliographystyle{bibstyle}
\bibliography{funPolpot}

\end{document}